\documentclass[twocolumn,aps,pra,10pt,floatfix,nofootinbib]{revtex4-1}

	\usepackage[usenames,dvipsnames]{xcolor}
	\usepackage{amsmath}
	\usepackage{amsfonts}
	\usepackage{amssymb}

	\usepackage{graphicx}
	\usepackage{bbold}		
	\usepackage[makeroom]{cancel}	
	\usepackage{multirow}			
	\usepackage[normalem]{ulem}     
	\usepackage{array}
	
	\usepackage{siunitx}
    \usepackage{verbatim}
    \usepackage{placeins}
    
    \usepackage[colorlinks=true,citecolor=blue,linkcolor=red]{hyperref}    
    
    \usepackage{bibunits}

	\newcolumntype{x}[1]{>{\centering\let\newline\\\arraybackslash\hspace{0pt}}p{#1}}

	\DeclareMathOperator{\tr}{tr}  		
	\DeclareMathOperator{\diag}{diag}  	
	\DeclareMathOperator*{\ordprod}{\overline{\prod}}
	
	\DeclareMathAlphabet{\mathbbold}{U}{bbold}{m}{n}
	\def\abs#1{\left|{#1}\right|}  	
	\def\bs#1{\boldsymbol{#1}}		
	\def\imi{\mathrm{i}}			
	\def\imj{\mathrm{j}}			
	\def\imk{\mathrm{k}}			
	\def\e#1{\mathrm{e}^{#1}}		
	\def\de{\mathrm{d}}
	
	\def\eps{\varepsilon}			
 
	\def\mcH{\mathcal{H}}		
	\def\mcbsA{\boldsymbol{\mathcal{A}}}	
	\def\mcA{\mathcal{A}}		
	\def\mcF{\mathcal{F}}		
	\def\mcT{\mathcal{T}}		
	\def\mcP{\mathcal{P}}		
	\def\mcC{\mathcal{C}}		
	\def\mcI{\mathcal{I}}		
	\def\mcK{\mathcal{K}}		

	\def\mcW{\mathcal{W}}		
	
	\def\intg{\mathbbold{Z}}	
	\def\ztwo{\mathbbold{Z}_2}	
	\def\triv{\mathbbold{0}}	
	\def\unit{\mathbbold{1}}	
	\def\reals{\mathbbold{R}}	
	\def\cmplx{\mathbbold{C}}	
	
	\def\bra#1{\left<{#1}\right|}	
	\def\ket#1{\left|{#1}\right>}	

	\DeclareMathOperator{\im}{im}
	
	\newcounter{subeqn} %
	\makeatletter
	\@addtoreset{subeqn}{equation}
	\makeatother

\definecolor{TB}{rgb}{1,0.5,0}

\def\AS#1{{}}
\def\TB#1{{}}
\def\orig#1{{}}

\def\edit#1{{\color{Black}#1}}

\defaultbibliography{bib}
\defaultbibliographystyle{apsrev4-1}

\begin{document}
\begin{bibunit}
\title{Non-Abelian band topology in noninteracting metals}

\author{QuanSheng Wu$^{1,2}$}
\author{Alexey A. Soluyanov$^{3,4}$}
\author{Tom\'{a}\v{s} Bzdu\v{s}ek$^{5,6}$}\email[Corresponding author: ]{bzdusek@stanford.edu}   

\affiliation{$^{1}$Institute of Physics, \'{E}cole Polytechnique F\'{e}d\'{e}rale de Lausanne, CH-1015 Lausanne, Switzerland}
\affiliation{$^{2}$National Centre for Computational Design and Discovery of Novel Materials MARVEL, Ecole Polytechnique F\'{e}d\'{e}rale de Lausanne (EPFL), CH-1015 Lausanne, Switzerland}
\affiliation{${^3}$Physik-Institut, Universit\"at Z\"urich, Winterthurerstrasse 190, CH-8057 Zurich, Switzerland}
\affiliation{$^{4}$Department of Physics, St. Petersburg State University, St. Petersburg, 199034 Russia}
\affiliation{$^{5}$Department of Physics, McCullough Building, Stanford University, Stanford, CA 94305, USA}
\affiliation{$^{6}$Stanford Center for Topological Quantum Physics, Stanford University, Stanford, CA 94305, USA}

\date{\today} 

\begin{abstract}
\vspace{-0.1cm}
Electron energy bands of crystalline solids generically exhibit degeneracies called band-structure nodes. Here, we introduce non-Abelian topological charges that characterize line nodes inside the momentum space of crystalline metals with space-time inversion (\texorpdfstring{$\mathcal{PT}$}{PT}) symmetry and with weak spin-orbit coupling. We show that these are quaternion charges, similar to those describing disclinations in biaxial nematics. Starting from two-band considerations, we develop the complete many-band description of nodes in the presence of \texorpdfstring{$\mathcal{PT}$}{PT} and mirror symmetries, which allows us to investigate the topological stability of nodal chains in metals. The non-Abelian charges put strict constraints on the possible nodal-line configurations. Our analysis goes beyond the standard approach to band topology and implies the existence of one-dimensional topological phases not present in existing classifications.
\end{abstract}

\maketitle

\textbf{\textsf{Introduction.}} ---  Nodal-line metals~\cite{Chen:2015,Kim:2015,Yu:2015,Bian:2015a,Chan:2016,Schoop:2016,Fang:2016,Fang:2015,Bzdusek:2017,Ahn:2018,Burkov:2011} and nodal-chain metals~\cite{Bzdusek:2016,Wang:2017,Yu:2017,Feng:2018,Yi:2018,Gong:2018,Heikkila:2015b,Zhu:2016,Yan:2018} are crystalline solids that exhibit line degeneracies of electron energy bands near the Fermi energy. Although a large variety of such metals have been discussed to date, many of their properties remain unknown. Here, we find that a non-Abelian charge (\emph{i.e.}~topological invariant) of nodal lines (NLs) in the momentum ($\bs{k}$) space of metals with weak spin-orbit coupling (SOC) in the presence of composed time-reversal ($\mcT$) and inversion ($\mcP$) symmetry. This non-Abelian topology in $\bs{k}$-space is fundamentally different from the non-Abelian exchange statistics of anyon quasiparticles in the coordinate space. It arises in the absence of interactions and superconductivity, and governs the evolution of NLs in $\bs{k}$-space. In particular, the non-Abelian charge implies constraints on admissible NL compositions, including \emph{chains} of intersecting NLs. We further find that materials hosting these topological excitations provide a $\bs{k}$-space analog of biaxial nematic liquid crystals, which exhibit non-Abelian vortex lines in coordinate space~\cite{Madsen:2004,Kleman:1977,Mermin:1979}.
Based on \emph{ab initio} calculations, we predict that the discussed phenomena can be observed in existing materials, where the NL locations in $\bs{k}$-space are manipulated by strain. This is illustrated with the example of elemental scandium ($\textsf{Sc}$). The manuscript covers our main results, while additional details and mathematical considerations are provided in the Supplemetary Information File (SIF)~\cite{Supp}.

\textbf{\textsf{Nodal lines in two-band models.}} --- We first consider NLs in two-band models. While this problem was already addressed in the works of Refs.~\cite{Chen:2015,Kim:2015,Yu:2015,Bian:2015a,Chan:2016,Schoop:2016,Fang:2016,Fang:2015,Bzdusek:2017,Ahn:2018,Burkov:2011,Bzdusek:2016,Wang:2017,Yu:2017,Feng:2018,Yi:2018,Gong:2018,Heikkila:2015b,Zhu:2016,Yan:2018}, we use it here to set the stage for the discussion of non-Abelian topology in later sections.

Two-band Hamiltonians can be decomposed using the identity $\mathbb{1}$ and the Pauli matrices $(\sigma_x,\sigma_y,\sigma_z) = \bs{\sigma}$ as
\begin{equation}
    {\cal{H}}_2({\bs{k}})= h_0(\bs{k}) \mathbb{1} + {\bs{h}}({\bs{k}})\cdot {\bs{\sigma}} 
\label{ham2PTM}
\end{equation}
where $\bs{h}(\bs{k}) = (h_x(\bs{k}),h_y(\bs{k}),h_z(\bs{k}))$ are real functions of $\bs{k}=(k_x,k_y,k_z)$. In the absence of SOC we set $\mcP\mcT = \mcK$ (complex conjugation), which removes $\sigma_y$ from Eq.~(\ref{ham2PTM}), making both $\mcH_2(\bs{k})$ and its eigenstates \emph{real}~\cite{Yu:2015}. The Hamiltonian exhibits a NL at $\bs{k}$ if two conditions are fulfilled~\cite{Fang:2015}, namely $h_z(\bs{k}) =0$ and  $h_x(\bs{k}) = 0$.

\begin{figure*}[t]
	\includegraphics[width=0.64\textwidth]{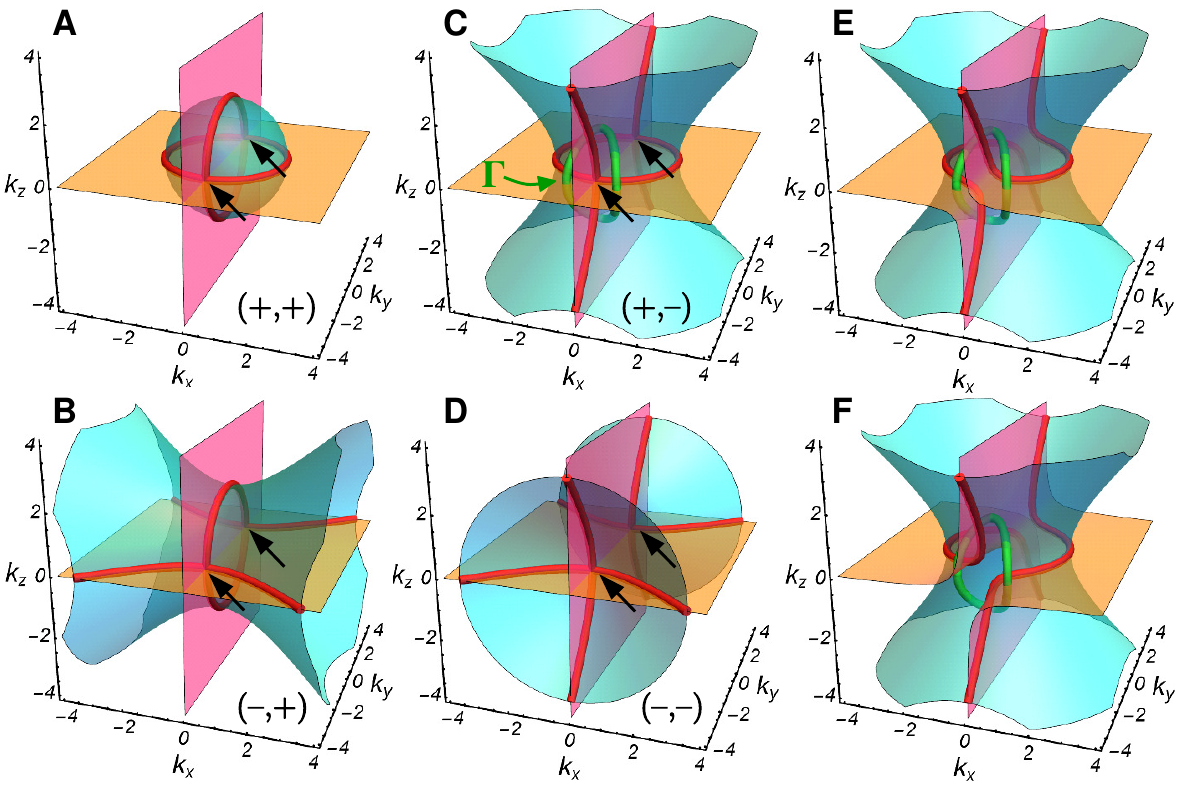}
	\caption{\textsf{\textbf{Nodal lines (NLs) in two-band models.} 
	Red~li\-nes indicate NLs of Hamiltonians of the form in Eq.~(\ref{ham2PTM}). (\textbf{A--D})~For the Hamiltonian in Eq.~(\ref{eqn:gallery-Ham}) (with the indicated choice of $\pm$ signs~in the expression for $h_z(\bs{k})$), NLs intersect at crossing points (black arrows), protected by a winding number $\abs{w_\Gamma} \!=\! 2\!$ on path $\Gamma$ (green) shown in panel (C). (\textbf{E} and \textbf{F}) Upon breaking of the~mirror symmetry, invariant $w_\Gamma\!$ enforces a non-trivial separation of NLs into ones that bend from vertical to horizontal.
	}}
	\label{fig:gallery}
\end{figure*} 

To uncover the topological structure stabilizing these NLs, one needs to consider the space of available Hamiltonians~\cite{Bzdusek:2017}. To emphasize an analogy with the theory of defects in ordered media~\cite{Mermin:1979}, here we call it the \emph{order-parameter space}. For later convenience, it is useful to encode Hamiltonians using their eigenstates. Assuming a $\bs{k}$-point that does not lie on a NL, we normalize the spectrum of the Hamiltonian in Eq.~(\ref{ham2PTM}) to $\pm 1$ by taking
\begin{equation}
{\cal{H}}_2({\bs{k}})=\mathbb{1}-2|u^\mathrm{o}_{\bs{k}}\rangle \langle u^\mathrm{o}_{\bs{k}}|.
\label{ham2PT}
\end{equation}
where $|u^\mathrm{o}_{\bs{k}}\rangle$ is the cell-periodic amplitude of the lower-energy Bloch state.~Since this is a normalized two-component real vector, the order-parameter space is a \emph{circle} ($\,S^1$).~To be precise, one should note that both $\pm|u^\mathrm{o}_{\bs{k}}\rangle$ encode the \emph{same} Hamiltonian.~However, removing this redundancy by identifying antipodal points of the $S^1\!$ still produces an $S^1\!$.~Closed paths $\Gamma$ in $\bs{k}$-space that avoid the NLs are characterized by the elements of the \emph{fundamental group}~\cite{Bzdusek:2017,Mermin:1979,Supp} of the order-parameter space,
\begin{equation}
    w_\Gamma \in \;\pi_1(S^1)=\mathbb{Z}.
\label{eq:Z}
\end{equation}
By taking a path $\Gamma$ that tightly encircles a NL, we can assign an integer \emph{winding number} $w_\Gamma$~\cite{Burkov:2011} to the NL. Topological charges described by homotopy groups are \edit{calculated directly from the Hamiltonian, and thus do not depend on the gauge of the eigenstates. However, in contrast to Wilson operators, the calculation of these topological charges requires fixing the basis of the underlying Hilbert space~\cite{Note1}.} 

\textbf{\textsf{Nodal chains in two-band models.}} ---  Before generalizing to models with more bands, we discuss the effects of one mirror symmetry on NLs in two-band models. While reproducing some findings of Refs.~\cite{Bzdusek:2016,Wang:2017,Yu:2017,Feng:2018,Yi:2018,Gong:2018,Heikkila:2015b,Zhu:2016,Yan:2018}, our discussion also contains original results concerning the topological stability of intersecting NLs. We will observe in the next section that some of these results are substantially altered in the presence of additional bands.

The presence of mirror symmetry $m_z \!:\! z\!\rightarrow \!-z$ represented by $\hat{m}_z \!=\! \sigma_z$ in the model of Eq.~(\ref{ham2PTM}) forces $h_{x(z)}(\bs{k})$ to be an odd (even) function of $k_z$~\cite{Yan:2018}. For example~\cite{a_Note2}
\begin{equation}
h_x(\bs{k}) = k_x k_z \quad\textrm{and}\quad h_z(\bs{k}) = \pm k_x^2 +k_y^2 \pm k_z^2 - b^2,\label{eqn:gallery-Ham}
\end{equation}
with $b = 2$ produces the NLs shown in Fig.~\ref{fig:gallery}(A--D)~\cite{Feng:2018}. They all exhibit \emph{crossing points} (CPs) of intersecting NLs. Expanding Eq.~(\ref{eqn:gallery-Ham}) around the CPs gives
\begin{equation}
h_x(\bs{k}) \approx k_x  k_z \qquad\textrm{and}\qquad h_z( \bs{k}) \approx \pm  k_y, 
\label{eqn:Ham-kp-expansion}
\end{equation}
which describe a pair of mutually perpendicular intersecting NLs along $ k_x =  k_y = 0$ and $ k_y =   k_z = 0$.

\begin{figure*}[t]
	\includegraphics[width=0.63\textwidth]{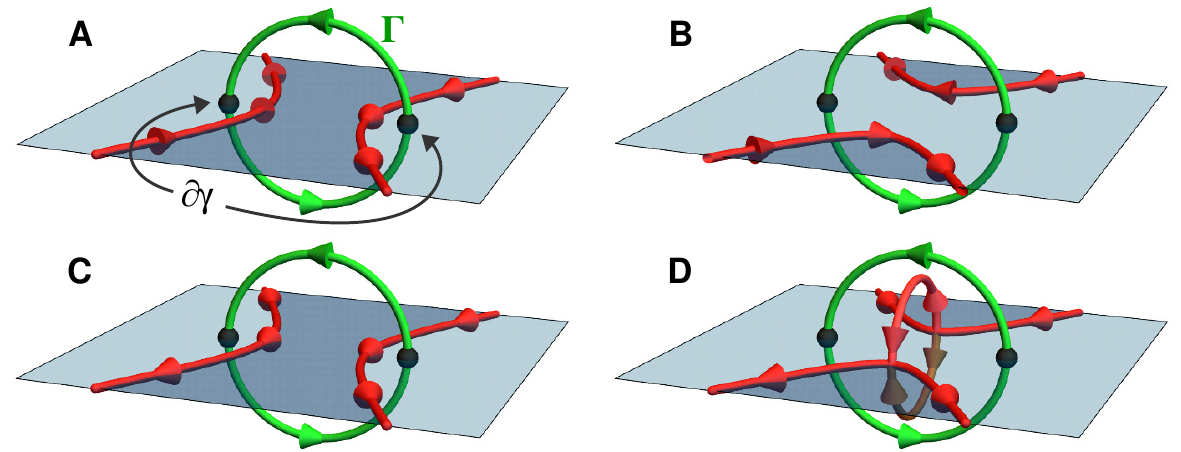}
	\caption{\textsf{\textbf{Formation of intersecting nodal lines (NLs).} Red~lines indicate NLs of two-band Hamiltonians with a horizontal mirror-symmetry plane. Two shades of blue, separated by in-plane NLs, indicate regions with different mirror eigenvalue $\lambda^\textrm{o}_{\bs{k}} \!=\! \pm 1$  of the lower band, allowing us to define a $\ztwo$ charge. (\textbf{A}) Mirror-symmetric path $\Gamma$ (green) encircles two in-plane NLs with \emph{opposite} orientation, and crosses the plane at two points $\partial \gamma$ (black dots) with the same $\lambda^\textrm{o}_{\bs{k}}$. The pair of points $\partial \gamma$ carry a trivial value of the $\ztwo$ charge, and $\Gamma$ carries zero value of the $\intg$ winding. Therefore, (\textbf{B}) in-plane NLs can trivially reconnect. (\textbf{C}) For an analogous setting with two \emph{parallel} in-plane NLs, the $\intg$ winding becomes $\abs{w_\Gamma}\!=\! 2$, indicating that an obstruction must remain for shrinking path $\Gamma$ to a point even \emph{after} the reconnection of in-plane NLs takes places. (\textbf{D}) This implies the appearance of two out-of-plane NLs that are connected to the in-plane NLs.}}
	\label{fig:windings-2} 
\end{figure*}

Geometrically, we interpret the formation of intersecting NLs by looking at Fig.~\ref{fig:gallery}: a NL is produced when a cyan sheet defined by $h_z(\bs{k}) \!=\! 0$ crosses the orange sheet of $h_x(\bs{k}) \!=\! 0$, which due to $m_z$ is the plane $k_z\!=\! 0$. Intersections of cyan and orange sheets produce \emph{in-plane} NLs. The existence of additional NLs that vertically cross the plane $k_z \!=\! 0$ requires that $h_x(\bs{k}) \!=\!  0$ also on a pink sheet orthogonal to the plane. Such a pink sheet appears in the model of Eq.~(\ref{ham2PTM}) when the product $k_z h_x(\bs{k})$ changes sign. The \emph{out-of-plane} NLs correspond to intersections of cyan and pink sheets. The CPs of in-plane and out-of-plane NLs correspond to \emph{three-sheet} intersections. Breaking $m_z$ leads to {mixing} of the orange and pink sheets, and causes a non-trivial separation of the CPs, as previously observed in Refs.~\cite{Ahn:2018,Yan:2018}. We illustrate this for $h_x(\bs{k}) \!=\! k_x k_z \!-\! \tfrac{1}{10}$ and $h_x(\bs{k}) \!=\! k_x k_z \!-\! \tfrac{k_y}{20}$ in Fig.~\ref{fig:gallery}(E--F).

In this work, we explain the stability of CPs \emph{topologically} using the relative homotopy approach of Ref.~\cite{Sun:2018}. The $m_z$ symmetry reduces the space of available Hamiltonians (with normalized spectrum) inside the $k_z \!=\! 0$ plane to only two points, namely $\pm \sigma_z$.~In-plane NLs separate regions with different $\hat{m}_z$ eigenvalues of $\ket{u^\mathrm{o}_{\bs{k}}}$~\cite{Chan:2016}, denoted $\lambda_{\bs{k}}^\textrm{o}$. This allows us to define a $\{+1,-1\} \cong \ztwo$ topological number $\nu_{\bs{k}_1,\bs{k}_2} = \lambda^\textrm{o}_{\bs{k}_1} \!\cdot \lambda^\textrm{o}_{\bs{k}_2}$ for any pair of in-plane momenta $\bs{k}_{1,2}$~\cite{Fang:2016}. However, tracking the sign of $h_{x,z}(\bs{k})$ on a loop encircling an in-plane NL reveals that it also carries a non-trivial $\intg$ charge defined by Eq.~(\ref{eq:Z}), \emph{i.e.} it has an orientation. Importantly, \emph{open-ended} paths $\gamma$ with end-points $\partial \gamma$ inside the plane $k_z \!=\! 0$ can be assigned a winding number $w_\gamma \in \intg$ too, provided that $\nu_{\partial\gamma} = +1$~\cite{Sun:2018}. The closed composition of path $\gamma$ with its mirror image $(m_z \gamma)^{-1}$, denoted $\Gamma$, carries an even winding $w_\Gamma = 2 w_\gamma$. In the presence of mirror-symmetry, CPs are protected by $\abs{w_\gamma} = 1$ on a semicircular path $\gamma$ enclosing the CP. The winding number $\abs{n_\Gamma} = 2$ on the \emph{closed} path $\Gamma$ remains meaningful even when the mirror symmetry is broken, and enforces the non-trivial separation of the CP seen in Fig.~\ref{fig:gallery}(E--F). The mismatch~\cite{Sun:2018} between the $\intg$ winding number  $w_\Gamma$ and the $\ztwo$ invariant $\nu_{\partial \gamma}$ leads to conditions for creating a nodal chain by colliding in-plane NLs, as illustrated in Fig.~\ref{fig:windings-2}. The relative orientation of NLs near a CP always follows the pattern observed in Fig.~\ref{fig:windings-2}(D).

\textbf{\textsf{Nodal chains in many-band models.}} --- As a first step towards a general many-band description, we demonstrate that the presence of additional bands modifies the topological stability of CPs. We illustrate this phenomenon by studying the spectrum of a Hamiltonian
\begin{equation}
\mcH_3(\bs{k}) = \left(\begin{array}{ccc}
E_0 		     	&	tk_x	& tk_z		\\
tk_x				&	k_y		& k_x k_z	\\
tk_z				&	k_x k_z	& -k_y
\end{array}\right),\label{eqn:3-band-detach}
\end{equation}
which augments the two-band model of Eq.~(\ref{eqn:Ham-kp-expansion}) with an additional orbital with energy $E_0$, coupled to the two original orbitals  with amplitudes proportional to $t$ (we set $t = \tfrac{1}{2}$ throughout). The Hamiltonian of Eq.~(\ref{eqn:3-band-detach}) has a mirror symmetry $\hat{m}_z = \diag(1,1,-1)$~\cite{a_Note2}.

Assuming first that $E_0 > 0$, the NLs formed by the \emph{lower} two bands (i.e.~the original ones) of the Hamiltonian of Eq.~(\ref{eqn:3-band-detach}) take the form of the red intersecting lines in Fig.~\ref{fig:preserved-TP}(A). These NLs interesect at a CP, similar to the two-band discussion around Eqs.~(\ref{eqn:gallery-Ham}--\ref{eqn:Ham-kp-expansion}). However, decreasing $E_0$ to a \emph{negative} value results in a \emph{trivial separation} of these NLs (red lines in Fig.~\ref{fig:preserved-TP}(C)). Such a separation is impossible in two-band models. Nevertheless, the CP did not vanish completely, but it now connects NLs formed by the \emph{upper} two bands (blue lines in Fig.~\ref{fig:preserved-TP}(A--C)). The transfer of the CP to another pair of bands occurs at a topological transition at $E_0 = 0$ (Fig.~\ref{fig:preserved-TP}(B)). The description of such a process requires a mathematical framework capable of capturing NLs between \emph{both} pairs of bands simultaneously, and goes beyond the established ``tenfold way'' classification of topological insulators and superconductors~\cite{Kitaev:2009,Ryu:2010} based on $K$-theory~\cite{Horava:2005}.

\textbf{\textsf{Quaternion charges in many-band models.}} --- Motivated by the observations of the previous section, we develop the complete many-band description of NLs in $\mcP\mcT$-symmetric models with weak SOC. We first formally derive the non-Abelian topology, while the next section discusses the implications to NL compositions.

For $\bs{k}$-points with non-degenerate spectrum, we deform $N$-band Hamiltonian $\mcH_N(\bs{k})$ such that it exhibits some standard set of band energies (e.g. $\eps_j = j$ for $1 \leq j \leq N$; we assume $N \geq 3$). This generalizes Eq.~(\ref{ham2PT}) to
\begin{equation}
\mcH_N(\bs{k}) = \sum_{j=1}^N \eps_j  |{u^j_{\bs{k}}}\rangle\langle{u^j_{\bs{k}}}|.
\label{eqn:russkiy-standart} 
\end{equation}
Such a Hamiltonian is uniquely encoded by a \emph{frame} $\{|u^j_{\bs{k}}\rangle\!\}_{\!j=1\!}^N$ of orthonormal $N$-component vectors, modulo transformations $|u^j_{\bs{k}}\rangle \!\! \mapsto \!- |u^j_{\bs{k}}\rangle$. The order-parameter space $M_N$ can be expressed as the space of right-handed frames (isomorphic to orthogonal group $\mathsf{SO}(N)$), modulo a point group of $\pi$ rotations flipping the sign of an even number of the frame elements. Geometrically, this is the space of all orientations of a generic $N$-dimensional ellipsoid. Remarkably, the case of $N\!=\! 3$ bands is mathematically equivalent to the order-parameter space of biaxial nematic liquid crystals~\cite{Madsen:2004}. In these materials, molecules with an approximate ellipsoid symmetry have random positions but a frozen orientation.

\begin{figure*}[t]
	\includegraphics[width=0.98\textwidth]{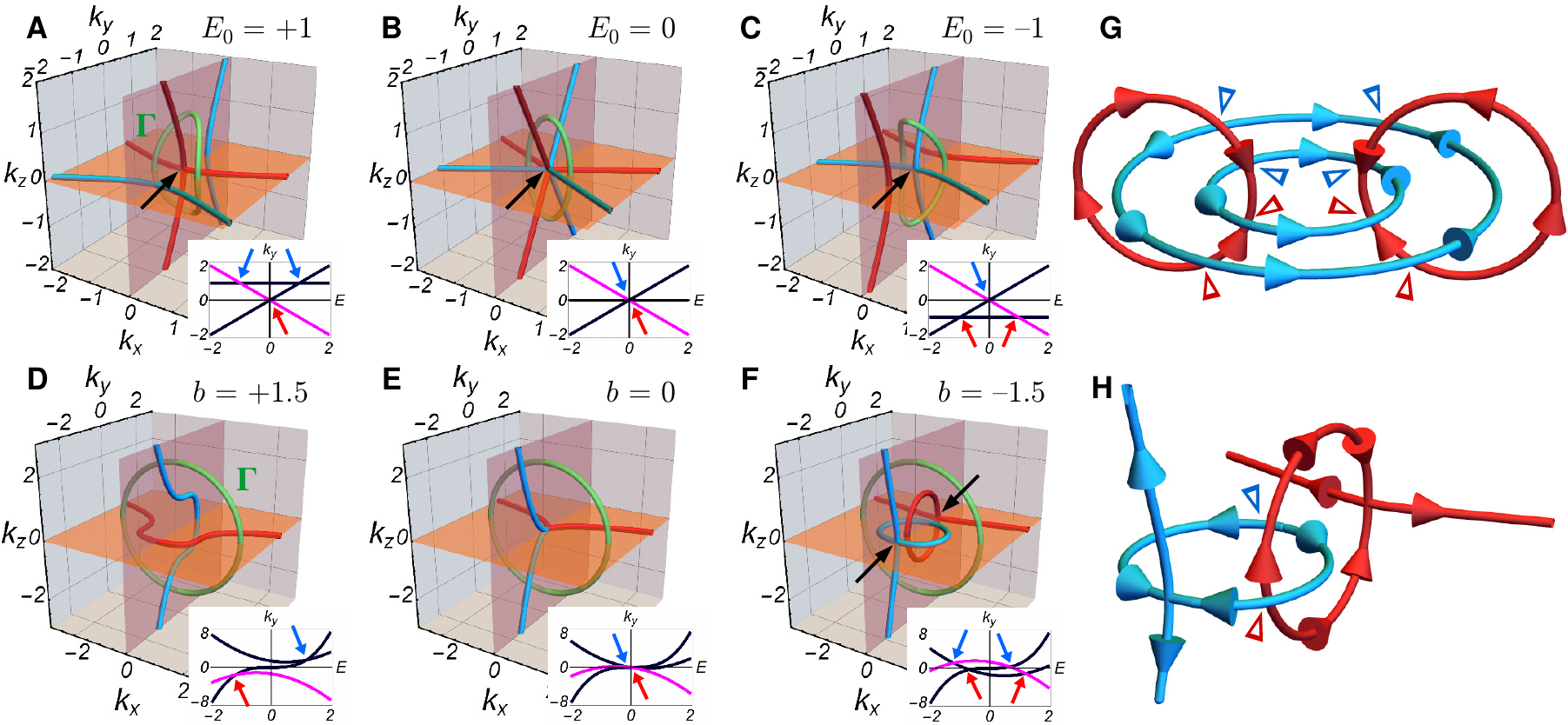}
	\caption{\textsf{\textbf{Nodal lines (NLs) in three-band models.} For three-band Hamiltonians, we plot NLs formed by the lower (upper)~two bands with red (blue) lines. (\textbf{A}--\textbf{C}) NLs of the model in Eq.~(\ref{eqn:3-band-detach}) for various values of $E_0$. A crossing point (black arrows) is transferred between the two species of NLs when $E_0$ changes sign. The transfer follows from a quaternion charge $n_\Gamma \!=\! -1$ on the green path. Insets show the spectrum along the $k_x \!=\! k_z \!=\! 0$ line, where black (magenta) bands carry positive (negative) $\hat{m}_z$ eigenvalue. (\textbf{D}--\textbf{F}) Analogous plots for the model in Eq.~(\ref{eqn:nodal-ears}) for various values of $b$. Moving NLs of different species across each other results in the formation of ``earring'' NLs in panel (F). Such behavior is imposed by the quaternion charge $n_\Gamma\!=\!-1$ on the green path. (\textbf{G}) The orientation of NLs in the composition with linked ``earrings''. Due to the non-Abelian group in Eq.~(\ref{eqn:suicide-quats}), the orientation of a NL is reversed (indicated by triangular arrowheads) each time it goes under a NL of the other color~\cite{Supp}. (\textbf{H}) An example of admissible NL composition involving linked rings. Each NL must enclose an even number of NLs of the other color.}}
	\label{fig:preserved-TP}
\end{figure*}

Let us explicitly discuss the case $N \!=\! 3$.~The order-parameter space is $M_3 \!=\! \mathsf{SO}(3)/\mathsf{D}_{2}$,~where $\mathsf{D}_{2}$ is the three-dimensional ``dihedral'' crystallographic point group, which contains $\pi$ rotations around three perpendicular axes, and the identity. Closed paths $\Gamma$ in $M_3$ are characterized by the fundamental group of the order-parameter space, which is the quaternion group~\cite{Kleman:1977,Mermin:1979,Supp,Note1}
\begin{equation}
    n_\Gamma \in \; \pi_1(M_\mathrm{3})=\mathsf{Q} = \{ \pm 1,\pm \imi, \pm \imj, \pm \imk \}  \label{eqn:suicide-quats}
\end{equation}
with anticommuting imaginary units $\imi^2\!=\!\imj^2\!=\!\imk^2 \!=\! -1$. The result in Eq.~(\ref{eqn:suicide-quats}) is related to $\pi$ rotation operators for spin-$\tfrac{1}{2}$ particles. The generalization of Eq.~(\ref{eqn:suicide-quats}) to $N \geq 3$, discussed in SIF~\cite{Supp}, is closely related to the symmetry of higher-dimensional Euclidean spaces~\cite{Atiyah:1964,Salingaros:1983}.

The topological charge of Eq.~(\ref{eqn:suicide-quats}) divides into five conjugacy classes: $\{+1\}$, $\{ \pm \imi\}$, $\{ \pm \imj\}$, $\{ \pm \imk\}$, $\{ -1\}$. In analogy with the discussion below Eq.~(\ref{eq:Z}), considering a tight loop $\Gamma$ encircling a line defect allows us to label it by one of the five conjugacy classes. In biaxial nematics, the conjugacy classes involving the imaginary units describe three different species of vortex lines~\cite{Mermin:1979}. In three-band crystalline solids, $\{\pm\imi\}$ ($\{\pm\imk\}$) characterize \edit{closed paths in $\bs{k}$-space that encircle} a NL between the upper (lower) two bands, while $\{\pm\imj\}$ corresponds to paths enclosing one of each species of NLs~\cite{Supp,Note1}. The \emph{sign} of the charge assigns an \emph{orientation} to the NLs. Two NLs of the same orientation between the same pair of bands are described by $\{-1\}$. The green path in Fig.~\ref{fig:preserved-TP}(A--C) belongs to this last conjugacy class, and the discussed transfer of CP from the lower to the upper band gap corresponds to reinterpreting $-1 \!=\! \imk^2$ as $-1  \!= \!\imi^2$~\cite{Supp}. Note also that while $\imi^2 \!=\! -1$ is non-trivial, $\imi^4 \!=\! +1$ is \emph{trivial}. We show in SIF~\cite{Supp} that, indeed, a path enclosing four NLs of the same type and orientation can be continuously shrunk to a point without crossing a NL.

\textbf{\textsf{Constraints on nodal-line compositions.}} ---  The quaternion group of Eq.~(\ref{eqn:suicide-quats}) is non-Abelian. For example, the two species of NLs for $N\!=\! 3$ follow the rule $\imi \cdot \imk \!=\! -\imk \cdot \imi \!\neq\! \imk \cdot \imi$. We prove in SIF~\cite{Supp} that such anticommutation property survives for $N\!\geq \!3$, where it applies to NL pairs formed inside two consecutive band gaps. In this section, we show that the non-Abelian property poses constraints on admissible NL compositions.

Let us first consider a model with two mirror planes, $\hat{m}_z = \diag(1,1,-1)$ and $\hat{m}_x = \diag(1,-1,1)$, namely
\begin{equation}
\!\mcH_3(\bs{k}) \!=\! \left(\begin{array}{ccc}
k_y^3        &   t k_x         &   t k_z         \\
t k_x         &   \!-k_y \!+\! (b+k_y^2)\!  &   ck_x k_z   \\
t k_z &  ck_x k_z  &  \!-k_y \!-\! (b+k_y^2) \!\! 
\end{array}\right).\label{eqn:nodal-ears}
\end{equation}
We set $t = 2$ and $c=\tfrac{1}{4}$, and study how the NLs (displayed as red and blue lines in Fig.~\ref{fig:preserved-TP}(D--F)) change when varying $b$. For $b\!>\!0$, we find an extended NL formed by the lower (upper) pair of bands inside the $k_z \!=\! 0$ ($k_x \!=\! 0$) plane (Fig.~\ref{fig:preserved-TP}(D)).~Decreasing the value of $b$ moves the two NLs towards each other, until they meet for $b=0$ (Fig.~\ref{fig:preserved-TP}(E)). The anticommutation relation implies that on the indicated green path $n_\Gamma \!= \imi \cdot (-\imk) \cdot (-\imi)\cdot \imk =\! -1$~\cite{Mermin:1979,Supp}. 
The non-trivial value of $n_\Gamma$ forbids moving the NLs across each other~\cite{Poenaru:1977}. Instead, we find that the extended NLs remain tangled for $b < 0$ via a link of two ``earring'' NLs (Fig.~\ref{fig:preserved-TP}(F)), i.e. ring-shaped NLs attached to other NLs with only one CP. Such earring NLs were not previously reported.

\begin{figure*}[t]
	\includegraphics[width=0.64\textwidth]{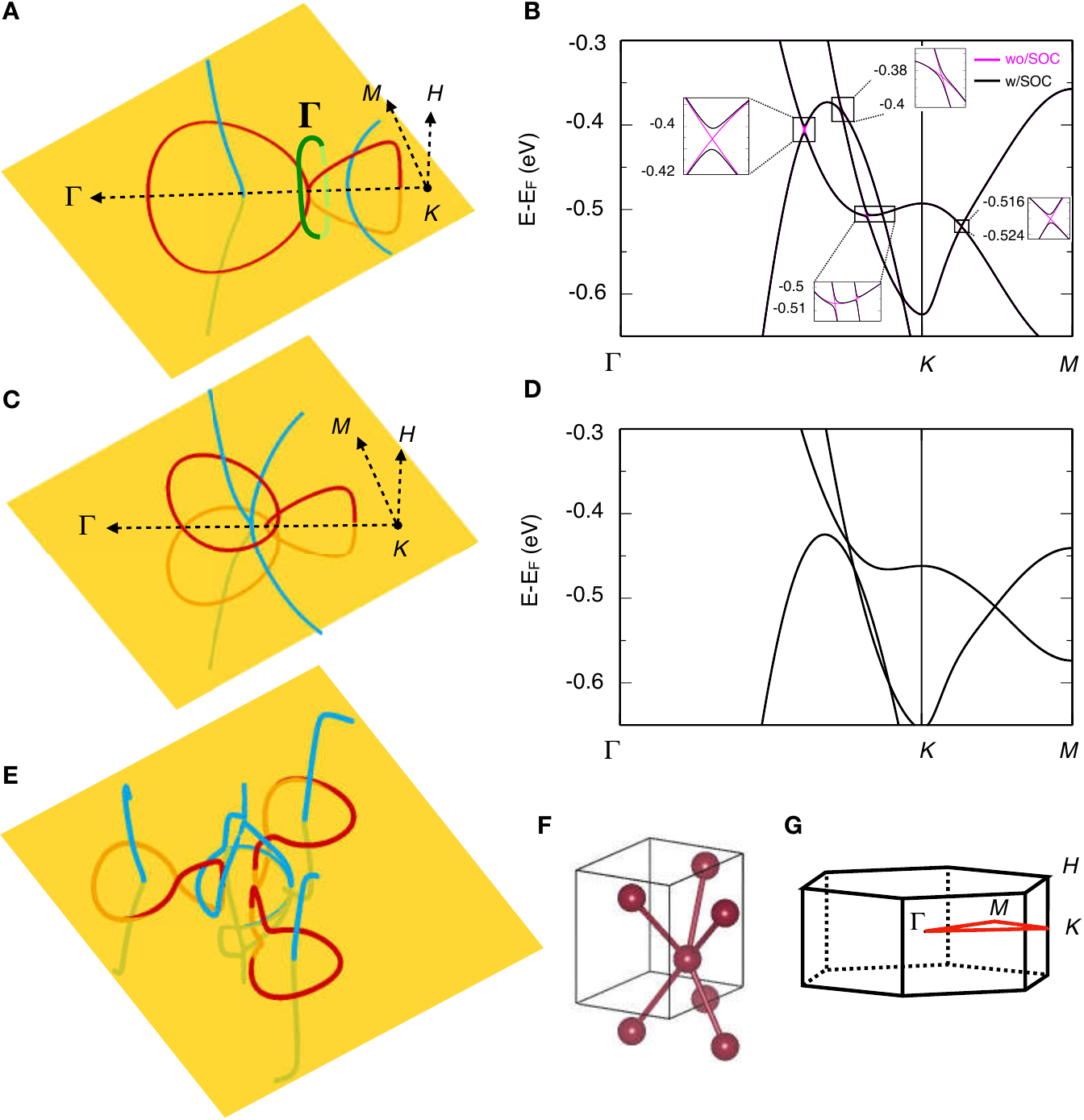}
	\caption{\textsf{\textbf{Nodal lines (NLs) of scandium (Sc).} 
	(\textbf{A})  Red (blue) NLs formed by the two lower (upper) of the three valence bands of $\textsf{Sc}$ without SOC in the $k_z = 0$ plane (orange sheet). The red NLs exhibit a crossing point (CP) along the $\Gamma\textrm{K}$ line, stabilized by the quaternion charge $n_\Gamma\!=\!-1$ on the green path $\bs{\Gamma}$ (not to be confused with the BZ center point $\Gamma$). (\textbf{B}) The band structure of the three (six) valance bands of $\textsf{Sc}$ along $\Gamma\textrm{K}\textrm{M}$ without (with) SOC shown with magenta (black) colors. The SOC-induced splitting of the nodal lines (shown in insets) is less than $10$ meV, and is thus irrelevant for a comparison with ARPES results. (\textbf{C} and \textbf{D}) Evolution of the NLs and the band structure caused by a \edit{$2.25\%$ biaxial tensile strain in the $x,y$-plane} in the absence of SOC. The CP is now formed by the blue NLs, while the red NLs exhibit a pair of out-of-plane earrings. (\textbf{E}) NLs in the case of mirror-breaking compressive $1\%$ strain in the $[10\bar{1}1]$ direction. The CP has disconnected non-trivially. (\textbf{F}) The atomic unit cell of Sc. (\textbf{G}) The BZ of hexagonal \textsf{Sc} with its high-symmetry points. \edit{See SIF~\cite{Supp} for a more detailed discussion and for additional \emph{ab initio} data}.}}
	\label{fig:nodalline-Sc}
\end{figure*}

The anticommutation relation $\imi \cdot \imk \!=\! -\imk \cdot \imi$ can be interpreted as reversing the \emph{orientation} of a NL each time it goes under a NL of the other species~\cite{Supp}. For example, orientations of NLs in Fig.~\ref{fig:preserved-TP}(F) follow the scheme in Fig.~\ref{fig:preserved-TP}(G). Consistency requires that NLs that are closed and isolated (\emph{i.e.}~without CP) enclose an \emph{even} number of NLs of the other species, such as the example in Fig.~\ref{fig:preserved-TP}(H). Careful considerations of orientation reversals in many-band models can be used~\cite{Tiwari:2019} to relate the monopole charges of NLs~\cite{Fang:2015,Bzdusek:2017} to their linking structure~\cite{Ahn:2018}. The orientation reversals also imply that the ability of two NLs to annihilate depends on the \emph{trajectory} used to bring the NLs together, which we illustrate in SIF~\cite{Supp}.

\textbf{\textsf{Topology in 1D beyond the tenfold way.}} ---  Given an $N$-band crystalline solid (assumed $\mcP\mcT$-symmetric with weak SOC), the discussed non-Abelian topology ascribes to every closed path in $\bs{k}$-space (assumed to have non-degenerate spectrum) an element of $\pi_1(M_N)$. For 1D systems, the Brillouin zone (BZ) itself forms such a closed path, allowing us to consider topological phases in 1D distinguished by a non-Abelian topological invariant $n_\textrm{BZ}\in\pi_1(M_N)$. Band structures with $n_\textrm{BZ} \neq +1$ cannot be adiabatically deformed into the Hamiltonian of uncoupled atomic orbitals (the \emph{atomic limit}) unless we form a degeneracy between some pair of bands somewhere in the 1D BZ. These topological obstructions are stable under adding trivial bands~\cite{Supp}.

Assuming $N\!\geq\!3$ of bands, the group $\pi_1(M_N)$ has $2^N = 2 \times 2^{N-1}$ elements, corresponding to ``doubled'' point-symmetry group of an $N$-dimensional ellipsoid~\cite{Supp}. The $\mcP\mcT$ symmetry quantizes the Berry phase~\cite{Berry:1984} of each band to $0$ vs.~$\pi$~\cite{Zak:1989,Fang:2015}. However, only $(N-1)$ of these phases are independent, because they must add up to $0 \;(\textrm{mod} \; 2\pi)$. Therefore, only $2^{N-1}$ elements of $\pi_1(M_N)$ are distinguished by Berry phases. The other $2^{N-1}$ elements correspond to composing the original $2^{N-1}$ elements with a (generalized) quaternion charge $n_\Gamma \!=\! -1$. We show in SIF~\cite{Supp} that for 1D paths with $n_\Gamma \!=\! -1$, the Berry phase of each band is trivial~\cite{a_Note2}. The computation of the generalized quaternion charges requires an augmentation of the Wilson loop technique~\cite{Yu:2011,Soluyanov:2011,Gresch:2017} to spin bundles, which we outline in SIF~\cite{Supp,Baez:1994}. 

\textbf{\textsf{Experimental signatures in scandium.}} ---  The predicted properties of NLs derived from non-Abelian topology can be experimentally verified using angle-resolved photoemission spectroscopy (ARPES) of elemental scandium ($\textsf{Sc}$) under strain. $\textsf{Sc}$ has weak SOC and crystallizes in the hexagonal close-packed structure. Neglecting SOC~\cite{Supp}, our \textit{ab initio} calculations reveal that three valence bands of $\textsf{Sc}$ form NLs plotted in Fig.~\ref{fig:nodalline-Sc}(A--B) near the $\mathrm{K}$-point of the BZ. Neglecting SOC, we observe a CP of two red ``earring'' NLs along the $\Gamma\textrm{K}$ line, both threaded by a blue NL, in Fig.~\ref{fig:nodalline-Sc}(A). As explained before, earring NLs are stabilized by the quaternion charge $n_\Gamma \!=\! \imi \cdot (-\imk)\cdot (-\imi) \cdot \imk \!= \! -1$ which cannot be detected by Berry phase nor by Wilson loops. Therefore, experimental observation of earring NLs would provide an indirect evidence of the non-Abelian topology. Although the weak SOC present in $\textsf{Sc}$ gaps the NLs, as shown in Fig~\ref{fig:nodalline-Sc}(B), the induced NL splittings are smaller than $10$~meV. Such small value is below the resolution of the best existing ARPES instruments, and thus the ARPES image of \textsf{Sc} does not depend on the actual presence of SOC. For this reason it is safe to neglect SOC in our discussion of $\textsf{Sc}$.

Applying a symmetry-preserving \edit{$2.25\%\!$ biaxial tensile strain to $\textsf{Sc}$ in the $x,y$-plane~\cite{Supp}} moves the blue NLs together. In this case, we observe a transfer of the CP from the red to the blue NLs, shown in Fig.~\ref{fig:nodalline-Sc}(C and D). This transfer is also governed by $n_\Gamma\!=\!-1$, analogously to Fig.~\ref{fig:preserved-TP}(A--C). Two additional out-of-plane earrings are developed by the red NLs in Fig.~\ref{fig:nodalline-Sc}(C), which are accidental, \emph{i.e.}~not imposed by the quaternion charge on any path. \edit{Elastic biaxial strain exceeding $2\%$ is commonly achieved in epitaxially grown thin films of even more complicated compounds using a lattice mismatch with the substrate, while admitting probation of the films using ARPES spectroscopy~\cite{King:2014,Burganov:2016} or X-ray scattering~\cite{Catalano:2014, Ivashko:2019}. In order to verify that ARPES measurements done on the biaxially stretched $\textsf{Sc}$ thin film would illustrate our claims made for $\textsf{Sc}$ bulk, we performed an \emph{ab initio} calculation of sub-band splitting for a $30$~nm thick $\textsf{Sc}$ thin film~\cite{Supp}. We find this splitting to be approximately $5$~meV, which means that from the experimental point of view this thin film represents the bulk band structure.}

Finally, we consider a compressive $1\%$ strain of $\textsf{Sc}$ in the $[10\bar{1}1]$ direction, which breaks all the mirror symmetries~\cite{Supp}. Such a distortion results in a non-trivial separation of CP, compatible with the quaternion charge on path $\Gamma$ in Fig.~\ref{fig:nodalline-Sc}(A). After the separation, the two red earring NLs of the unstrained case have merged into a \emph{single} closed NL that encircles two blue NLs, compatible with the constraints discussed below Eq.~(\ref{eqn:nodal-ears}). \edit{ARPES observation of the NL compositions in $\mathsf{Sc}$ under strain, in addition to the unstrained case, would provide a solid experimental support for the theoretical predictions made in this work.}

\textbf{\textsf{Note added.}} --- After the submission of this manuscript, a nontrivial exchange of band nodes in momentum space was conjectured for a class of 2D models using the mathematical technique of characteristic classes~\cite{Ahn:2018b}. A very recent work~\cite{Slager:2019}, which calls the quaternion charges introduced here “frame rotations,” has generalized the non-Abelian topology to a class of 3D Weyl semimetals and has also related the technique of characteristic classes to the fundamental groups discussed in the present work.

\textbf{\textsf{Acknowledgements.}} --- We thank A.~Broido, \edit{J.~Chang, M.~Gibert, G.~De Luca}, F.~Valach, X.-Q.~Sun, A.~Tamai, G. E.~Volovik and S.-C.~Zhang for valuable discussions. We also thank the reviewers for their useful comments, which helped us improve the clarity of the manuscript. \textbf{\textsf{Funding:}} Q.~W. acknowledges the support of NCCR MARVEL. A.~A.~S. acknowledges the support of NCCR MARVEL, NCCR QSIT and SNSF Professorship grants, and of Microsoft Research. T.~B. was supported by the Gordon and Betty Moore Foundation’s EPiQS Initiative, Grant GBMF4302. \textbf{\textsf{Author contributions:}} T.~B. had the initial idea, carried out the theoretical analysis, and led the project. Q.~W. identified $\textsf{Sc}$ as an illustrative material and performed the first-principle studies. T.~B. and A.~A.~S. wrote the manuscript. \textbf{\textsf{Competing interests:}} The authors declare that they have no competing interests. \textbf{\textsf{Data and materials availability:}} Schematic illustrations and plots for the two-band and three-band models were generated using Wolfram Mathematica (version 11.3). To obtain the ab initio
study of Sc with and without strain, we performed first-principle calculations as implemented in software Vienna Ab initio Simulation Package (VASP)~\cite{PhysRevB.54.11169}. The nodal-line configurations are obtained by an open-source software, WANNIERTOOLS~\cite{wanniertools}, based on the tight-binding models constructed by VASP+WANNIER90, where WANNIER90~\cite{wannier90} is an open-source
software. Details are discussed in Sec.~VI of~\cite{Supp}. We have made the Wolfram Mathematica files; the VASP, WANNIER90, and WANNIERTOOLS input files for the computation of band structures and nodal-line configurations of Sc; and the numerically obtained band-structure data publicly available on Materials Cloud Archive~\cite{b_Suppdata}.


\let\oldaddcontentsline\addcontentsline
\renewcommand{\addcontentsline}[3]{}
\putbib
\let\addcontentsline\oldaddcontentsline

\end{bibunit}

\begin{bibunit}

\onecolumngrid
\pagebreak
\phantom{
\color{white}\cite{Chen:2015,Kim:2015,Yu:2015,Bian:2015a,Chan:2016,Schoop:2016,Fang:2016,Fang:2015,Bzdusek:2017,Ahn:2018,Burkov:2011,Bzdusek:2016,Wang:2017,Yu:2017,Feng:2018,Yi:2018,Gong:2018,Heikkila:2015b,Zhu:2016,Yan:2018,Madsen:2004,Kleman:1977,Mermin:1979,Supp,Note1,a_Note2,Sun:2018,Kitaev:2009,Ryu:2010,Horava:2005,Atiyah:1964,Salingaros:1983,Poenaru:1977,Tiwari:2019,Berry:1984,Zak:1989,Yu:2011,Soluyanov:2011,Gresch:2017,Baez:1994,King:2014,Burganov:2016,Catalano:2014,Ivashko:2019,Ahn:2018b,Slager:2019,PhysRevB.54.11169,wannier90,wanniertools,b_Suppdata}
\pagebreak}
\twocolumngrid

\newpage

\setcounter{equation}{0}
\setcounter{page}{1}
\setcounter{figure}{0}
\renewcommand\thefigure{{S-\arabic{figure}}}


\title{
Supplemental Information for:
\texorpdfstring{\vspace{0.5cm}}{}
\texorpdfstring{\\}{}
Non-Abelian band topology in noninteracting metals}

\author{QuanSheng Wu$^{1,2}$}
\author{Alexey A. Soluyanov$^{3,4}$}
\author{Tom\'{a}\v{s} Bzdu\v{s}ek$^{5,6}$}

\affiliation{$^{1}$Institute of Physics, \'{E}cole Polytechnique F\'{e}d\'{e}rale de Lausanne, CH-1015 Lausanne, Switzerland}
\affiliation{$^{2}$National Centre for Computational Design and Discovery of Novel Materials MARVEL, Ecole Polytechnique F\'{e}d\'{e}rale de Lausanne (EPFL), CH-1015 Lausanne, Switzerland}
\affiliation{${^3}$Physik-Institut, Universit\"at Z\"urich, Winterthurerstrasse 190, CH-8057 Zurich, Switzerland}
\affiliation{$^{4}$Department of Physics, St. Petersburg State University, St. Petersburg, 199034 Russia}
\affiliation{$^{5}$Department of Physics, McCullough Building, Stanford University, Stanford, CA 94305, USA}
\affiliation{$^{6}$Stanford Center for Topological Quantum Physics, Stanford University, Stanford, CA 94305, USA}

\date{\today}

\maketitle

\onecolumngrid
\vfill
\begin{center}
\begin{minipage}{0.65\textwidth}
\tableofcontents
\end{minipage}
\end{center}
\vfill
\vfill
\newpage
\twocolumngrid


\section{Homotopic characterization of band-structure nodes}\label{eqn:homo-aproach}

In this section, we provide an overview of how to describe nodes in metallic band structures using \emph{homotopy theory}. The discussion is structured as follows: In Sec.~\ref{sec:PT-nodes} we review the theory of band-structure nodes protected by symmetries local in $\bs{k}$-space, as originally worked out in Ref.~\cite{Bzdusek:2017}. Furthermore, we also present here our generalization of Ref.~\cite{Bzdusek:2017} that simultaneously describes nodal compositions formed inside multiple band gaps. The development of such a generalized theory is central to the main text of our manuscript. We continue in Sec.~\ref{sec:mirror-nodes} by reviewing the homotopic description of nodes in the presence of additional crystalline symmetries, such as mirror symmetry, as recently worked out by Ref.~\cite{Sun:2018}. 

After summarizing the key notions, we include two mathematical subsections. In Sec.~\ref{eqn:homotopy-theory} we summarize the basics of homotopy theory, while in Sec.~\ref{eqn:homotopy-cosets} we derive a pair of useful identities for homotopy groups of coset spaces, which appear frequently in our exposition. The theory summarized in the present section is employed throughout the remainder of the Supplementary Information file (SIF). It allows us to analyze and to understand nodal compositions that appear in $\mcP\mcT$-symmetric systems with various number of bands in the presence or absence of mirror symmetry.


\subsection{Nodes protected by \texorpdfstring{$\mcP\mcT$}{PT} symmetry}\label{sec:PT-nodes}

The description of band-structure nodes in our work follows the method developed in Ref.~\cite{Bzdusek:2017}, which has been further generalized in Ref.~\cite{Sun:2018}. The method is based on studying the Hamiltonian on a manifold surrounding the node, and is ussually applied to nodes occurring between the highest occupied (HO) and the lowest unoccupied (LU) band. More generally, one can split the total number of bands into the $n$ lower ones which we call ``occupied'' and the $\ell$ upper ones which we call ``unoccupied''. (We apply this terminology even if it does not reflect the actual location of the chemical potential.) We have this latter case in mind when discussing nodal lines in \emph{metals}.

In this work, we further generalize the method of Refs.~\cite{Bzdusek:2017,Sun:2018} to describe nodes occurring between \emph{all} pairs of bands (including among fully occupied/unoccupied pairs of bands) simultaneously. Both the original method and its present generalizations require the identification of the appropriate \emph{space of Hamiltonians} $M$ and of its \emph{homotopy groups}.

For simplicity, we consider here only systems with $\mcP\mcT$ symmetry (i.e.~the composition of spatial inversion $\mcP$ and time reversal $\mcT$) and with negligible spin-orbit interaction (SOC), which correspond to nodal class $\mathrm{AI}$ of Ref.~\cite{Bzdusek:2017}. For this symmetry class, a suitable rotation of the basis leads to a representation $\mcP\mcT = \mcK$ (complex conjugation), implying that the Hamiltonian is \emph{real} at all momenta $\bs{k}$ inside the Brillouin zone (BZ). More generally, the method discussed in the present subsection can be used to classify nodes protected solely by symmetries local in $\bs{k}$-space, i.e.~$\mcP\mcT$, $\mcP\mcC$ and $\mcC\mcT$ ($\mcC$ is charge conjugation), which constrain the space of admissible Hamiltonians equally at all momenta $\bs{k}$ inside the Brillouin zone (BZ)~\cite{Bzdusek:2017}.

We first identify the appropriate space of Hamiltonians, which we also call the \emph{order-parameter space} in the main text. To achieve this goal, we consider a system with $n$ occupied and $\ell$ unoccupied bands ($n,\ell \geq 1$). We write $n + \ell = N$ for the total number of bands. At every $\bs{k}\in\textrm{BZ}$, we can label the bands by increasing energy as
\begin{equation}
\eps^1_{\bs{k}}\leq\eps^2_{\bs{k}} \leq \ldots \leq \eps^{n+\ell}_{\bs{k}}. \label{eqn:eig-ordering}
\end{equation}
We further write $\{|u^j_{\bs{k}}\rangle \}_{j=1}^{n+\ell}$ for the corresponding eigenstates (taking into account only the cell-periodic part of the Bloch functions). 
Since the Hamiltonian is real in the presence of $\mcP\mcT$, we can locally gauge the eigenstates to be real too. The nodes formed between HO and LU bands correspond to a set
\begin{subequations}
\begin{equation}
\mathcal{N}_\mcH^n = \{\bs{k}\in\textrm{BZ}| \eps^n_{\bs{k}} = \eps^{n+1}_{\bs{k}}\}\subset \textrm{BZ}.\label{eqn:nodes-positions-set}
\end{equation}
For momenta $\bs{k}\in\textrm{BZ}\backslash \mathcal{N}_\mcH^n$ it is possible to move the energy of all the occupied (unoccupied) states to ``standard'' values $-1$ ($+1$) without closing the gap between HO and LU bands, i.e. one can continuously deform the Hamiltonian into a ``standard'' form
\begin{equation}
\mcH(\bs{k}) \!=\! -\!\sum_{j=1}^n |{u^j_{\bs{k}}}\rangle\langle{u^j_{\bs{k}}}| +\!\!\!
\sum_{j=n+1}^{n+\ell} \! |{u^j_{\bs{k}}}\rangle\langle{u^j_{\bs{k}}}|.\label{eqn:gen-ham-form}
\end{equation}
We use $M_{(n,\ell)}$ to denote the space of all distinct $\mcP\mcT$-symmetric Hamiltonians of the form in Eq.~(\ref{eqn:gen-ham-form}). \edit{We remark that in two dimensions (2D) the same symmetry class also describes $C_{2z}\mcT$-symmetric models ($C_{2z}$ is a $\pi$ rotation of the 2D plane) both \emph{with} or \emph{without} SOC~\cite{Ahn:2018b}.}

We want to understand the topology of space $M_{(n,\ell)}$. First, note that any Hamiltonian is completely fixed once we provide the list of all its eigenstates, which can be collected into an orthogonal $\mathsf{O}(n+\ell)$ matrix. However, rotations by $\mathsf{O}(n)$ matirces among the occupied states as well as rotations by $\mathsf{O}(\ell)$ matirces among the unoccupied states clearly keep the form of the Hamiltonian in Eq.~(\ref{eqn:gen-ham-form}) unchanged. Therefore, we identify the space of Hamiltonians as the coset space
\begin{equation}
M_{(n,\ell)} = \mathsf{O}(n+\ell)/\mathsf{O}(n)\!\times\!\mathsf{O}(\ell)\label{eqn:M-n-l-real-Gr}
\end{equation}
\end{subequations}
which is also called the \emph{real Grassmannian}. The momenta in Eq.~(\ref{eqn:nodes-positions-set}) correspond to discontinuities for the spectral projection in Eq.~(\ref{eqn:gen-ham-form}). 

More generally, momenta that support a degeneracy between \emph{some} pair of bands build up a set
\begin{subequations}
\begin{equation}
{\mathcal{N}_\mcH} = \{\bs{k}\in\textrm{BZ}|\exists j: \eps_{\bs{k}}^j = \eps_{\bs{k}}^{j+1}\}.\label{eqn:generalized-node-positions}
\end{equation} 
Clearly, $\mathcal{N}_\mcH = \bigcup_{j=1}^{N-1} \mathcal{N}_\mcH^j$. For $\bs{k}\in\textrm{BZ}\backslash {\mathcal{N}_\mcH}$ we can move the energies to ``standard'' values 
\begin{equation}
\forall j: \eps^j_{\bs{k}} = j\label{eqn:standrad-energies}
\end{equation}
without encountering a spectral degeneracy in the process. Such a spectral projection brings the Hamiltonian to a ``standard'' form
\begin{equation} 
\mcH(\bs{k}) = \sum_{j=1}^N j |{u^j_{\bs{k}}}\rangle\langle{u^j_{\bs{k}}}|.\label{eqn:Hamiltonian-AI-3-N}
\end{equation}
We use $M_{N}$ to denote the space of all distinct $\mcP\mcT$-symmetric Hamiltonians of the form in Eq.~(\ref{eqn:Hamiltonian-AI-3-N}). Such Hamiltonians are uniquely described by an orthogonal $\mathsf{O}(N)$ matrix of their ordered eigenstates. However, multiplying any of the eigenstates by $\pm 1 \equiv \mathsf{O}(1)$ does not change the form of the Hamiltonian. We therefore identify the space of Hamiltonians as the coset space
\begin{equation}
M_{N} = \mathsf{O}(N)/\mathsf{O}(1)^N.\label{eqn:M-N-space-gen}
\end{equation}
\end{subequations}
The momenta in Eq.~(\ref{eqn:generalized-node-positions}) correspond to discontinuities of the spectral projection in Eq.~(\ref{eqn:Hamiltonian-AI-3-N}). We will look into the topology of the spaces in Eq.~(\ref{eqn:M-N-space-gen}) more carefully later in Sec.~\ref{sec:multi-homotopy}.

Knowing the space of Hamiltonians $M$ (with the appropriate subscript), we characterize band-structure nodes using homotopy groups $\pi_p(M)$. To give the construction a rigorous mathematical footing, we write $\iota(S^p)$ for a chosen embedding of a $p$-dimensional sphere $S^p$ inside $\textrm{BZ}\backslash \mathcal{N}_\mcH$ (with the appropriate superscript on $\mathcal{N}_\mcH$). We assume that $\iota(S^p)$ does not wind around the $\textrm{BZ}$ torus, such that it would become contractible to a point in the absence of nodes (i.e. when $\mathcal{N}_\mcH = \varnothing$). The embedding can be composed with the (spectrally projected) Hamiltonian $\mcH$, which assigns each $\bs{k}$ an element of $M$,~i.e.
\begin{subequations}
\begin{equation}
S^p \stackrel{\iota}{\hookrightarrow} \textrm{BZ}\backslash \mathcal{N}_\mcH \stackrel{\mcH}{\to} M,
\end{equation}
such that the composition 
\begin{equation}
\mcH\circ \iota : S^p \to M,
\end{equation}
\end{subequations}
goes directly from the $p$-sphere to the space of Hamiltonians. Small changes of $\iota$ or $\mcH$ (assuming we avoid the nodes, i.e. we demand $\iota(S^p)\cap \mathcal{N}_\mcH = \varnothing$ while performing the change) lead to continuous deformations of $\mcH\circ\iota$.~\footnote{For $p=1$ and a 3D $\bs{k}$-space, the embeddings $\iota$ themselves form an interesting mathematical structure studied in the knot theory~\cite{Prasolov:1997}. We don't pursue this direction in our work.}

We make two observations:
\begin{enumerate}
\item If the embedding $\iota(S^p)$ does not enclose a node, then it can be continuously shrunk to a single point $\bs{k}_0 \in \textrm{BZ}$ without encountering a singularity of $\mcH$. Such a deformation continuously evolves $\mcH\circ \iota$ to a \emph{constant} map $S^p \mapsto \mcH(\bs{k}_0)$.
\item On the other hand, if the image of the $p$-sphere $(\mcH\circ \iota)(S^p)$ \emph{cannot} be continuously deformed into a constant map in $M$, then there must be an obstruction for shrinking $\iota(S^p)$ to a point. This implies that $\iota(S^p)$ encloses a node. 
\end{enumerate}
We infer that the presence of a robust node inside $\iota(S^p)$ depends on whether $\mcH\circ\iota$ can be continuously deformed into a constant map. Such questions belong to the realm of \emph{homotopy theory}. 

Being somewhat informal for now (mathematical details appear in Sec.~\ref{eqn:homotopy-theory}), the \emph{homotopy group} $\pi_p(M)$ describes the \emph{equivalence classes} $[f]$ of continuous functions $f: S^p \to M$, where two functions are called equivalent if one can be changed into the other using only continuous deformations. By taking $\iota(S^p)$ that tightly encloses just a single node (i.e.~one connected component of $\mathcal{N}_\mcH$), we can use the corresponding element $[\mcH\circ \iota]\in \pi_p(M)$ to assign a \emph{topological charge} to that node. Codimension counting implies that a topologically stable node of dimension $d$ inside a $D$-dimensional BZ is revealed by a non-trivial homotopy group $\pi_{D-d-1}(M)$~\cite{Bzdusek:2017}. Especially, nodal lines (points) in 3D (2D) are protected by the \emph{first homotopy group} $\pi_1(M)$ of space $M$. The same mathematical object is often called the \emph{fundamental group of $M$} -- as we do in the main text.

The derivation of the relevant homotopy groups, and their use in explaining the topological stability of nodal lines in $\mcP\mcT$-symmetric systems, are discussed further in the SIF. Here, we summarize the results which are (explicitly or implicitly) used in the main text, namely
\begin{subequations}
\begin{eqnarray}
\pi_1 (M_{(n,\ell)}) &=& \left\{ \begin{array}{ll} \intg & \textrm{for $n=\ell =1$} \\ \ztwo & \textrm{for $n+\ell \geq 3$}\end{array}\right. \\
\pi_1 (M_{N}) &=& \left\{ \begin{array}{ll} \intg & \textrm{for $N = 2$} \\ \mathsf{Q} & \textrm{for $N = 3$} \\ \overline{\mathsf{P}}_N & \textrm{for $N \geq 3$}\end{array}\right. 
\end{eqnarray}
\end{subequations}
where $\mathsf{Q}$ is the quaternion group discussed in Sec.~\ref{subsec:3-deriv}, and $\overline{\mathsf{P}}_N$ is the Salingaros vee group of Clifford algebra $C\!\ell_{0,N-1}$ (called \emph{generalized quaternions} in the main text), which we describe in Sec.~\ref{subsec:gen-many}~\cite{Salingaros:1983}. Importantly, groups $\pi_1(M_{N})$ are \emph{non-Abelian} for $N \geq 3$. 

\subsection{Nodes protected by mirror symmetry}~\label{sec:mirror-nodes}

The discussion in the previous subsection only applies to nodes protected by $\mcP\mcT$, $\mcP\mcC$ and $\mcC\mcT$~\cite{Bzdusek:2017} which map every $\bs{k}\in\textrm{BZ}$ to itself. Such local-in-$\bs{k}$ symmetries constrain the space of admissible Hamiltonians uniformly throughout BZ. On the other hand, additional symmetries lead to the presence of \emph{invariant subspaces} inside $\textrm{BZ}$ where the Hamiltonian has to fulfill additional constraints. The description of nodes stabilized by such additional symmetries requires a generalization of the theory summarized in Sec.~\ref{sec:PT-nodes}. Such a generalization has been recently developed by Ref.~\cite{Sun:2018}. Below, we present the generalized theory for the special case of mirror symmetry $m_z: z \to -z$, i.e. one that flips the sign of the $z$ coordinate. Such symmetry setting appears repeatedly in the main text of our work.

The $m_z$ symmetry leads to the appearance of $m_z$-invariant planes, by which we mean the collection 
\begin{subequations}
\begin{equation}
\Pi = \{\bs{k} \in \textrm{BZ} | m_z \bs{k} = \bs{k} \} \subset \textrm{BZ}
\end{equation}
of $\bs{k}$-points. An $m_z$-symmetric Hamiltonian $\mcH(\bs{k})$ fulfills
\begin{equation}
\hat{m}_z \mcH(\bs{k}) \hat{m}_z  = \mcH(m_z \bs{k})
\end{equation}
where $\hat{m}_z$ represents the appropriate \emph{operator} of the mirror symmetry. Especially, inside the invariant plane 
\begin{equation}
\forall \bs{k}\in \Pi: [\hat{m}_z, \mcH(\bs{k})]  = 0,
\end{equation}
i.e. the Hamiltonian commutes with the mirror operator. This property motivates us to define the subspace
\begin{equation}
X_{m_z} = \{\mcH\in M | [\hat{m}_z , \mcH ] = 0\} \subset M \label{eqn:m-z-subspace-H}
\end{equation}
\end{subequations}
of $m_z$-symmetric Hamiltonians (limiting our attention only to those with the standard energy spectrum). 

We know that mirror symmetry can protect nodal lines (NLs) inside $\Pi$, which are produced by the crossing of bands with different $m_z$ eigenvalue. To generalize the description of Sec.~\ref{eqn:homo-aproach} to nodes inside symmetric planes, note that such nodes are naturally enclosed by \emph{$m_z$-symmetric} $p$-spheres~\cite{Sun:2018}. Furthermore, all the information about the Hamiltonian on such a $m_z$-symmetric $p$-sphere is contained on the hemisphere on \emph{one side} of the symmetric plane (allowing us to drop the other hemisphere), which is topologically a \emph{$p$-dimensional disc} $D^p$. The boundary of the $p$-disc (i.e. the equator of the original $S^p$) lies inside $\Pi$, where the Hamiltonian is constrained by Eq.~(\ref{eqn:m-z-subspace-H}). We are thus led to study continuous maps 
\begin{equation}
f: D^p \to M\qquad\textrm{such that} \qquad \partial D^p \to X_{m_z} \label{eqn:why-rel-hom}
\end{equation}
where by $f$ we already mean the composition of the embedding with the Hamiltonian, i.e. $f=\mcH\circ\iota$, as explained in Sec.~\ref{sec:PT-nodes}. We show in the next Sec.~\ref{eqn:homotopy-theory} that equivalence classes of such maps for $p\geq 2$ are classified by \emph{relative homotopy group} $\pi_p(M,X_{m_z})$. To guarantee group structure also for $p=1$, we further need to require that both endpoints $\partial D^1 \simeq S^0$ (a \emph{zero-dimensional sphere}) are mapped to the same connected component of $X_{m_z}$. The presence of a robust node inside a $m_z$-symmetric sphere $\iota(S^p)$ is revealed by a \emph{non-trivial} equivalence class $[\mcH\circ \iota] \in \pi_p(M,X_{m_z})$.

The relative homotopy groups that are relevant for our discussion of nodal lines and nodal chain are
\begin{subequations}\label{eqn:summary:rel-hom}
\begin{eqnarray}
\pi_1(M_{(1,1)},X_{m_z}) &=& \intg \quad \textrm{for  $\hat{m}_z = \diag(\mathbf{+},\mathbf{-})$} \label{eqn:Z-TP-inv}\\ 
\pi_1(M_{(1,2)},X_{m_z}) &=& \triv \quad \textrm{for $\hat{m}_z = \diag({\color{red}\mathbf{+}},{\color{NavyBlue}\mathbf{+}},{\color{NavyBlue}\mathbf{-}})$} \label{eqn:Z-TP-no-inv} \\
\pi_1(M_{(1,2)},X_{m_z}) &=& \ztwo \quad \textrm{for $\hat{m}_z = \diag({\color{NavyBlue}\mathbf{+}},{\color{NavyBlue}\mathbf{+}},{\color{red}\mathbf{-}})$} \\
\pi_1(M_{3},X_{m_z}) &=& \ztwo \quad 
\textrm{for $\hat{m}_z = \diag(\mathbf{+},\mathbf{+},\mathbf{-})$}. \label{eqn:Z2-TP-inv}
\end{eqnarray}
\end{subequations}
where the red (blue) font indicates the $\hat{m}_z$ eigenvalue of the occupied (unoccupied) band if the result depends on this informaiton. In all cases, we assume that both endpoints $\partial D^1$ are mapped to the same connected component of $X_{m_z}$. For example, the winding number $n_\gamma$ on an open-ended semicircle $\gamma$ with end-points at $k_z \!=\! 0$ that is considered in the main text corresponds to Eq.~(\ref{eqn:Z-TP-inv}). The derivation of these results appears in Secs.~\ref{sec:NLs-NCs-stability-2} and~\ref{sec:multiband+mirror} of SIF.

Finally, one may also consider $p$-spheres located entirely \emph{inside} the invariant plane. Such $p$-spheres correspond to boundaries of the $p$-discs discussed above. The equivalence classes of maps $\mcH\circ\iota \equiv f$ on such spheres are classified by homotopy groups $\pi_p(X_{m_z})$. Especially, the groups relevant for our discussion are
\begin{subequations}\label{eqn:summary:hom-bound}
\begin{eqnarray}
\pi_1(X_{m_z}\subset M_{(1,1)}) &=& \triv \quad \textrm{for  $\hat{m}_z = \diag(\mathbf{+},\mathbf{-})$}\;\; \\ 
\pi_1(X_{m_z}\subset M_{(1,2)}) &=& \intg \quad \textrm{for  $\hat{m}_z = \diag({\color{red}\mathbf{+}},{\color{NavyBlue}\mathbf{+}},{\color{NavyBlue}\mathbf{-}})$}\;\; \\ 
\pi_1(X_{m_z}\subset M_{(1,2)}) &=& \triv \quad \textrm{for  $\hat{m}_z = \diag({\color{NavyBlue}\mathbf{+}},{\color{NavyBlue}\mathbf{+}},{\color{red}\mathbf{-}})$}\;\; \\ 
\pi_1(X_{m_z}\subset M_{(3)}) &=& \intg \quad \textrm{for  $\hat{m}_z = \diag({\mathbf{+}},{\mathbf{+}},{\mathbf{-}})$}\;\;
\end{eqnarray}
\end{subequations}
with the same color scheme as in Eqs.~(\ref{eqn:summary:rel-hom}). 


\subsection{Basics of homotopy theory}\label{eqn:homotopy-theory}

In this subsection, we more formally introduce some basic definitions and properties related to homotopy theory. An expanded introduction has recently appeared in SIF to Ref.~\cite{Sun:2018}. For a rigorous mathematical treatment, we refer the reader to Refs.~\cite{Hatcher:2002} and~\cite{Nakahara:2003}. A physically motivated exposition of the same mathematical framework appears in Ref.~\cite{Mermin:1979}, where it is applied to the study of topological defects in media with an ordered parameter.

For a topological space $M$ with a basepoint $\mathfrak{m}\in M$ and for integer $p \geq 1$, a (pointed) \emph{homotopy group} $\pi_p(M,\mathfrak{m})$ is the set of equivalence classes $[f]$ of continuous functions (sometimes called \emph{maps})
\begin{subequations}
\begin{equation}
f: I^p \to M\qquad\textrm{such that} \quad f:\partial I^p \to \mathfrak{m}\label{eqn:homo-funct-cond}
\end{equation}
where $I^p \equiv [0,1]^p$ is a $p$-dimensional hypercube, and $\partial$ is the boundary operator. Two functions $f,g$ are equivalent, $[f] = [g]$, if one can be continuously deformed into the other while preserving the boundary condition, i.e. if there exists a \emph{homotopy} $\mcF:I^p\times I\to M$ such that
\begin{eqnarray}
\mcF(\bs{x},0) &=& f(\bs{x}) \nonumber \\ 
\forall\bs{x}\in I^p,\;\;
\forall\bs{y}\in \partial I^p,\;\;
\forall t\in I:\quad \mcF(\bs{x},1) &=& g(\bs{x})\;\;\;\;\;\; \label{eqn:homotopy-map}\\
\mcF(\bs{y},t) &=& \mathfrak{m}. \nonumber  
\end{eqnarray}
The group structure on $\pi_p(M,\mathfrak{m})$ is obtained by introducing the composition rule $[f]\circ[g] = [f\circ g]$, where
\begin{equation}
(f\circ g)(x_1,x_2,\ldots) \!=\!\! \left\{\begin{array}{ll}
\! f(2x_1,x_2,\ldots)  & \! \textrm{for $x_1 \!\leq\! \tfrac{1}{2}$}  \\
\! g(2x_1\!-\! 1,x_2,\ldots)& \! \textrm{for $x_1 \!>\! \tfrac{1}{2}$}
\end{array}\right.\label{eqn:homotopy-composition-rule}
\end{equation}
\end{subequations}
and $\{x_i\}_{i=1}^p\in [0,1]^p\equiv I^p$ are coordinates along the hypercube dimensions. The inverse of $f(x_1,x_2,\ldots)$ is $f(1-x_1,x_2,\ldots)$, i.e. the same function with inverted first argument. The identity element corresponds to the equivalence class of the constant function $f_{\mathfrak{m}}:I^p\mapsto \mathfrak{m}$. The first homotopy group $\pi_1(M)$ is also called the \emph{fundamental group of $M$}.

Let us summarize some properties of homotopy groups. The group $\pi_p(M,\mathfrak{m})$ is Abelian (commutative) for $p\geq 2$, while it can be non-Abelian for $p=1$. The \emph{zeroth} homotopy set $\pi_0(M,\mathfrak{m})$, which describes mappings from the zero-dimensional sphere $S^0$ (a collection of two points) to $M$, does not in general have a group structure~\footnote{
Instead, the zeroth homotopy set forms a \emph{groupoid}.}. However, an important exception occurs if $M = \mathsf{G}$ is a \emph{group}. Then one defines $\pi_0(\mathsf{G}) = \mathsf{G}/\mathsf{G}^0$ where $\mathsf{G}^0 \leq \mathsf{G}$ is the connected component of $\mathsf{G}$ containing the identity~\cite{Mermin:1979}. A topological space $M$ is called \emph{path-connected} (or just \emph{connected} for short) if it consists of just one connected component, and $\emph{simply connected}$ if further $\pi_1(M,\mathfrak{m}) = \triv$, i.e. if all closed loops in $M$ are continuously contractible to a point. If $M$ is path-connected, then groups $\pi_p(M,\mathfrak{m})$ for various $\mathfrak{m}$ are mutually isomorphic, so one typically simplifies the notation by writing only $\pi_p(M)$. 

Note that the definition in Eq.~(\ref{eqn:homo-funct-cond}) requires that $f$ is constant on the hypercube boundary. This allows one to interpret the entire boundary $\partial I^p$ as a single point by constructing the quotient space $I^p/\partial I^p \simeq S^p$, while preserving the continuity of $f$. This construction leads to an equivalent definition of homotopy groups using \emph{$p$-spheres}. In this formulation, the group consists of equivalence classes $[f]$ of continuous functions
\begin{equation}
f: S^p \to M\qquad\textrm{such that} \quad f: \mathrm{N} \mapsto \mathfrak{m}\label{eqn:homotop-spheres}
\end{equation}
where the point $\mathrm{N}\in S^p$ [the ``North pole'' with coordinates $(0,\ldots,0,1)$] is the image of $\partial I^p$ in the quotient space. The equivalence relation and the group structure of functions in Eq.~(\ref{eqn:homotop-spheres}) should be understood by first expanding $\mathrm{N}$ back into $\partial I^p$, then applying the rules formulated for hypercubes, and finally quotienting back to $p$-spheres. In Sec.~\ref{eqn:homo-aproach}, we use the $p$-sphere formulation of homotopy groups to describe band-structure nodes.

\begin{figure}[t!]
\includegraphics[width=0.45\textwidth]{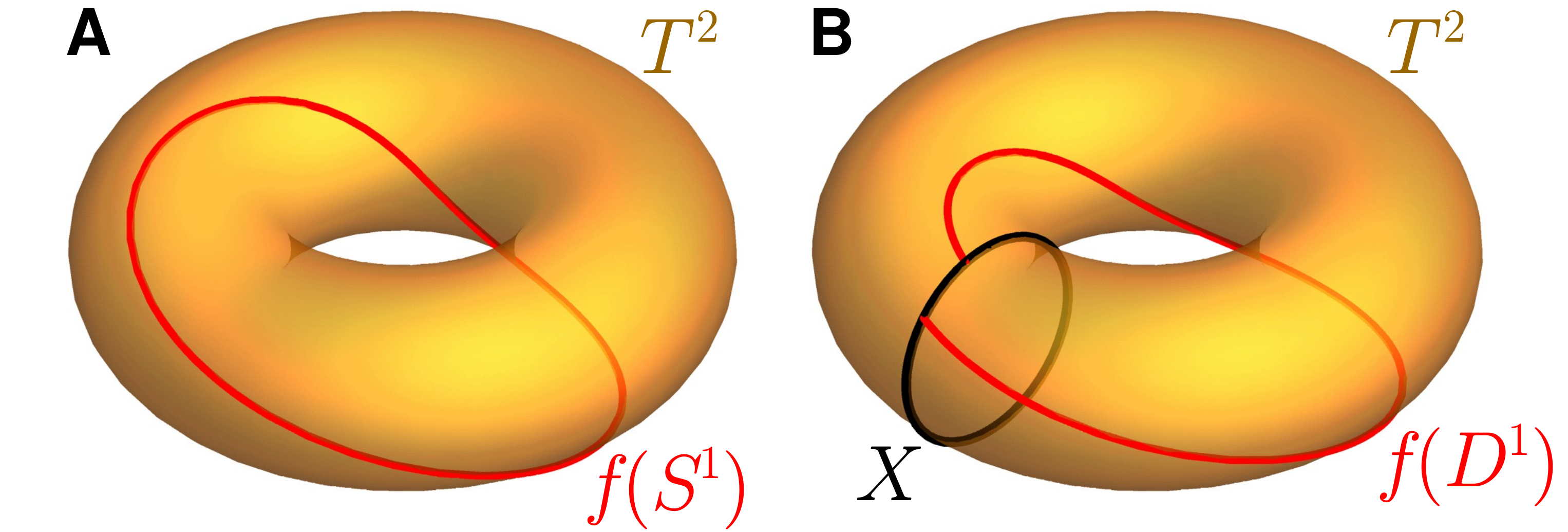}
 \caption{\textsf{(A) The first homotopy group of a torus is $\pi_1(T^2) = \intg\oplus\intg$, indicating that the map $f:S^1\to T^2$ is characterized by one winding number in the toroidal and one in the poloidal direction. The illustrated $f$ (red) winds once in both directions. (B) Taking $X\subset T^2$ to be a circle (black) winding around the torus in the poloidal direction, the resulting first \emph{relative} homotopy group is $\pi_1(T^2,X) = \intg$, corresponding to the toroidal direction only. The reason is that the endpoints of $[f]\in\pi_1(T^2,X)$ are free to move along $X$, allowing to remove any winding in the poloidal direction. Generally speaking, the relative homotopy group $\pi_p(M,X)$ with $X\subseteq M$ relaxes the conditions for two maps in $\pi_p(M)$ to be equivalent.}}
\label{fig:torus}
\end{figure}

Another useful concept is that of \emph{relative} homotopy group $\pi_p(M,X,\mathfrak{m})$ where $\mathfrak{m}\in X \subset M$ and $p \geq 2$ is integer. The relative group consists of equivalence classes $[f]$ of continuous functions $f: I^p \to M$ such that
\begin{subequations}
\begin{equation}
f:\partial I^p\to X\qquad\textrm{and}\qquad f: J^{p-1} \to \mathfrak{m} \label{eqn:rel-hom-condt}
\end{equation}
where 
\begin{equation}
J^{p-1} = I^{p-1}\!\times\!\{0\} \cup (\partial I^{p-1}) \!\times\! I
\end{equation}
is the boundary of $I^p$ without the top face $\{x_p = 1\}$. The equivalence $[f]=[g]$ is defined by the existence of homotopy $\mcF: I^p\times I\to M$ that preserves the boundary condition, i.e. $\forall\bs{x}\in I^p$, $\forall\bs{y}\in \partial I^p$, $\forall\bs{z}\in J^p$, and $\forall t\in I:$
\begin{eqnarray}
\mcF(\bs{x},0) = f\quad &\quad&  
\quad \mcF(\bs{x},1) = g(\bs{x}) \nonumber \\
\mcF(\bs{y},t) \in X \quad&\quad&\quad
\mcF(\bs{z},t) = \mathfrak{m}. \label{eqn:homotopy-map-2}
\end{eqnarray}
\end{subequations}
The group structure is achieved with the same composition rule as in Eq.~(\ref{eqn:homotopy-composition-rule}). To provide basic intuition about the concept of relative homotopy, we explicitly discuss a simple example in Fig.~\ref{fig:torus}.

Relative homotopy groups $\pi_p(M,X,\mathfrak{m})$ are Abelian for $p \geq 3$, while they can be non-Abelian for $p=2$. If $X$ is path-connected, then $\pi_p(M,X,\mathfrak{m})$ for various $\mathfrak{m}\in X$ are isomorphic, hence one typically simplifies the notation by writing just $\pi_p(M,X)$. The first relative homotopy \emph{set}, which consists of equivalence classes of function
\begin{equation}
f: I \to M\qquad\textrm{such that} \quad f:\begin{array}{rcl} \{0\} & \mapsto  & \mathfrak{m} \\ \{1\} &\to & X \end{array} 
\end{equation}
does \emph{not} in general have a group structure (instead, it forms a \emph{groupoid}), although an exception occurs when $X$ itself is a group.

Relative homotopy groups can also be formulated using spheres $S^p$ and \emph{discs} $D^p$. By a $p$-disc we mean a contractible region inside $\reals^p$ with boundary $\partial D^p \simeq S^{p-1}$. First, note that $I^p \simeq D^p$ are homeomorphic, i.e. one can be continuously deformed into the other. Furthermore, $J^{p-1}\subset \partial I^p$ can be identified as a point inside $I^p$ since $f|_{J^{p-1}} = \mathfrak{m}$ is a constant. This allows us to continuously deform the triad $(I^p\supset \partial I^p \supset J^{p-1})$ into $(D^p \supset S^{p-1}\supset \mathrm{N})$. Therefore, the relative homotopy group $\pi_p(M,X,\mathfrak{m})$ also describes equivalence classes $[f]$ of continuous functions $f: D^p \to M$ subject to 
\begin{equation}
f: S^p \to X\qquad\textrm{and} \qquad f: \mathrm{N} \mapsto \mathfrak{m}
\end{equation}
The equivalence relation and the group structure in this formulation are analogous to that of non-relative homotopy groups. We remark that a hemi-sphere (one half of a $p$-sphere) is topologically equivalent to a $p$-disc with boundary located at the equator of the original sphere. This property makes relative homotopy groups useful to classify band-structure nodes in the presence of certain crystalline symmetries~\cite{Sun:2018}, as we explained in Sec.~\ref{sec:mirror-nodes}.

The relative homotopy groups of a pair $(M,X)$ can be often explicitly constructed if the homotopy groups of $M$ and of $X$ are already known. To show this, note that a choice of $X \subset M$ induces homomorphisms $\pi_p(X)\stackrel{i_p\;}{\to} \pi_p(M) \stackrel{j_p\;}{\to} \pi_p(M,X)$, simply because the elements of the former are automatically elements of the latter. Furthermore, by restricting $f$ fulfilling Eqs.~(\ref{eqn:rel-hom-condt}) to the top face $\{x^p = 1\}$, one obtains a function, denoted $\partial f$, that fulfils conditions in Eqs.~(\ref{eqn:homo-funct-cond}) with $p\mapsto (p-1)$. This \emph{boundary map} induces a homomorphism $\pi_p(M,X)\stackrel{\partial_p\;}{\to} \pi_{p-1}(X)$. One can thus build a \emph{long sequence} of homomorphisms
\begin{equation}
\!\!\! . . .\!\stackrel{i_{p}\;}{\to}\!\pi_p(\!M\!)\!\stackrel{j_p\;}{\to} \!\pi_p(\!M,\!X\!) \!\stackrel{\partial_p\;}{\to}\! \pi_{p-1}(\!X\!)\!\!\stackrel{i_{p-1}\;}{\to}\!\!\!\pi_{p-1}(\!M\!)\!\!\stackrel{j_{p-1}\;}{\to}\!\!\!\! ... \!\!\!\!\label{eqn:LES}
\end{equation}
which can be shown to be \emph{exact}~\cite{Hatcher:2002}, meaning that the image of any arrow in the sequence equals the kernel of the next one. In practice, one combines the exactness of the sequence in Eq.~(\ref{eqn:LES}) with the relation between the image and the kernel of a group homomorphism. For example, if $\varphi: \mathsf{G}_1 \to \mathsf{G}_2$ is a group homomorphism, then there exists an isomoprhism
\begin{equation}
\im\varphi \cong \mathsf{G}_1/\ker\varphi.\label{eqn:group-homomorph}
\end{equation}
Two example derivations of relative homotopy groups appear below in Sec.~\ref{eqn:homotopy-cosets}. More technical examples which are directly relevant for the discussion of NLs in the presence of mirror symmetry appear later in Secs.~\ref{sec:NLs-NCs-stability-2} and~\ref{sec:multiband+mirror}.


\subsection{Homotopy groups of coset spaces}~\label{eqn:homotopy-cosets}

In many situations, it is possible to treat the space of Hamiltonians as a coset space~\footnote{An element of the left coset space $\mathsf{G} / \mathsf{H}$ (right coset space $\mathsf{H} \!\setminus\!\!  \mathsf{G}$) is a set obtained by acting on the group $\mathsf{H}$ from the left (from the right) by an element $g\in\mathsf{G}$, i.e. $g \mathsf{H} = \{g h | h\in\mathsf{H}\}$ (or $\mathsf{H}g = \{h g | h\in\mathsf{H}\}$). If $\mathsf{H}$ is a \emph{normal} subsgroup of $\mathsf{G}$ (see the next footnote), then the left and right coset spaces coincide, i.e. $\mathsf{G}/\mathsf{H} = \mathsf{H}\!\setminus\!\! \mathsf{G}$.} 
\begin{equation}
M = \mathsf{G}/\mathsf{H}\label{eqn:coset-space-of-h}
\end{equation}
where $\mathsf{G}$ and $\mathsf{H}$ are groups. Such a situation arises naturally if one can generate the entire space $M$ of Hamiltonians by acting on a selected one with a \emph{group of transformations} $\mathsf{G}$, while for every selected Hamiltonian $\mcH\in M$ there exists a \emph{normal subgroup} $\mathsf{H}\triangleleft \mathsf{G}$~\footnote{By definition, normal subgroup $\mathsf{H}\triangleleft\mathsf{G}$ commutes (as a set) with all elements  of $\mathsf{G}$, i.e. formally $\forall g\in\mathsf{G}:g\mathsf{H} g^{-1}= \mathsf{H}$. The interpretation in the present context is that all the Hamiltonians $\mcH \in M$ should have the same isotropy subgroup. This latter property further follows from the assumption that the space of Hamiltonians $M$ can be obtained by acting on a chosen element $\mcH\in M$ transitively by a group of transformations $\mathsf{G}$, i.e. by definition $M$ is a \emph{homogeneous space}.} which keeps $\mcH$ invariant (also called \emph{stabilizer group} or \emph{isotropy subgroup} of $\mcH \in M$). Spaces of the form in Eq.~(\ref{eqn:coset-space-of-h}) allow us to derive certain simple results for homotopy groups $\pi_p(M)$ by virtue of the long exact sequence in Eq.~(\ref{eqn:LES})~\cite{Mermin:1979}. 

More specifically, let us assume that $\mathsf{G}$ is a Lie group which is \emph{path-connected} [$\pi_0(\mathsf{G}) = \triv$] and \emph{simply connected} [$\pi_1(\mathsf{G}) = \triv$]. If we start with a group $\mathsf{G}$ that does not satisfy these conditions, the remedy is to first replace it by its connected component containing identity $\mathsf{G}^0$, and then to take its universal cover $\widetilde{\mathsf{G}}^0$, which by construction \emph{is} simply connected. Both constructions are uniquely defined for an arbitrary Lie group $\mathsf{G}$. We encounter examples of such constructions in Secs.~\ref{subsec:A21}, ~\ref{subsec:3-deriv} and~\ref{subsec:gen-many}. Furthermore, a theorem by Cartan~\cite{Cartan:1936,8961} guarantees that for all {Lie groups} also $\pi_2(\mathsf{G}) = \triv$. We show that the simultaneous triviality of $\pi_{0,1,2}(\mathsf{G})$ together imply that
\begin{subequations}\label{eqn:coset-homo}
\begin{eqnarray}
\pi_1(\mathsf{G}/\mathsf{H}) &=& \pi_0(\mathsf{H}) = \mathsf{H}/\mathsf{H}^0\label{eqn:fundamental-theorem} \\ 
\pi_2(\mathsf{G}/\mathsf{H}) &=& \pi_1(\mathsf{H} ^0)\label{eqn:fundamental-theorem-2}
\end{eqnarray}
\end{subequations}
where $\mathsf{H}^0$ is the connected component of $\mathsf{H}$ containing the identity. The derivation of Eqs.~(\ref{eqn:coset-homo}) is somewhat technical and fills the rest of the present section.

\begin{figure*}[t!]
\includegraphics[width=0.99 \textwidth]{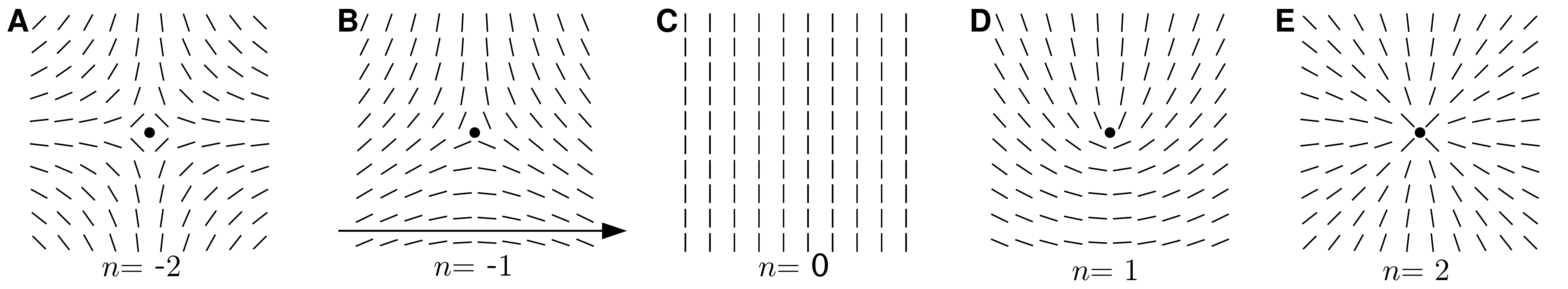}
 \caption{\textsf{Planar directors with the indicated winding number around the defect in the center (black dot). If the directors are allowed to rotate out of the plane, as is the case for $\mcP\mcT$-symmetric Hamiltonians with $n=1$ occupied and $\ell=2$ unoccupied bands (or vice versa) or for unixial nematics, then only the \emph{$\ztwo$ parity} remains well defined. For example, the $n=2$ singularity (panel E) clearly goes away after a continuous transformation that orients all the directors to be perpendicular to the plane of the paper. Alternatively, rotating all the directors of the $n=-1$ defect (panel b) by $\pi$ around the indicated axis (while keeping the \emph{positions} of the directors unchanged) transforms the picture into the situation with $n=1$ (panel D).}}
\label{fig:windings}
\end{figure*}

We first derive the result in Eq.~(\ref{eqn:fundamental-theorem}). Let us consider a circle $S^1$ continuously embedded in $\mathsf{G}/\mathsf{H}$ with a basepoint at coset $\mathsf{H}$ (which is a \emph{point} inside the coset space). This embedding is described by a map $f:I\to \mathsf{G}/\mathsf{H}$ subject to $f(0)=f(1)=\mathsf{H}$. There exists a \emph{lift} of $f$ into a continuous map $g:I\to \mathsf{G}$ such that $g(0) = {e}$ is the identity element and $g(1)\in\mathsf{H}$. We can thus express $\pi_1(\mathsf{G}/\mathsf{H}) = \pi_1(\mathsf{G},\mathsf{H})$, i.e. as the \emph{first relative homotopy group} of the pair $\mathsf{G}\geq \mathsf{H}$ (which is a \emph{stronger} statement than the subspace relation $\mathsf{G}\supseteq \mathsf{H}$). We obtain from the long exact sequence in Eq.~(\ref{eqn:LES}) and from the triviality $\pi_{1,2}(\mathsf{G}) = \triv$ that
\begin{subequations}
\begin{equation}
\triv\stackrel{j_1\;}{\to} \pi_1(\mathsf{G}/\mathsf{H}) \stackrel{\partial_1\;}{\to} \pi_0(\mathsf{H})\stackrel{i_0\;}{\to}\triv.\label{eqn:LES-p1}
\end{equation}
It follows from the exactness that $\ker\partial_1 = \triv$ and that $\im \partial_1 = \pi_0(\mathsf{H})$, which together imply that $\partial_1$ is a \emph{group isomorphism}. Recalling the definition of zeroth homotopy group of a group in Sec.~\ref{eqn:homotopy-theory}, we arrive at the result in Eq.~(\ref{eqn:fundamental-theorem}).

To derive Eq.~(\ref{eqn:fundamental-theorem-2}), let us consider a map $f: I^2\to \mathsf{G}/\mathsf{H}$ subject to $f(\partial I^2) = \mathsf{H}$. Such a map $f$ can be lifted into a map $g: I^2 \to \mathsf{G}$ fulfilling $g: \partial I^2 \to \mathsf{H}$. Because of the continuity of $g$, the boundary map $\partial g$ must lie in a \emph{single connected component} of $\mathsf{H}$. Therefore, using an appropriate multiplication within the group and homotopy, we are able to continuously deform the lift such that it fulfills $g: \partial I^2\to \mathsf{H}^0$ and $g(J^1) = {e}$. We can thus express $\pi_2(\mathsf{G}/\mathsf{H}) = \pi_2(\mathsf{G},\mathsf{H}^0)$, i.e. as the \emph{second relative homotopy group} of pair $\mathsf{G}\geq \mathsf{H}^0$. Considering the exact sequence~(\ref{eqn:LES}) for $p=2$ together with the trivial homotopy groups $\pi_{0,1}(\mathsf{G}) = \triv$, we obtain
\begin{equation}
\triv\stackrel{j_2\;}{\to} \pi_2(\mathsf{G}/\mathsf{H}) \stackrel{\partial_2\;}{\to} \pi_1(\mathsf{H}^0)\stackrel{i_1\;}{\to}\triv \label{eqn:LES-p2}
\end{equation}
\end{subequations}
Similar to our discussion of Eq.~(\ref{eqn:LES-p1}), it follows from the exactness of Eq.~(\ref{eqn:LES-p2}) that $\partial_2$ is a group isomorphism, such that Eq.~(\ref{eqn:fundamental-theorem-2}) automatically follows.


\section{\texorpdfstring{$\mcP\mcT$}{PT}-symmetric nodal lines}~\label{sec:NLs-NCs-stability}

In this section, we apply homotopy theory presented in Sec.~\ref{eqn:homo-aproach} to derive the topological charge and the properties of nodal lines (NLs) in three-dimensional (3D) $\mcP\mcT$-symmetric systems with negligible spin-orbit coupling (SOC). The discussion here ignores the possible presence of a mirror symmetry, which will be explicitly included in the next Sec.~\ref{sec:NLs-NCs-stability-2}. Furthermore, we only consider here the description of nodes between the highest occupied (HO) and the lowest unoccupied (LU) band~\footnote{In the case metals, we apply the terminology of HO and LU bands to denote two consecutive bands even if these labels do not reflect the actual occupation of the bands by electrons.} as developed by Ref.~\cite{Bzdusek:2017}. The generalized multi-band description of NLs appears later in Sec.~\ref{sec:multi-homotopy}.

Our discussion is structured as follows. We begin in Sec.~\ref{subsec:A11} with a brief overview of NLs in two-band models, which are protected by a $\intg$-valued winding number. Subsequently, we discuss in Sec.~\ref{subsec:A21} NLs of three-band models, which are protected by a $\ztwo$-valued topological charge related to quantization of Berry phase. We will see that NLs in three-band models are mathematically related to $\pi$-vortex lines in (uniaxial) nematic liquid crystals.


\subsection{Two-band models}\label{subsec:A11}

We consider a two-band model with one occupied and one unoccupied band. The most general (spectrally flattened) Hamiltonian in Eq.~(\ref{eqn:gen-ham-form}) simplifies to
\begin{subequations}
\begin{equation}
\mcH_{2}(\bs{k})=\unit - 2|{u^\textrm{o}_{\bs{k}}}\rangle\langle{u^\textrm{o}_{\bs{k}}}|\label{eqn:H-A11}
\end{equation}
where $|{u^\textrm{o}_{\bs{k}}}\rangle$ is the cell-periodic component of the two-component Bloch function corresponding to the lower energy (i.e.~``occupied'') state. The commutation with $\mcP\mcT = \mcK$ allows us to locally gauge $|{u^\textrm{o}_{\bs{k}}}\rangle$ to be \emph{real}. It follows from Eq.~(\ref{eqn:H-A11}) that this vector uniquely specifies the Hamiltonian, so we can visualize the \emph{Hamiltonian itself} as a normalized two-component (i.e.~planar) vector. However, this description contains a redundancy in the overall sign of the vector, meaning that the Hamiltonian in Eq.~(\ref{eqn:H-A11}) can be represented as a planar \emph{unoriented director}. Therefore, the order-parameter space is topologically a circle $S^1$ with identified antipodal points. But such an identification $S^1/\ztwo \cong S^1$ is still a circle, i.e.
\begin{equation}
M_{(1,1)}=\mathsf{O}(2)/\mathsf{O}(1)\!\times\!\mathsf{O}(1)\cong S^1/\ztwo \cong S^1.\label{eqn:AI-1-1-op-space}
\end{equation} 
The first homotopy group is easy to find, namely
\begin{equation}
\pi_1(M_{(1,1)}) = \intg. \label{eqn:A11-pi1-value}
\end{equation}
\end{subequations}
The defects in two (three) spatial dimensions are nodal points (lines) characterized by an integer winding number $n\in\intg$. We call such defects $n\pi$-vortices, since the director rotates by $n\pi$ when carried around the defect. A few simple examples of fields exhibiting defects with $n\in{0,\pm 1,\pm 2}$ are plotted in Fig.~\ref{fig:windings}. As mentioned in the main text, the integer character of the charge assigns these topological defects a well-defined \emph{orientation}.


\subsection{Three-band models}\label{subsec:A21}

Spectrally flattened Hamiltonians $\mcH_{(1,2)}(\bs{k})$ of a system with $n=1$ occupied and $\ell = 2$ unoccupied bands can be expressed again using the right-hand side of Eq.~(\ref{eqn:H-A11}), where $|{u^\textrm{o}_{\bs{k}}}\rangle$ is now a \emph{three}-component real normalized vector. Since the vector is well-defined only up to an overall sign, the Hamiltonian is encoded by an unoriented director in \emph{three} dimensions, 
\begin{subequations}
\begin{equation}
M_{(1,2)} \cong {S^2}/{\ztwo} \equiv \reals P^2 \label{eqn:RP2-double-cover-1}
\end{equation}
where $\reals P^2$ is called the \emph{real projective plane}. It can be visualized as a hemisphere with identified antipodal points along the equator, see Fig.~\ref{fig:RP2}(A). By symmetry, the space of Hamiltonians with switched number of occupied and unoccupied states is the same, i.e. $M_{(1,2)} \simeq M_{(2,1)}$. A mathematically equivalent situation to Eq.~(\ref{eqn:RP2-double-cover-1}) arises in (uniaxial) nematic liquid crystals, where the unoriented director describes the configuration of long molecules with approximate cylindrical symmetry~\cite{Chaikin:1995,Sethna:2006}. 

\begin{figure*}[t!]
\includegraphics[width=0.99 \textwidth]{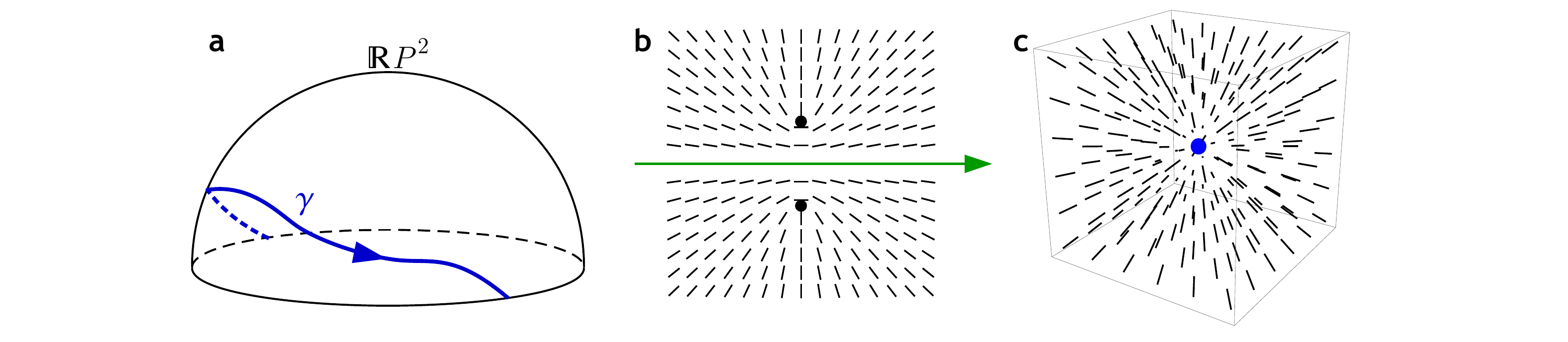}
 \caption{\textsf{(A) Real projective plane $\reals P^2$ in Eq.~(\ref{eqn:RP2-double-cover-1}) can be represented as a hemisphere with identified antipodal points on the equator. The blue path $\gamma$ is \emph{closed} and represents a non-trivial element of the homotopy group in Eq.~(\ref{eqn:AI-1-2-1st-homotopy}). The particular choice of $\gamma$ in (A) encodes an order-parameter field similar to that of Fig.~\ref{fig:windings}(D). (B) A node with a non-trivial value of the second-homotopy charge in Eq.~(\ref{eqn:AI-1-2-2nd-homotopy}) can be obtained by revolving the indicated field of directors with two point defects (black dots) around the green line in the centre. Revolution of the two point nodes produces a nodal loop in 3D. (C) Shrinking the nodal loop to a single point (blue dot) creates a ``hedgehog'' defect characterized by a non-trivial element of the second homotopy group in Eq.~(\ref{eqn:AI-1-2-2nd-homotopy}).}}
\label{fig:RP2}
\end{figure*}

To derive the homotopy groups of $M_{(1,2)}$, it is convenient to rewrite the expression in Eq.~(\ref{eqn:RP2-double-cover-1}) as a coset space~\cite{Mermin:1979}
\begin{equation}
M_{(1,2)} \cong {\mathsf{SO}(3)}/{\mathsf{D}_\infty} \cong {\textsf{SU}(2)}/{\overline{\mathsf{D}_\infty}}\label{eqn:RP2-double-cover}
\end{equation}
\end{subequations}
with a simply connected Lie group $\textsf{SU}(2)$ in the numerator, such that one can apply the theorems derived in Sec.~\ref{eqn:homotopy-cosets}. The relations in Eq.~(\ref{eqn:RP2-double-cover}) are understood as follows. First, $\mathsf{SO}(3)$ is a Lie group of transformations (rotations) acting on the director, while the \emph{infinite dihedral group} $\mathsf{D}_\infty < \mathsf{SO}(3)$  is the isotropy subgroup that keeps a chosen director invariant. It comprises rotations by arbitrary angle around the director as well as all $\pi$-rotations around axes perpendicular to the director. The group $\mathsf{D}_\infty$ is non-Abelian, and it can be expressed as a semidirect product $\mathsf{D}_\infty = \mathsf{SO}(2)\rtimes \ztwo$. Assuming Cartesian coordinates with $\pm \hat{z}$ along the director, the isotropy subgroup consists of matrices 
\begin{subequations}
\begin{equation}
R_0(\alpha) = \e{\alpha L_3} = \left(\begin{array}{ccc}
\cos \alpha    & -\sin\alpha  & 0     \\
\sin\alpha     & \cos\alpha   & 0     \\
0               &  0            & 1
\end{array}\right)
\end{equation}
\begin{center}and\end{center}
\begin{equation}
R_1(\alpha) = \e{\alpha L_3}\e{\pi L_1} = \left(\begin{array}{ccc}
\cos \alpha    & \sin\alpha   & 0     \\
\sin\alpha    & -\cos\alpha  & 0     \\
0               &  0            & -1
\end{array}\right)
\end{equation}
\end{subequations}
where $\alpha \in [0,2\pi)$ and matrices 
\begin{subequations}\label{eqn:replace-generators}
\begin{equation}
(L_{i})_{jk}=-\epsilon_{ijk}\label{eqn:gens-so-3}
\end{equation}
($\eps_{ijk}$ is the fully antisymmetric tensor) form the basis of Lie algebra $\mathfrak{so}(3)$. As a topological space, $\mathsf{D}_\infty$ looks like the \emph{disjoint union} of two circles (denoted $S^1\amalg S^1$). We set the subscript $a\in\{0,1\}$ of $R_a(\alpha)$ according to the corresponding element of the zeroth homotopy group $\pi_0(\mathsf{D}_\infty) = \ztwo$, which captures the two-component structure of $\mathsf{D}_\infty$. The connected component containing the identity is $\mathsf{D}_\infty^0 \cong \mathsf{SO}(2)\cong S^1$.

The second transformation in Eq.~(\ref{eqn:RP2-double-cover}) corresponds to taking the double cover in order to obtain a \emph{simply connected} group of transformations in the numerator. The lift of a matrix $R\in\mathsf{SO}(3)$ into the double cover $\mathsf{Spin}(3) = \mathsf{SU}(2)$ is achieved by replacing the generators $L_j$ in  $R = \e{\alpha_i L_i}$ by $\mathfrak{su}(2)$ basis elements
\begin{equation}
t_j = -\tfrac{\imi}{2}\sigma_j. \label{eqn:gens-su-2}
\end{equation}
\end{subequations}
which obey the same structure constants~\cite{Fecko:2006}. We use the symbol $\sigma_{i}$ to denote the Pauli matrices. 

The same substitution is performed on the subgroup $\mathsf{D}_\infty$, which yields the lift $\overline{\mathsf{D}_\infty} < \mathsf{SU}(2)$ consisting of
\begin{subequations}
\begin{equation}
\overline{R}_0(\alpha) = \e{\alpha t_3} = \left(\begin{array}{cc}
\e{-\imi\frac{\alpha}{2}}    & 0     \\
0     & \e{\imi\frac{\alpha}{2}}      
\end{array}\right)
\end{equation}
\begin{center}and\end{center}
\begin{equation}
\overline{R}_1(\alpha) = \e{\alpha t_3}\e{\pi t_1} = \left(\begin{array}{cc}
0               & -\imi \e{-\imi\frac{\alpha}{2}}       \\
-\imi \e{\imi\frac{\alpha}{2}}    & 0
\end{array}\right)
\end{equation}
\end{subequations}
where now $\alpha \in [0,4\pi)$ because of the double covering. Here as well as in the following sections, we denote the lift of a point group $\mathsf{H}\leq \mathsf{SO}(N)$ into the double cover $\mathsf{Spin}(N)$ with a horizontal bar, i.e. as $\overline{\mathsf{H}}$.

As a topological space, the group $\overline{\mathsf{D}_\infty }$ is still homeomorphic to $S^1\amalg S^1$, and the subscript of $\overline{R}_a(\alpha)$ is again chosen according to the corresponding element of $\pi_0(\overline{\mathsf{D}_\infty})=\ztwo$. It follows by applying theorems in Eqs.~(\ref{eqn:fundamental-theorem}) and~(\ref{eqn:fundamental-theorem-2}) from Sec.~\ref{eqn:homotopy-cosets} that
\begin{subequations}
\begin{eqnarray}
\pi_1(M_{(1,2)}) &=& \pi_0(\overline{\mathsf{D}_\infty}) = \ztwo \label{eqn:AI-1-2-1st-homotopy} \\
\pi_2(M_{(1,2)}) &=& \pi_1(\overline{\mathsf{D}_\infty}\;\!^0) = \intg \label{eqn:AI-1-2-2nd-homotopy}
\end{eqnarray}
\end{subequations}
These results describe the topological charge of nodes in nodal class $\textrm{AI}$ with $n=1$ and $\ell =2$ bands. In particular, Eq.~(\ref{eqn:AI-1-2-1st-homotopy}) implies the possibility of nodal points (lines) in 2D (3D) systems. The $\ztwo$ character of the topological charge implies that, unlike the two-band case, one \emph{cannot} assign nodal lines in three-band models a well-defined orientation. It can be shown that the $\ztwo$ character of the fundamental group does not change if the number of bands grows to $n+\ell > 3$~\cite{Fang:2015,Bzdusek:2017}. Furthermore, Eq.~(\ref{eqn:AI-1-2-2nd-homotopy}) states that closed nodal \emph{loops} (which can be enclosed by $S^2$) can carry an \emph{additional} integer charge, so-called \emph{monopole charge}. Although this property of nodal lines~\cite{Fang:2015,Bzdusek:2017} is not directly relevant to our exposition in the main text, Ref.~\cite{Tiwari:2019} uncovers some deeper connections between the monopole charge and the non-Abelian topology derived in Sec.~\ref{sec:multi-homotopy}. 

An example of a non-trivial map from $S^1$ to $M_{(1,2)}$ is shown in Fig.~\ref{fig:RP2}(A). The defect can be called a $\pi$-vortex, since the director performs a $\pi$-rotation when carried around the defect. Importantly, there is no difference between a $+\pi$ and a $-\pi$ vortex. Indeed, rotating all the directors in Fig.~\ref{fig:windings}(B) by $\pi$ around the indicated horizontal axis (while keeping the \emph{positions} of the individual directors fixed) produces the field in Fig.~\ref{fig:windings}(D). Therefore, the winding number $n$ is now defined only modulo $2$, thus manifesting the underlying $\ztwo$ character stated by Eq.~(\ref{eqn:AI-1-2-1st-homotopy}). Consequently, any pair of non-trivial vortices can annihilate when brought together. The realization of a node carrying a non-trivial value of the \emph{second}-homotopy charge~(\ref{eqn:AI-1-2-2nd-homotopy}) requires three spatial dimensions. An example of such a ``hedgehog'' defect is illustrated in Fig.~\ref{fig:RP2}(B and C).


\section{\texorpdfstring{Nodal chains protected by $\mcP\mcT$}{PT} and mirror symmetry}~\label{sec:NLs-NCs-stability-2}
With a solid understanding of NLs protected by $\mcP\mcT$ in systems with negligible SOC, we turn our attention to the formation of nodal \emph{chains} in the presence of an additional mirror symmetry. This requires the application of the \emph{relative} homotopy description developed in Ref.~\cite{Sun:2018} and reviewed in Sec.~\ref{sec:mirror-nodes}. 

We begin the discussion in Sec.~\ref{sec:NCs-two-bands} by considering two-band models. We show that the presence of a single mirror symmetry enables the formation of \emph{intersecting} NLs, and we identify a topological invariant responsible for the stability of the NL crossing point (CP), corresponding to Eq.~(\ref{eqn:Z-TP-inv}). Near the CP, two NLs always approach each other at right angle. Removing the mirror symmetry (e.g. through the application of strain) \emph{detaches} the NLs. However, the detachment is \emph{never trivial} towards a horizontal and a vertical NL. Instead, the separated NLs are ``\emph{mixed}'', and perform a sharp overturn around the former position of the CP. We show that this behavior also follows from the computed homotopy groups.

In Sec.~\ref{sec:NCs-three-bands} we show that including a third band \emph{removes} the mirror-induced topological invariant protecting the CP of two-band models. This is reflected by the fact that the homotopy group in Eq.~(\ref{eqn:Z-TP-no-inv}) is \emph{trivial}. Indeed, we demonstrate that touching NLs can be trivially separated through a process of hybridization with the additional band. However, we observe that such a separation is only possible at the cost of connecting together \emph{another pair} of bands. This observation of CP \emph{transfer} eventually leads us to consider the multi-band description outlined in Sec.~\ref{sec:PT-nodes}. Armed with more powerful mathematical tools, we explain in Sec.~\ref{sec:multi-homotopy} the transfer of CPs using a $\ztwo$ invariant corresponding to Eq.~(\ref{eqn:Z2-TP-inv}).


\subsection{Two-band models}~\label{sec:NCs-two-bands}

An extensive discussion of two-band $\mcP\mcT$-symmetric Hamiltonians with additional $m_z$ mirror symmetry appears in the main text. Rather than repeating the geometric arguments presented in the main text, we reformulate here the topological arguments explaining the stability of CPs using the homotopy language of Sec.~\ref{eqn:homo-aproach}. Although more abstract, the advantage of the homotopy approach is that it provides a striaghtforward generalization to the case of multiple bands. When an explicit Hamiltonian example is necessary, we suggest the reader to consider the Hamiltonian in Eq.~(4) of the main text, which creates the NL compositions plotted in Fig.~1 of the main text. 

For a generic momentum $\bs{k}\in\textrm{BZ}$, the space of (spectrally projected) Hamiltonians is $M_{(1,1)} = S^1$, cf. Eq.~(\ref{eqn:AI-1-1-op-space}). NLs lying \emph{outside} the symmetric plane are still stabilized by the integer winding number in Eq.~(\ref{eqn:A11-pi1-value}), just like in the absence of the mirror symmetry. On the other hand, for momenta $\bs{k}$ \emph{inside} the $m_z$-invariant plane, the Hamiltonian has to commute with the mirror operator. Assuming $\hat{m}_z = \sigma_z$, the subspace of $m_z$-symmetric Hamiltonians reduces to
\begin{subequations}
\begin{equation}
X_{m_z} = \{\pm\sigma_z\} \simeq S^0.
\end{equation}
Therefore, NLs lying inside the $k_z = 0$ plane can be interpreted as \emph{domain walls} between regions with different mirror eigenvalue of the occupied band (denoted $\lambda^\textrm{o}_{\bs{k}}$ in the main text). Using the language of homotopy theory, we say that \emph{in-plane} NLs are protected by the zeroth homotopy group 
\begin{equation}
\pi_0(X_{m_z})=\ztwo. \label{eqn:AI-1-1-mz-pi0}
\end{equation}
\end{subequations}
The $\ztwo$ character implies that any pair of in-plane NLs can be mutually removed from the $m_z$-invariant plane. The value of the charge in Eq.~(\ref{eqn:AI-1-1-mz-pi0}) for a pair of points $\bs{k}_1,\bs{k}_2$ is labeled $\nu_{\bs{k}_1,\bs{k}_2}$ in the main text.  

While the invariant in Eq.~(\ref{eqn:AI-1-1-mz-pi0}) fixes the corresponding NL to the $m_z$-invariant plane, in-plane NLs also carry a non-trivial value of the winding number in Eq.~(\ref{eqn:A11-pi1-value}). To see this, recall from Eq.~(1) of the main text that we can decompose the Hamiltonian using the Pauli matrices $\sigma_{x,y,z}$ as
\begin{subequations}
\begin{equation}
\mcH(\bs{k}) = h_x (\bs{k}) \sigma_x + h_z(\bs{k}) \sigma_z,
\end{equation}
where $h_x(\bs{k})$ is an odd and $h_z(\bs{k})$ is an even function of $k_z$ when $m_z$ is present. The space of Hamiltonians in Eq.~(\ref{eqn:AI-1-1-op-space}) is related to the normalization of the band energies to $\pm 1$. Explicitly, 
\begin{equation}
M_{(1,1)} = \left\{(h_x,h_z)\in \reals^2 | h_x^2 + h_z^2 = 1\right\} \cong S^1. \label{AI11-why-s1} 
\end{equation}
\end{subequations}
We know that $h_x(\bs{k})$ changes sign across the $k_z = 0$ plane. Furthermore, an in-plane NL acts like a domain wall separating regions with positive vs.~negative $h_z(\bs{k})$. Since both $h_{x,z}(\bs{k})$ are \emph{odd} functions of $\bs{k}$ in the vicinity of an in-plane NL, it follows that a small loop enclosing the NL winds non-trivially around the space in Eq.~(\ref{AI11-why-s1}), i.e. it carries a non-trivial value of the winding number in Eq.~(\ref{eqn:A11-pi1-value}). The consequence is that in-plane NLs are not gapped upon breaking the $m_z$ symmetry, but are just released outside of the $m_z$-invariant plane.

To understand the stability of CPs, one needs to consider closed $m_z$-symmetric paths, which pass through the $m_z$-invariant plane at two points $\partial \gamma$ with the same value of  $\lambda_{\bs{k}}^\textrm{o}$~\cite{Sun:2018}. Because of the $m_z$ symmetry, all the information about the Hamiltonian $\mcH$ on such a closed path $\Gamma$ is contained on the open-ended segment $\gamma$ located in the upper half-space. Taking into account the orientation of the paths, we have
\begin{subequations}
\begin{equation}
\Gamma = (m_z \gamma)^{-1} \circ \gamma.\label{eqn:gamma-Gamma}
\end{equation}
When composing paths, we use the convention that we first go along the path indicated to the right of the composition symbol [i.e.~$\gamma$ in the case of Eq.~(\ref{eqn:gamma-Gamma})]. The inverse in the superscript indicates that we traverse the corresponding path in the reverse direction.

The segment lying in the upper half-space can be understood as an embedding of a one-dimensional \emph{disc} $D^1$, 
\begin{equation}
\gamma = \iota(D^1)
\end{equation}
\end{subequations}
If the spectrum on $\gamma$ is gapped, the endpoints $\partial D^1 \cong S^0$ are mapped by $\mcH\circ \iota$ to $X_{m_z}$, while all the intermediate points of $D^1$ are mapped to $M_{(1,1)}$. According to the discussion in Sec.~\ref{sec:mirror-nodes}, the equivalence classes of such constrained maps are captured by the \emph{first relative homotopy set} $\pi_1(M_{(1,1)},X_{m_z})$~\cite{Sun:2018}. 

We mentioned in Sec.~\ref{sec:mirror-nodes} that the first relative homotopy set does not in general have a group structure. To avoid this complication, we restrict our attention to situations where both endpoints $\partial D^1$ are mapped by $\mcH\circ\iota$ to the {same connected component} of $X_{m_z}$. This is sufficient to guarantee a group structure~\cite{Hatcher:2002}. In practice, this restriction implies that the endpoints $\partial \gamma$ should lie inside domains with the same $\hat{m}_z$ eigenvalue of the the occupied band. However, since connected components of $X_{m_z}$ are \emph{points}, the relative homotopy group is isomorphic to the non-relative (pointed) homotopy group in Eq.~(\ref{eqn:A11-pi1-value}), i.e. it is just a winding number, 
\begin{subequations}
\begin{equation}
\pi_1(M_{(1,1)},X^+) = \pi_1(M_{(1,1)},X^-) = \intg.\label{eqn:relative-homo-AI-1-1}
\end{equation}
In the previous expression, we use $X^+$ ($X^-$) to denote the connected component of $X_{m_z}$ with positive (negative) $m_z$ eigenvalue $\lambda_{\bs{k}}^\textrm{o}$ of the occupied band. The integer invariant in Eq.~(\ref{eqn:relative-homo-AI-1-1}) counts the number of turns performed by $(\mcH\circ\iota)(D^1)$ inside $M = S^1$. The image of the \emph{complete} path $\Gamma$ exhibits \emph{twice} that number of turns in $M$, corresponding to even integers in $\pi_1(M)$.

\begin{figure}[t]
	\includegraphics[width=0.48\textwidth]{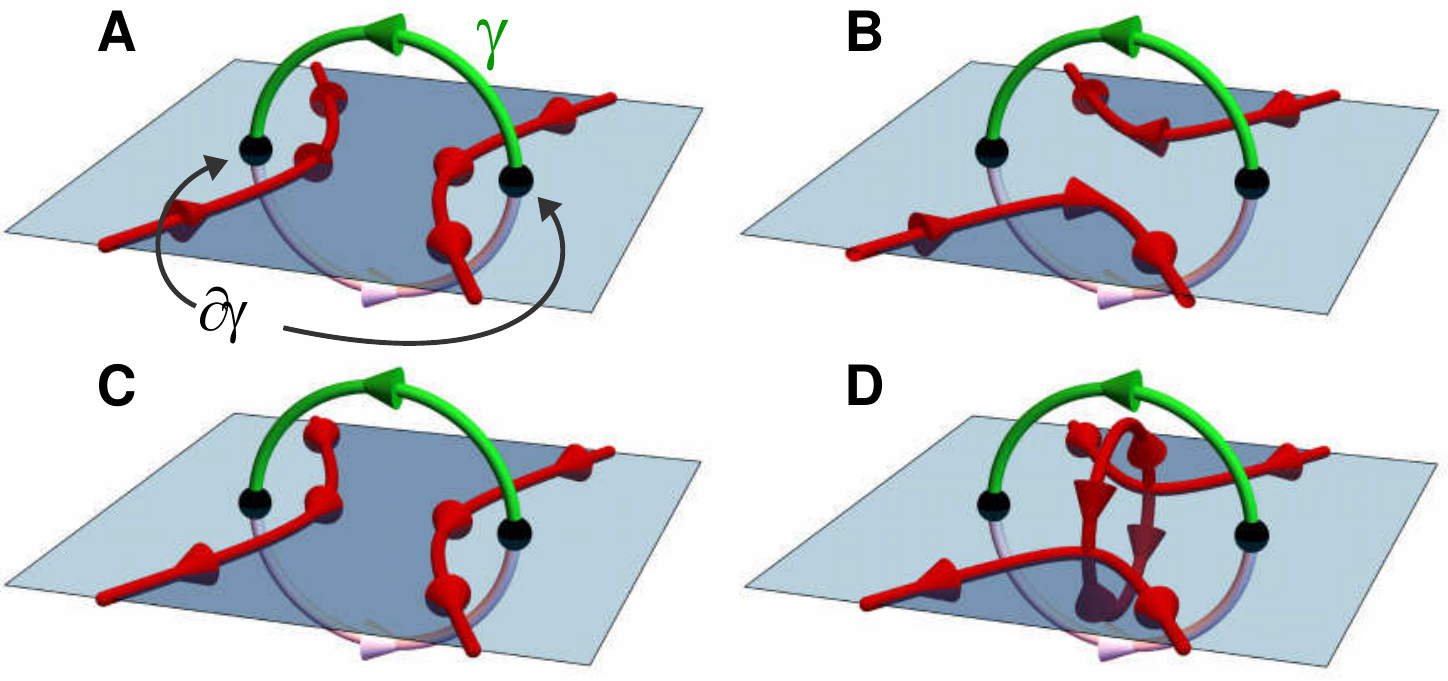}
	\caption{\textsf{(Similar to Fig.~2 of the main text.) A pair of in-plane NLs (red) is enclosed by a $m_z$-symmetric path $\Gamma$, which is decomposed into a segment $\gamma$ (green) lying in the upper half-space, and its mirror image $m_z \gamma$ (bright pink) lying in the lower half-space. The endpoints $\partial \gamma$ (black dots) lie inside the $m_z$-invariant plane. The in-plane NLs separate regions with a different $\hat{m}_z$ eigenvalue of the occupied band (shown in two shades of blue). The endpoints $\partial \gamma$ lie inside regions of the same $\hat{m}_z$ eigenvalue. (A and B) Colliding a pair of in-plane NLs with \emph{opposite} winding leads to a simple reconnection of the NLs. (C and D) Colliding in-plane NLs with \emph{parallel} winding leads to the appearance of two out-of-plane NL arcs. The crossing point (CP) connecting the out-of-plane NLs to the in-plane NLs is protected by the invariant in Eq.~(\ref{eqn:relative-homo-AI-1-1}) along the semi-circle $\gamma$.}}
	\label{fig:windings-2}
\end{figure}

To see how the result in Eq.~(\ref{eqn:relative-homo-AI-1-1}) facilitates robust crossing points (CPs) of NLs, consider a path $\Gamma$ encircling a pair of nodal lines with \emph{parallel} winding as illustrated in Fig.~\ref{fig:windings-2}(C). Clearly, the $\ztwo$ invariant of Eq.~(\ref{eqn:AI-1-1-mz-pi0}) carried by $\partial D^1$ is \emph{trivial} because both endpoints lie in regions with the same $\hat{m}_z$ eigenvalue of the occupied state. It is therefore possible to pairwise remove segments of the two in-plane NLs from the $m_z$-invariant plane using a symmetry-preserving deformation of the Hamiltonian, such as to make the two endpoints $\partial\gamma$ contractible to a single point without encountering a node. However, the winding number on $\Gamma$ is clearly non-vanishing, namely $\abs{n_\Gamma} = 2$ (or equivalently $\abs{n_\gamma} = 1$ for the segment in the upper half-space). This implies the presence of a node preventing the contraction of $\Gamma$ to a point even \emph{after} segments of the in-plane NLs have been pairwise removed. We thus deduce that a pair of \emph{out-of-plane} NL must appear which are connected to the in-plane NLs as in Fig.~\ref{fig:windings-2}(D). 

The existence of robust CPs follows from a non-trivial group $\pi_1(M,X_{m_z})$ in the presence of the $m_z$ symmetry. However, the charge $\abs{n_\Gamma} = 2$ is meaningful even in the \emph{absence} of the mirror. This implies that an obstruction for shrinking $\Gamma$ to a point persists even when $m_z$ is broken. Since the \emph{trivial} separation of the NL composition into the in-plane and the out-of-plane component would make $\Gamma$ contractible to a point, such a scenario is topologically prohibitted. Instead, breaking $m_z$ has to mix the in-plane and out-of-plane NLs as observed in Fig.~1(E and F) of the main text.

The discussion in the previous paragraph should be contrasted with the situation in Fig.~\ref{fig:windings-2}(A) when $\Gamma$ encircles in-plane NLs with \emph{opposite} winding. In that case, the winding number on $\Gamma$ vanishes, such that colliding a pair of in-plane NLs does \emph{not} produce out-of-plane NLs. Instead, only a simple reconnection of the in-plane NLs occurs as illustrated in Fig.~\ref{fig:windings-2}(B).

Before concluding this section, note that within the two-band approximation
\begin{equation}
\pi_1(X_{m_z}\subset M_{(1,1)}) = \triv.\label{eqn:A-1-1-in-plane-homotopy}
\end{equation}
\end{subequations}
This means that any in-plane closed path $\Gamma$ can be contracted to a point without encountering a band degeneracy (such process may require an appropriate continuous deformation of the Hamiltonian). Especially, the result in Eq.~(\ref{eqn:A-1-1-in-plane-homotopy}) implies that an out-of-plane NL can cross the symmetric plane \emph{only} along an in-plane NL, i.e~it automatically produces a CP. This statement is easily proved by contradiction: An out-of-plane NL crossing the symmetric plane \emph{away from} the in-plane NL would create a point singularity inside the symmetric plane. This point singularity could be enclosed by $S^1$ which should carry a non-trivial value of charge $\pi_1(X_{m_z})$. But within the two-band approximation, this homotopy group is trivial, leading to the contradiction.  


\subsection{Three-band models}~\label{sec:NCs-three-bands}

In this subsection, we show that the topological invariant $\pi_1(M,X_{m_z})$ protecting intersections of NLs becomes absent if  additional bands are present. We first demonstrate the instability of TPs explicitly on a simple model adapted from the main text, before rooting this fact in homotopy theory. However, we conclude this subsection with an observation that the TP does not vanish altogether, but is instead transferred in between \emph{another pair} of bands. This observation eventually motivates us to consider the multi-band generalization of the homotopic description of band-structure nodes outlined in Sec.~\ref{sec:PT-nodes}, which we further develop in Sec.~\ref{sec:multi-homotopy}.

\begin{figure}[t]
	\includegraphics[width=0.48\textwidth]{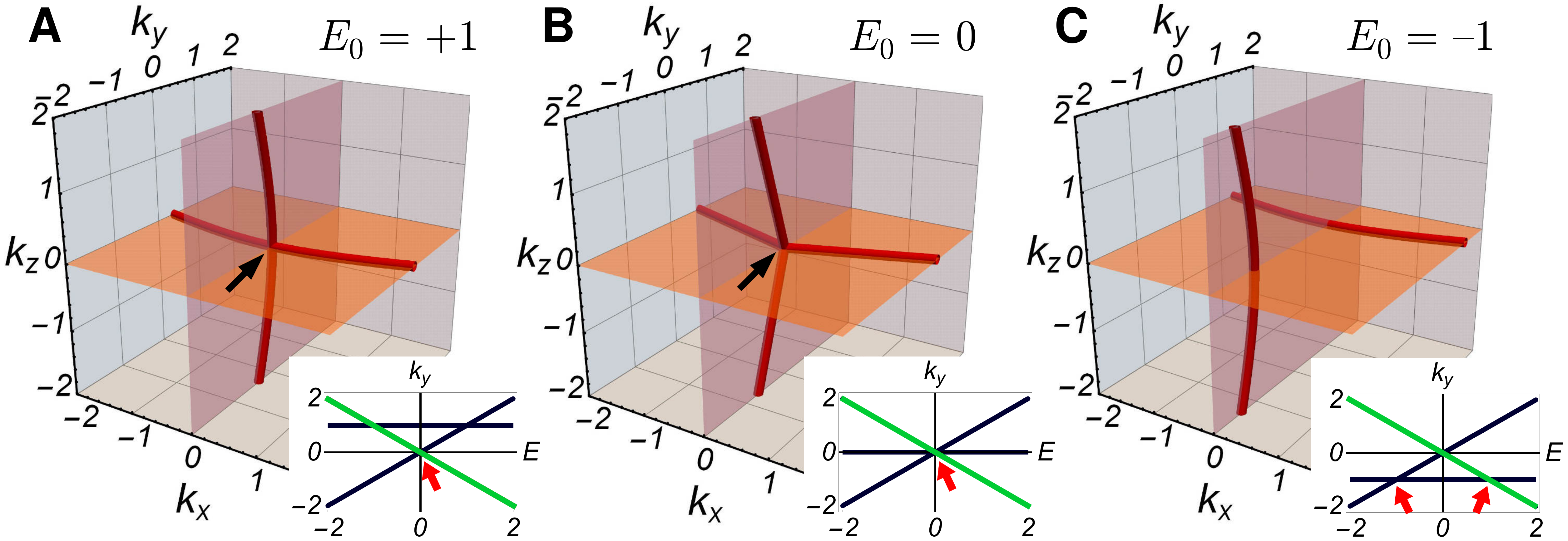}
	\caption{\textsf{(Related to Fig.~3(A--C) of the main text.) NLs (red) formed by the lower two bands of the model in Eq.~(\ref{eqn:3-band-detach}) for $t=\tfrac{1}{2}$ and indicated values of $E_0$. The crossing point (indicated by black arrow) disentangles trivially for $E_0 < 0$, indicating a violation of the two-band description presented in Sec.~\ref{sec:NCs-two-bands}. The insets show the band structure along the line of $k_x = 0$ and $k_z = 0$. Bands plotted in black (green) color have positive (negative) $\hat{m}_z$ eigenvalue.}}
	\label{fig:detaching-3-bands}
\end{figure} 

To observe the separability of the CP in a solvable model, we consider again the Hamiltonian in Eq.~(7) of the main text, i.e.
\begin{subequations}
\begin{equation}
\mcH(\bs{k}) = \left(\begin{array}{ccc}
E_0         	 	&	tk_x	& tk_z		\\
tk_x				&	k_y		& k_x k_z	\\
tk_z				&	k_x k_z	& -k_y
\end{array}\right)\label{eqn:3-band-detach}
\end{equation}
which respects two mirror symmetries, represented by
\begin{eqnarray}
\hat{m}_z &=& \diag(1,1,-1) \label{eqn:mirror-three-bands} \\
\hat{m}_x &=& \diag(1,-1,1). \label{eqn:mirror-three-bands-x}
\end{eqnarray} 
\end{subequations}
Only the presence of the $m_z$ mirror is relevant for the subsequent discussion, while the role of $m_x$ is only to simplify the Hamiltonian. We set the hybridization amplitude to $t=\tfrac{1}{2}$, and we keep the orbital energy $E_0$ as a tunable parameter. Assuming first that only one band is occupied, we find a pair of NLs formed by the lower (labelled $\downarrow$) pair of bands at
\begin{subequations}\label{eqn:subeqn-3-band-NLs}
\begin{eqnarray}
\bs{k}_{\textrm{NL},z}^\downarrow(E_0,p) &=& \left(p,\tfrac{1}{2}(-E_0+ \sqrt{E_0^2 + 2(tp)^2}),0\right)\quad  \\
\bs{k}_{\textrm{NL},x}^\downarrow(E_0,p) &=& \left(0,\tfrac{1}{2}(+ E_0 - \sqrt{E_0^2 + 2(tp)^2}),p\right)\quad
\end{eqnarray}
\end{subequations}
where $p\in\reals$ is a parameter running along the NLs, and the subscript $z$ ($x$) indicates that the NL lies inside the $k_z \!=\! 0$ ($k_x \!=\! 0$) plane. For $E_0 \geq 0$, these NLs form a CP at $\bs{k}_\textrm{CP} \!=\! (0,0,0)$. However, the CP disentangles trivially for $E_0<0$, see Fig.~\ref{fig:detaching-3-bands}(C). 

We explain the separability of intersecting NLs within three-band models using homotopy theory. According to Eq.~(\ref{eqn:RP2-double-cover-1}), the space of spectrally flattened $\mcP\mcT$-symmetric Hamiltonians with one occupied and two unoccupied bands is $M_{(1,2)}\simeq\reals P^2$, with $\pi_1(M_{(1,2)})=\ztwo$. On the other hand, the most general $\mcP\mcT$-symmetric Hamiltonian commuting with $\hat{m}_z$ in Eq.~(\ref{eqn:mirror-three-bands}) is
\begin{subequations}
\begin{equation}
\mcH = \left(\begin{array}{cc}
d_0\unit + d_x \sigma_x + d_z \sigma_z  &   \triv   \\
\triv                                  &   \varepsilon
\end{array}\right)~\label{eqn:3-band-gen-mz-Ham}
\end{equation}
where $d_{0,x,z}$ and $\varepsilon$ are real parameters. The eigenvalues of the Hamiltonian in Eq.~(\ref{eqn:3-band-gen-mz-Ham}) are $d_0\pm\sqrt{d_x^2 + d_z^2}$ (with positive $\hat{m}_z$ eigenvalue) and $\varepsilon$ (negative $\hat{m}_z$ eigenvalue). The $m_z$-symmetric subspace $X_{m_z}\!\subset \!M_{(1,2)}$ consists of two disjoint components 
\begin{equation}
X_{m_z} = X^{+} \amalg X^{-}\label{eqn:AI-1-2-Mmz}
 \end{equation}
 \end{subequations}
where the superscript indicates the $\hat{m}_z$ eigenvalue of the occupied state. The existence of the two components can be restated by $\pi_0(X_{m_z}) = \ztwo$, i.e. in-plane NLs can still be interpreted as domain walls between regions with opposite sign of the $\hat{m}_z$ eigenvalue $\lambda^\textrm{o}_{\bs{k}}$. 

The connected components of $X$ are homotopic to
\begin{subequations}
\begin{equation}
X^- \simeq \textrm{point} \qquad \textrm{and}\qquad X^+ \simeq S^1 \label{eqn:AI-2-1-homeomorphs}
\end{equation}
where the first is obtained from Eq.~(\ref{eqn:3-band-gen-mz-Ham}) by setting $d_0 = 1$, $d_x = d_z = 0$ and $\varepsilon = -1$, while the latter corresponds to the choice $d_0 = 0$, $d_x^2 + d_z^2 = 1$ and $\varepsilon = 1$. For the following arguments, we further need to know the nature of the inclusion 
\begin{equation}
\iota: X^{+} \hookrightarrow M_{(1,2)}\label{eqn:iota-mz-A21}
\end{equation}
\end{subequations}
Parametrizing $X^{+}$ using $d_x \!=\! -\sin\theta$ and $d_z \!=\! \cos\theta$ with $\theta\in[0,2\pi)$, we find that the wave-function of the occupied band in a globally continuous gauge (it is impossible to find a globally continuous \emph{real} gauge) is
\begin{equation}
\ket{u^\textrm{o}(\theta)} \!=\! \frac{\e{-\imi{\theta}/{2}}}{\sqrt{2}}\!\left(\begin{array}{ccc}
\!\!\sin\tfrac{\theta}{2},\!   &\! \cos\tfrac{\theta}{2} ,& \!\!0\!\!\!
\end{array}\right)\!\!^\top.\label{eqn:calculate-phase}
\end{equation}
The Berry connection~\cite{Berry:1984,Xiao:2010} is $\mcA_\theta^\textrm{o} = \bra{u^\textrm{o}(\theta)}\partial_\theta\ket{u^\textrm{o}(\theta)} = -\tfrac{\imi}{2}$ in this gauge. The corresponding Berry phase $\varphi_\textrm{B} =  \imi \oint_0^{2\pi} \tr \mcA_\theta^\mathrm{o} \de\theta = \pi \; \textrm{(mod 2$\pi$)}$ is non-trivial. This means that $X^{+}\simeq S^1$ winds non-trivially around $M_{m_z}$, i.e. it corresponds to the non-trivial element of the first homotopy group in Eq.~(\ref{eqn:AI-1-2-1st-homotopy}).

\begin{figure}[t]
	\includegraphics[width=0.45\textwidth]{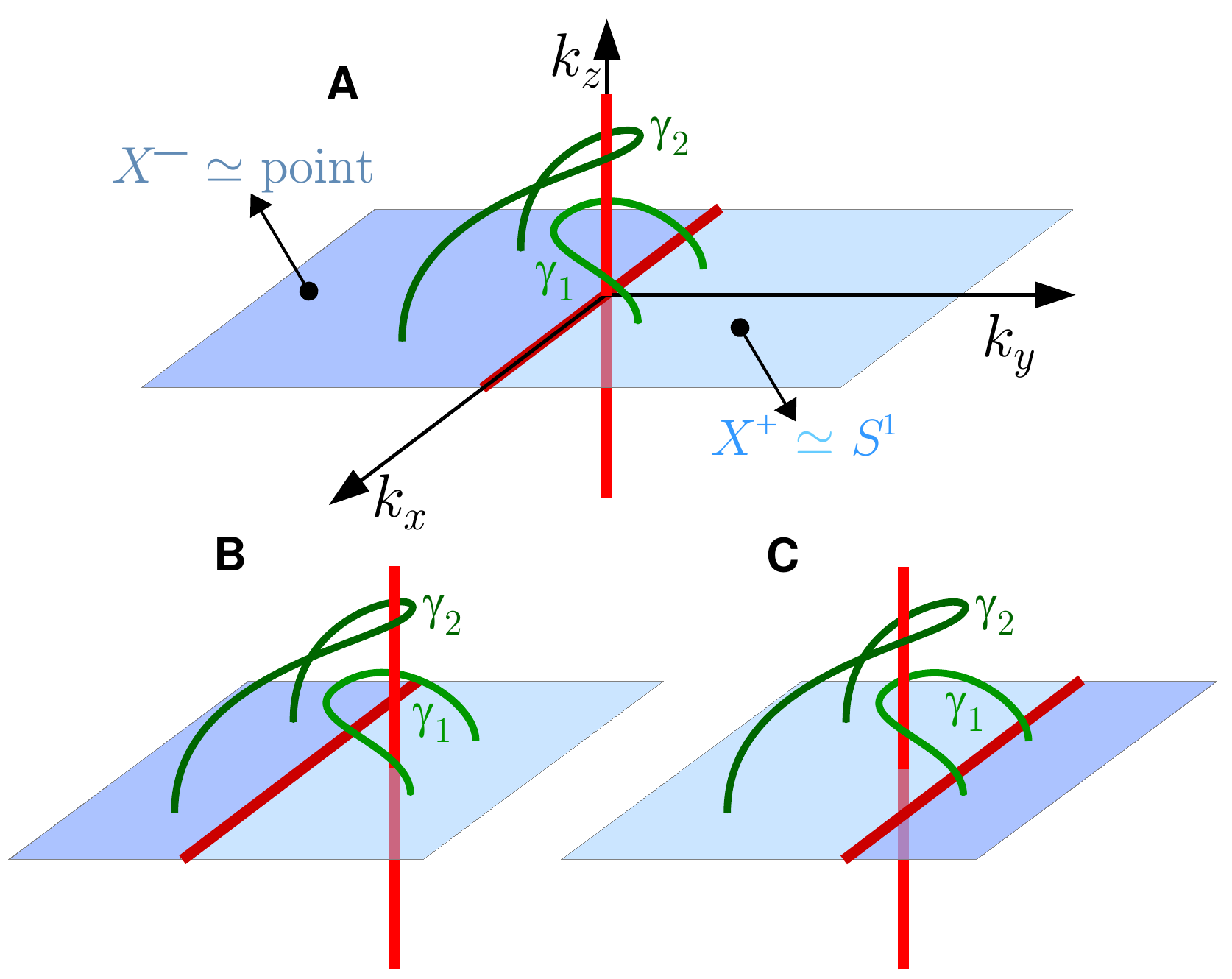}
	\caption{\textsf{(A) Schematic illustration of nodal lines (NLs, red) of the three-band Hamiltonian in Eq.~(\ref{eqn:3-band-detach}) for $E_0 = 1$. The NLs intersect at a crossing point (CP). For simplicity, we have only considered terms to linear order in $p$ on the right-hand side of Eqs.~(\ref{eqn:subeqn-3-band-NLs}). The horizontal $m_z$-invariant plane is colored with two shades of blue according to the $\hat{m}_z$ eigenvalues of the occupied band. The endpoints of open-ended green path $\gamma_{1}$ ($\gamma_{2}$) are located inside region $X^+$ ($X^-$). To each path we can assign an element of the relevant relative homotopy group in Eq.~(\ref{eqn:A-2-1-rel-homotop}). In the presence of $m_z$ symmetry, path $\gamma_1$ ($\gamma_2$) carries (does not carry) a topological obstruction. (B) By decreasing $E_0$ to negative values, the CP disentangles while making $\gamma_2$ (but not $\gamma_1)$ contractible to a point, compatible with the absence/presence of a topological obstruction on the path. (C) Flipping the locations of the $X^+$ and $X^-$ regions \emph{reverses} the directions in which the NLs detach for $E_0<0$.}}
	\label{fig:straight-NLs}
\end{figure} 

We use the obtained information to study the first relative homotopy group of pair $(M_{(1,2)},X_{m_z})$, assuming again that the endpoints $\partial D^1$ lie inside the same connected component of $X_{m_z}$. The result depends on the choice of the component, namely
\begin{subequations}\label{eqn:A-2-1-all-rel-hom}
\begin{equation}
\pi_1(M_{(1,2)},X^{-}) = \ztwo\quad\textrm{and}\quad \pi_1(M_{(1,2)},X^{+}) = \triv.\label{eqn:A-2-1-rel-homotop}
\end{equation}
These results follow from the exactness of sequences
\begin{equation}
\!\!\!\genfrac{}{}{0pt}{}{\pi_1(X^{-})}{\triv}\!\genfrac{}{}{0pt}{}{\stackrel{i_1\;}{\rightarrow}}{\rightarrow}\!\genfrac{}{}{0pt}{}{M_{(1,2)}}{\ztwo}\!\genfrac{}{}{0pt}{}{\stackrel{j_1\;}{\rightarrow}}{\rightarrow}\!\genfrac{}{}{0pt}{}{\pi_1(M_{(1,2)},X^{-})}{??}\!\genfrac{}{}{0pt}{}{\stackrel{\partial_1\;}{\rightarrow}}{\rightarrow}\!\genfrac{}{}{0pt}{}{\pi_0(X^{-})}{\triv}\!\!\!
\end{equation}
\begin{center}and \end{center}
\begin{equation}
\!\!\!\genfrac{}{}{0pt}{}{\pi_1(X^{+})}{\intg}\!\genfrac{}{}{0pt}{}{\stackrel{i_1\;}{\rightarrow}}{\rightarrow}\!\genfrac{}{}{0pt}{}{M_{(1,2)}}{\ztwo}\!\genfrac{}{}{0pt}{}{\stackrel{j_1\;}{\rightarrow}}{\rightarrow}\!\genfrac{}{}{0pt}{}{\pi_1(M_{(1,2)},X^{+})}{??}\!\genfrac{}{}{0pt}{}{\stackrel{\partial_1\;}{\rightarrow}}{\rightarrow}\!\genfrac{}{}{0pt}{}{\pi_0(X^{+})}{\triv}\!\!\! \label{eqn:Les-second-line}
\end{equation}
where in the lower rows we inserted the known homotopy groups of a point, $S^1$, and $\reals P^2$. To derive the latter result in Eq.~(\ref{eqn:A-2-1-rel-homotop}), one has to put in by hand that $\im i_1 =\ztwo$ in the sequence of Eq.~(\ref{eqn:Les-second-line}), which is a consequence of the non-trivial inclusion~(\ref{eqn:iota-mz-A21}). The reason that $\pi_1(M_{(1,2)},S^1)$ is smaller than $\pi_1(M_{(1,2)},\textrm{point})$ is ultimately the same as for the poloidal circle on a torus in Fig.~\ref{fig:torus}. Furthermore, \begin{equation}
\pi_1(X^{-}) = \triv\quad\textrm{and}\quad \pi_1(X^{+}) = \intg,\label{eqn:A-2-1-nonrel-homotop}
\end{equation}
\end{subequations}
meaning that only the latter admits out-of-plane NLs crossing the symmetric plane at points \emph{away} from in-plane NLs, see Fig.~\ref{fig:straight-NLs}(B and C).

Let us discuss the physical consequences of homotopy groups in Eqs.~(\ref{eqn:A-2-1-all-rel-hom}). We first consider the Hamiltonian of Eq.~(\ref{eqn:3-band-detach}) for $E_0 = 1$, which exhibits the intersecting NLs illustrated in Fig.~\ref{fig:detaching-3-bands}(A). For simplicity, we approximate Eq.~(\ref{eqn:subeqn-3-band-NLs}) to linear order in $p$, obtaining straight NLs shown in Fig.~\ref{fig:straight-NLs}(B). The in-plane NL separates regions with positive/negative $\hat{m}_z$ eigenvalue of the occupied band. To discuss the stability of the CP, we consider two open-ended paths $\gamma_{1,2}$ winding around the CP with both endpoints $\partial \gamma_1$ ($\partial \gamma_2$) lying inside the region with positive (negative) $\hat{m}_z$ eigenvalue of the occupied band, as shown in Fig.~\ref{fig:straight-NLs}(A). According to Eq.~(\ref{eqn:A-2-1-rel-homotop}), symmetry-preserving deformations of the Hamiltonian that keep the spectrum along the paths gapped can make $\gamma_2$ contractible to a point (there is no topological charge), while this is not possible for $\gamma_1$ (because of a $\ztwo$ obstruction). One thus deduces that the CP can be \emph{disconnected} by moving the vertical NL inside the $X^+$ region, as illustrated in Fig.~\ref{fig:straight-NLs}(B). However, it is not possible to detach the TP by moving the vertical NL into the $X^-$ region. (More generally, it can be shown by considering models with multiple bands that the CP can be separated in both directions if there is at least one band with positive and at least one band with negative $\hat{m}_z$ eigenvalue among both the occupied as well as the unoccupied states.)

The conclusions from the previous paragraph are also compatible with the homotopy groups in Eq.~(\ref{eqn:A-2-1-nonrel-homotop}). Because of the triviality of $\pi_1(X^{-})$, an isolated out-of-plane NL \emph{cannot} cross the symmetric plane through the $X^-$ region. One thus immediately concludes that if CP separation takes place, the vertical NL must shift into the $X^+$ region, where $\pi_1(X^{+})$ is non-trivial.

\begin{figure}[t]
	\includegraphics[width=0.48\textwidth]{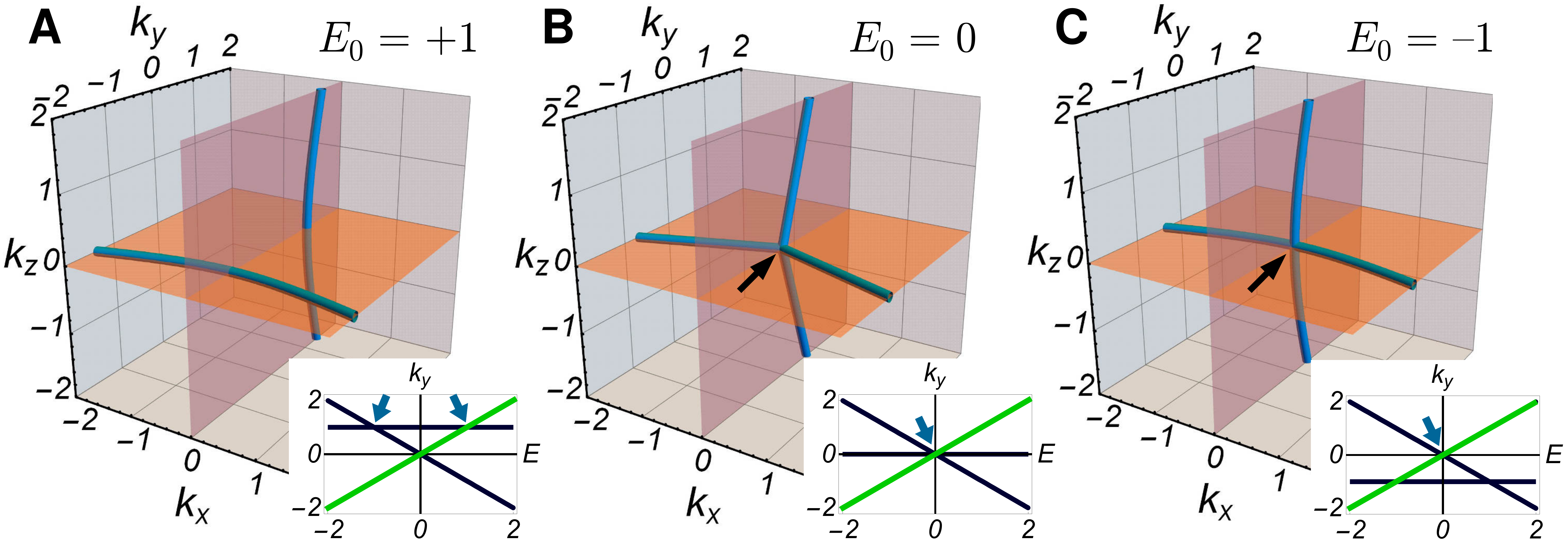}
	\caption{\textsf{(Related to Fig.~3(A--C) of the main text.) NLs (blue) formed by the \emph{upper} two bands of the model in Eq.~(\ref{eqn:3-band-detach}) for the indicated values of $E_0$. We observe the \emph{appearance} of a crossing point (CP, indicated by black arrow) of the NLs at $E_0 = 0$, simultaneously with the disappearance of the CP in Fig.~\ref{fig:detaching-3-bands}. The insets show the band structure along the line of $k_x = 0$ and $k_z = 0$. Bands plotted in black (green) color have positive (negative) $m_x$ eigenvalue.}}
	\label{fig:NC-model-switch}
\end{figure} 

Before concluding this section, let us briefly discuss NLs formed between the pair of \emph{unoccupied} bands (labelled $\uparrow)$ of the Hamiltonian in Eq.~(\ref{eqn:3-band-detach}). These are described by
\begin{subequations}\label{eqn:subeqn-3-band-NLs-upper}
\begin{eqnarray}
\bs{k}_{\textrm{NL},z}^\uparrow(E_0,p) &=& \left(p,\tfrac{1}{2}(-E_0- \sqrt{E_0^2 + 2(tp)^2}),0\right)\quad  \\
\bs{k}_{\textrm{NL},x}^\uparrow(E_0,p) &=& \left(0,\tfrac{1}{2}(+ E_0 + \sqrt{E_0^2 + 2(tp)^2}),p\right)\quad,
\end{eqnarray}
\end{subequations}
with $p\in\reals$, and are plotted in Fig.~\ref{fig:NC-model-switch}. We observe that simultaneously with separating the intersecting NLs formed by the lower two bands in Fig.~\ref{fig:detaching-3-bands}, a pair of NLs formed by the upper two bands \emph{become connected by a CP}. In other words, a nodal chain between one pair of bands is separated, while at the same time a nodal chain appears between a neighboring pair of bands. The CP thus does not disappear altogether, but is transferred in between another pair of bands. 

It turns out that none of the charges in Eqs.~(\ref{eqn:A-2-1-all-rel-hom}) is capable of explaining the observed transfer of the CP. The reason is that the information about the dispersion of the unoccupied bands is lost after performing the spectral projection of the Hamiltonian in Eq.~(\ref{eqn:gen-ham-form}), since it brings both of these states to energy $+1$. In the next section, we build upon the generalized homotopic description of band-structure nodes outlined in Sec.~\ref{sec:PT-nodes}, which is based on the less coarse spectral projection of Eq.~(\ref{eqn:Hamiltonian-AI-3-N}). We find that the CP transfer observed in Figs.~\ref{fig:detaching-3-bands} and~\ref{fig:NC-model-switch} (plotted jointly as Fig.~3(A--C) of the main text) can be explained using such an extended theory. Surprisingly, the generalized theory assigns nodal lines in $\mcP\mcT$-symmetric systems a non-Abelian charge, leading to non-trivial ``braiding rules'' in $\bs{k}$-space.


\section{Multi-band description}~\label{sec:multi-homotopy}

In the previous Sec.~\ref{sec:NCs-three-bands} we found that certain properties of NLs in $\mcP\mcT$-symmetric systems cannot be explained using the standard approach of Refs.~\cite{Bzdusek:2017,Sun:2018} which projects all the occupied (unoccupied) bands to energy $-1$ ($+1$). However, we argued that the multi-band description of NLs based on the spectral projection in Eq.~(\ref{eqn:Hamiltonian-AI-3-N}) should have the power to also describe the novel properties of NLs reported in our manuscript. In this section, we further develop this generalized theory.

The present section is organized as follows. We begin in Sec.~\ref{subsec:3-deriv} with explicitly discussing the generalized theory in the case of $N=3$ bands and in the absence of mirror symmetry. We find that NLs of such $\mcP\mcT$-symmetric systems can be described using the quaternion group, which leads to certain non-Abelian properties. In the next Sec~\ref{subsec:biaxials} we show that the classification of NLs in three-band models is, in fact, mathematically equivalent to the classification of vortex lines in biaxial nematic liquid crystals -- a problem that has been theoretically analyzed four decades ago~\cite{Toulouse:1977,Kleman:1977,Poenaru:1977,Chaikin:1995}. Especially, we use Sec.~\ref{subsec:braid} to discuss properties of NLs under exchange (i.e.~``braiding rules''), following closely the discussion of vortex lines in biaxial nematics by Ref.~\cite{Mermin:1979}. In Sec.~\ref{subsec:compositions}, we formulate consistency criteria for admissible NL compositions, which arise as a consequence of the non-Abelian topological charge. 

In the remaining parts of this section, we generalize the simple non-Abelian three-band description in two ways. First, in Sec.~\ref{subsec:gen-many} we generalize the decription to an arbitrary number $N \geq 3$ of bands, which involves the application of Clifford algebras~\cite{Atiyah:1964} and Salingaros vee groups~\cite{Salingaros:1981,Brown:2015,Ablamowicz:2017,Salingaros:1983}. In the main text, we refer to the result as \emph{generalized quaternion charge}. Finally, in Sec.~\ref{sec:multiband+mirror} we include mirror symmetry into the three-band model, which in combination with the relative homotopy description presented in Sec.~\ref{sec:mirror-nodes} allows us to explain the transfer of CP observed in Sec.~\ref{sec:NCs-three-bands}.


\subsection{Quaternion charge in three-band models}\label{subsec:3-deriv}

We consider the multi-band theory of band-structure nodes in $\mcP\mcT$-symmetric systems as outlined in Sec.~\ref{sec:PT-nodes} for the specifc case $N=3$. We are not interested in the occupation of the individual bands by electrons, but only in the overall NL composition exhibited by the whole band structure. The notion of chemical potential is irrelevant for the discussion. The presented mathematical arguments are adapted from Ref.~\cite{Mermin:1979}.

First, we rewrite the Hamiltonian form in Eq.~(\ref{eqn:Hamiltonian-AI-3-N}) as
\begin{subequations}\label{eqn:Ham-frames}
\begin{equation}
\mcH(\bs{k}) = \mathsf{u}(\bs{k}) \mathcal{E} \mathsf{u}\!^\top\!(\bs{k})\label{eqn:H-uEu}
\end{equation}
\begin{center}where \end{center}
\begin{equation}
\mathcal{E} = \diag(1,2,\ldots,N)\label{eqn:standard-energies}
\end{equation}
is a diagonal matrix of the standard eigenvalues given by Eq.~(\ref{eqn:standrad-energies}), and $\mathsf{u}(\bs{k})$ is a matrix of (ordered) eigenstates of $\mcH(\bs{k})$. When discussing components of $\mathsf{u}(\bs{k})$, we stick to the notation
\begin{equation}
\mathsf{u}_{ij}(\bs{k}) = \langle{i}|{u^j_{\bs{k}}}\rangle \quad\textrm{and}\quad \bs{\mathsf{u}}_{j}(\bs{k}) = |{u^j_{\bs{k}}}\rangle,
\end{equation}
\end{subequations}
although the index gymnastics becomes important only in Sec.~\ref{sec:calculating-the-charge}.
The matrix $\mathsf{u}(\bs{k})$ is orthogonal, and we will often call it a \emph{frame}. Assuming we fix some basis $\{|i\rangle \}_{i=1}^N$ of the Hilbert space, we will call $\hat{\mathsf{u}} = \unit$ as the \emph{standard frame} corresponding to that basis.

\begin{figure*}[hbt!]
\includegraphics[width=0.99 \textwidth]{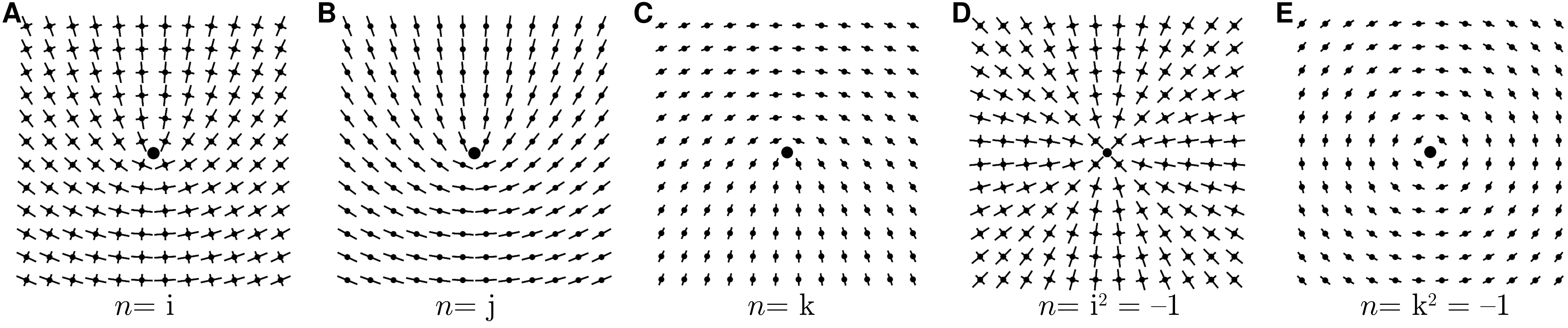}
 \caption{\textsf{(The figure is based on Figs.~35  and~36 of Ref.~\cite{Mermin:1979}.) We visualize the order-parameter field of biaxial nematics by portraying the prevalent orientation of the three symmetry axes of the $\mathsf{D}_{2h}$-symmetric molecules using arms of different length (long, short, dot) at regular intervals in space. The displayed order-parameter fields exhibit a singularity (the black disc) in the center. (A) A $\pi$-rotation of the order parameter around the dot axis corresponds to quaternion charge $n = \imi$, (B) a $\pi$-rotation around the short axis to charge $n = \imj$, and (C) a $\pi$-rotation around the long axis to  $n = \imk$. (D) A $2\pi$-rotation around the dot axis corresponds to charge $n = \imi^2 = -1$. As opposed to vortices of uniaxial nematics illustrated in Fig.~\ref{fig:windings}, a $2\pi$-vortex in biaxial nematics cannot be deformed into a trivial configuration using continuous deformations. Specifically, applying the same trick as in Fig.~\ref{fig:windings}(E), namely rotating all the molecules around the short axis by $\pi/2$ to make the long axis perpendicular to the plane of the illustration everywhere (while keeping the \emph{position} of the ellipsoids fixed), does not remove the singularity. Instead, (E) we obtain the $2\pi$-rotation around the long axis, which carries the same charge $n = \imk^2 = -1$. }}
\label{fig:windings-biaxial}
\end{figure*}

We rewrite the space of Hamiltonians in Eq.~(\ref{eqn:M-N-space-gen}) as~\cite{Mermin:1979}
\begin{equation}
M_{3} = \mathsf{O}(3)/\mathsf{D}_{2h} = \mathsf{SO}(3)/\mathsf{D}_2 = \mathsf{SU}(2)/\overline{\mathsf{D}}_2. \label{eqn:AI3-isomorphisms}
\end{equation}
The first equality in Eq.~(\ref{eqn:AI3-isomorphisms}) states that a Hamiltonian $\mcH\in M_3$ is identified with a (three-component) frame $\mathsf{u}$, which is obtained by an $\mathsf{O}(3)$ rotation of the standard frame $\hat{\mathsf{u}}$. However, flipping the sign of some vectors $\bs{\mathsf{u}}_{j}$ of the frame leaves the Hamiltonian in Eq.~(\ref{eqn:H-uEu}) invariant, which implies the $\mathsf{D}_{2h}$ quotient. The point group $\mathsf{D}_{2h}\equiv \mathsf{O}(1)^3 \equiv \ztwo^3$ is generated by three mutually perpendicular mirror symmetries. 

Furthermore, in the second step of Eq.~(\ref{eqn:AI3-isomorphisms}) we replace both groups in the coset expression by their special component (i.e. we keep only \emph{proper rotations} with positive determinant). The point group $\mathsf{D}_2$ consists of the identity and of three $\pi$-rotations around mutually perpendicular axes. In the last step of Eq.~(\ref{eqn:AI3-isomorphisms}), we replace $\mathsf{SO}(3)$ by its double cover $\mathsf{Spin}(3)\cong\mathsf{SU}(2)$, and we lift the dihedral group $\mathsf{D}_2$ into the double group $\overline{\mathsf{D}}_2$, similar to our discussion in Sec.~\ref{subsec:A21}. Importantly, the group $\overline{\mathsf{D}}_2$ is isomorphic to the \emph{quaternion group} generated by three anticommuting imaginary units, 
\begin{subequations}
\begin{equation}
\overline{\mathsf{D}}_2 \cong \mathsf{Q} = \left\{\pm 1,\pm\imi,\pm\imj,\pm\imk\right \}.\label{eqn:quaternions}
\end{equation}
This fact is best seen by explicitly calculating the lift of $\pm\pi$-rotations $\e{\pm \pi L_i}\in\mathsf{D}_2$ with $L_i$ defined in Eq.~(\ref{eqn:gens-so-3}), which is $\e{\pm \pi t_i} = \mp\imi  \sigma_i \in\overline{\mathsf{D}}_2$ with $t_i$ defined in Eq.~(\ref{eqn:gens-su-2}). It is straightforward to check that replacing
\begin{equation}
\unit \mapsto 1,\;\; -\imi\sigma_x \mapsto \imi,\;\; -\imi\sigma_y \mapsto \imj,\;\;\textrm{and} \;\; -\imi\sigma_z \mapsto \imk\label{eqn:quat-d-isomoprh}
\end{equation}
\end{subequations}
represents the isomorphism between $\overline{\mathsf{D}}_2$ and $\mathsf{Q}$. 

The final expression in Eq.~(\ref{eqn:AI3-isomorphisms}) is particularly useful for deriving the homotopy groups of $M_{3}$ through the application of the theorems derived in Sec.~\ref{eqn:homotopy-cosets}. It follows from Eqs.~(\ref{eqn:coset-homo}) that 
\begin{subequations}
\begin{eqnarray}
\pi_1(M_{3}) &=& \mathsf{Q} \label{eqn:quat-charge} \\ 
\pi_2(M_{3}) &=& \triv. \label{eqn:quat-charge-2}  
\end{eqnarray}
\end{subequations}
We find that band-structure nodes of $\mcP\mcT$-symmetric three-band models are described by the \emph{non-Abelian} quaternion group. Robust band-structure nodes are point-like in 2D and one-dimensional lines in 3D.

\begin{figure}[t!]
\includegraphics[width=0.24 \textwidth]{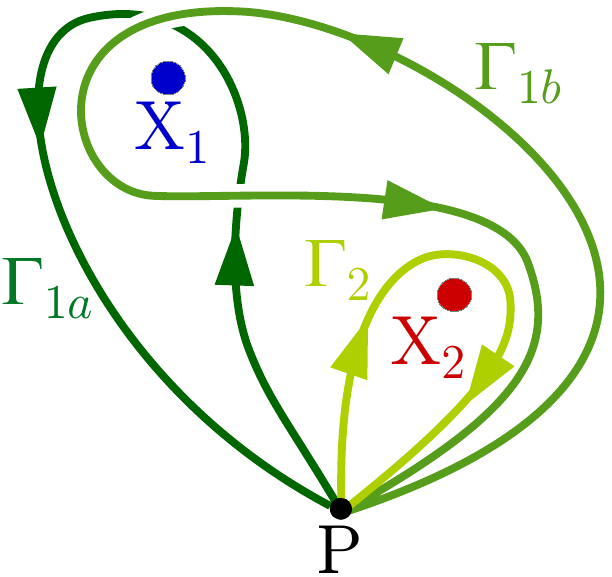}
 \caption{\textsf{In the presence of multiple topological defects $\mathrm{X}_i$, there may be multiple inequivalent ways to encircle a selected defect $\textrm{X}_1$ with a closed path $\Gamma \simeq  S^1$ based at point $\textrm{P}$. Two such paths $\Gamma_{1a}$ and $\Gamma_{1b}$ are shown. The charge of node $\textrm{X}_1$ determined using path  $\Gamma_{1a}$ and using path $\Gamma_{1b}$ are related by a conjugacy with the charge of defect $\mathrm{X}_2$ on path $\Gamma_2$, cf. Eqs.~(\ref{eqn:conjuga}).}}
\label{fig:conjugacy}
\end{figure}

We want to provide each element of group~(\ref{eqn:quat-charge}) a meaning. As explained later, see Sec.~\ref{subsec:biaxials} and Fig.~\ref{fig:conjugacy}, the first-homotopy charge $c\in\pi_1(M)$ of a topological defect is well-defined only up to conjugacy with other elements of the fundamental group, cf. Eq.~(\ref{eqn:up-to-conjug}). In the present case, there are five conjugacy classes
\begin{equation}
C_\mathsf{Q}=\left\{\{1\},\{\pm \imi\},\{\pm \imj\},\{\pm\imk\},\{-1\}\right\}.\label{eqn:conj-classes}
\end{equation}
Clearly, class $\{1\}$ corresponds to trivial loops, i.e. those that are contractible to a point without encountering a band degeneracy. 

To interpret the other conjugacy classes in Eq.~(\ref{eqn:conj-classes}), note that forming a node between bands $|{u^1_{\bs{k}}}\rangle$ and $|{u^2_{\bs{k}}}\rangle$ (while keeping the state $|{u^3_{\bs{k}}}\rangle$ constant) corresponds to rotating the standard frame as $\mathsf{u}(\alpha) = R(\alpha) \hat{\mathsf{u}}$ with rotation matrix
\begin{subequations}
\begin{equation}
R(\alpha) = \left(\begin{array}{ccc}
\cos\tfrac{\alpha}{2}&   -\sin\tfrac{\alpha}{2}  &   0   \\
\sin\tfrac{\alpha}{2} &   \cos\tfrac{\alpha}{2}  &   0   \\
0           &   0           &   1
\end{array}\right) = \e{\tfrac{\alpha}{2}L_3},\label{eqn:so3-node-1-2}
\end{equation}
where $\alpha\in[0,2\pi)$ parametrizes a closed loop $\Gamma: S^1 \to \textrm{BZ}\backslash {\mathcal{N}_\mcH^1}$. Note that $\mathsf{u}(2\alpha) \neq \mathsf\mathsf{u}({0})$, corresponding to the $\mathsf{D}_2$ redundancy in Eq.~(\ref{eqn:AI3-isomorphisms}). Indeed, the Hamiltonian
\begin{equation}
\mcH(\alpha) = R(\alpha)\mathcal{E}R(\alpha)^\top.
\end{equation}
fulfills $\mcH(2\pi) = \mcH(0)$, i.e.~it is a continuous and single-valued function on the closed path $\Gamma$. Following the logic of Eq.~(\ref{eqn:AI3-isomorphisms}), we consider the \emph{lift} of Eq.~(\ref{eqn:so3-node-1-2}) inside the covering group. Following the recipe from Eqs.~(\ref{eqn:replace-generators}) in Sec.~\ref{subsec:A21}, the lift at $\alpha = 2 \pi$ equals
\begin{equation}
\overline{R}(2\pi) = \e{-\imi \pi \sigma_z/2} = -\imi \sigma_z \mapsto \imk,
\end{equation}
\end{subequations}
where in the last step we used the assignment in Eq.~(\ref{eqn:quat-d-isomoprh}). We thus conclude that conjugacy class $\{\pm \imk\}$ corresponds to the presence of a node between bands $|{u^1_{\bs{k}}}\rangle$ and $|{u^2_{\bs{k}}}\rangle$ inside the loop. 

Similarly, one can easily check that $\{\pm\imi\}$ describes a node between bands $|{u^2_{\bs{k}}}\rangle$ and $|{u^3_{\bs{k}}}\rangle$. Using the product rule in the quaternion group, class $\{\pm\imj\}$ indicates that the chosen path encloses \emph{one of each} species of nodes. The $\pm$ sign indicates that the \emph{orientation} of the node depends on the specific choice of a path used to enclose the node(s). Finally, the element $\{-1\}$ indicates a pair of nodes of the same orientation between the same pair of bands. In the following text, we sometimes replace each conjugacy class by a representative element when we find the risk of confusing the reader to be sufficiently small. Note also that while $(\pm\imi)^2 = (\pm\imk)^2 = -1$ is non-trivial, $(\pm\imi)^4 = (\pm\imk)^4 = 1$ is trivial. Indeed, we will show in Sec.~\ref{subsec:compositions} that a path enclosing four NLs of the same type and orientation can be continuously shrunk to a point without encountering a NL in the process. Finally, we mathematical formalize the quaternion charge and present an algorithm for its numerical computation in Sec.~\ref{sec:calculating-the-charge}.


\subsection{Analogy with biaxial nematics}\label{subsec:biaxials}

In this section, we point out the mathematical analogy between NLs of $\mcP\mcT$-symmetric three-band models and vortex lines of biaxial nematic liquid crystals. Liquid crystals are a phase of matter that preserves the translational symmetry of a liquid, but which breaks the rotational symmetry like a crystal. More specifically, the $\mathsf{O}(3)$ rotation symmetry is lowered to $\mathsf{D}_{2h}$ for \emph{biaxial nematic} liquid crystals. A possible route to realize this phase is to consider molecules with approximately ellipsoid shape, provided that the three axes of the ellipsoid are all of different length. The inidividual molecules are \emph{randomly positioned} (leading to the preserved translational invariance), while they \emph{freeze in orientation} (breaking the rotational symmetry). Biaxial nematics have been theoretically considered since 1970's~\cite{Toulouse:1977,Kleman:1977,Poenaru:1977,Chaikin:1995} while their experimental realization was achieved only relatively recently~\cite{Madsen:2004,Prasad:2005} in so-called bent-core mesogens. 

The order-parameter space for biaxial nematics is given by Eq.~(\ref{eqn:AI3-isomorphisms}) where $\mathsf{O}(3)$ can be interpreted as a group of rotations acting on the ellipsoid, while $\mathsf{D}_{2h}$ is the isotropy subgroup preserving the orientation of a chosen ellipsoid. By repeating the arguments presented in Sec.~\ref{subsec:3-deriv}, the description of defects in biaxial nematics is isomorphic to the description of band-structure nodes in three-band $\mcP\mcT$-symmetric models. This isomorphism allows us to recall the known properties of vortex lines in biaxial nematics, and study their analogy in band structures. Theoretical properties of vortex lines in biaxial nematics have been beautifully reviewed by Ref.~\cite{Mermin:1979}.

To visualize the order parameter of biaxial nematics, we employ the method of Ref.~\cite{Mermin:1979}. We indicate the orientation of the ellipsoid by drawing the three major axes of the ellipsoid with different length: long, short and dot. To plot the order-parameter \emph{field}, we show the prevalent orientation of the ellipsoids at regular intervals in space, see Fig.~\ref{fig:windings-biaxial}. Note that while in uniaxial nematics, illustrated in Fig.~\ref{fig:windings}, there is only one type of a $\pi$-vortex, biaxial nematics support \emph{three} distinct types of a $\pi$-vortex, corresponding to $\pi$ rotations around the dot/short/long axis as shown in Fig.~\ref{fig:windings-biaxial}(A--C). These defects correspond to the three imaginary units $\imi,\imj,\imk$ of the quaternion group in Eq.~(\ref{eqn:quat-charge}). 

We have previously argued, recall Fig.~\ref{fig:windings}, that a $2\pi$-vortex in unixial nematics is removable by a continuous deformation of the order-parameter field, i.e.~it does not present a topologically stable singularity. However, the conclusion is different for biaxial nematics. We show in Fig.~\ref{fig:windings-biaxial}(D) an order-parameter field with a $2\pi$-vortex, corresponding to a $2\pi$ rotation around the dot axis, which is represented by charge $\imi^2 = -1 \in \mathsf{Q}$. We can perform a continuous deformation of the order-parameter field which rotates all the ellipsoids by $\pi/2$ around the short axis (while keeping the \emph{positions} of the ellipsoids fixed). This is the analog of the continuous deformation that removed the $2\pi$-vortex in unixial nematics. However, applying the same trick to biaxial nematics produces a $2\pi$ rotation around the long axis, plotted in Fig.~\ref{fig:windings-biaxial}(E). The resulting order-parameter field is described by the same quaternion charge $\imk^2 = -1$.

\subsection{Non-trivial exchange of non-Abelian nodes}\label{subsec:braid}

\begin{figure*}[t!]
\includegraphics[width=0.84 \textwidth]{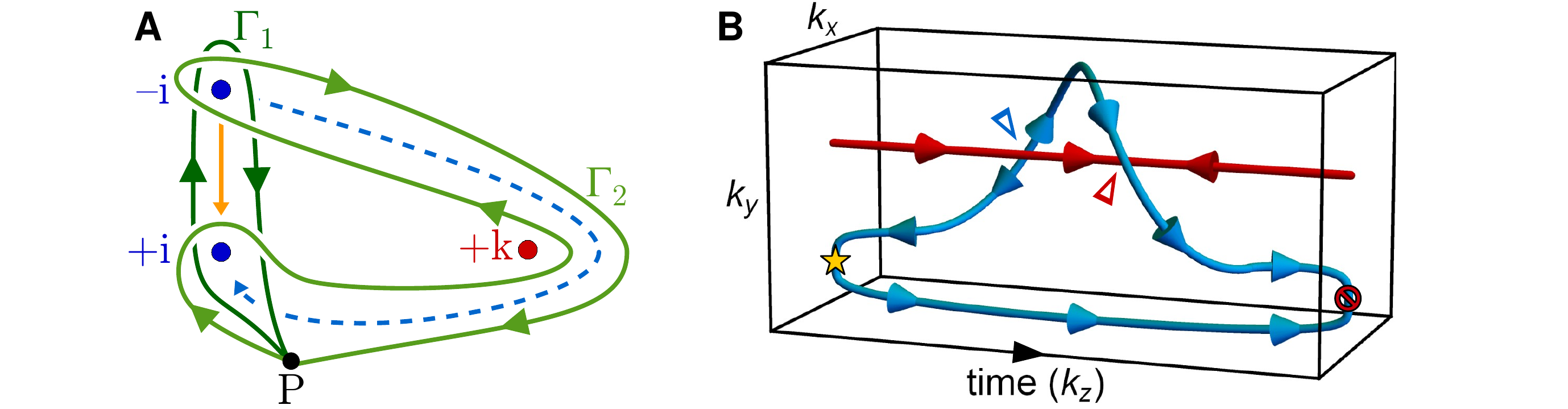}
 \caption{\textsf{(A) A pair of nodes (blue dots) are assumed to have been pairwise created inside path $\Gamma_1$ (based at $\textrm{P}$). The nodes carry opposite charge $\pm \imi$ on path $\Gamma_1$, but the same charge $+\imi$ on path $\Gamma_2$. As a consequence, the blue nodes will annihilate if moved together along  a trajectory inside $\Gamma_1$ (solid orange), and they do not annihilate if brought together along a trajectory inside $\Gamma_2$ (dashed blue). The trajectory-dependent outcome of the node collision manifests the non-Abelian topological charge. (B) World lines of the described exchange of nodes. The two blue nodes are created at an early time (indicated by yellow star), and separated. The opposite value of their $\pm\imi$ charge is indicated by the opposite orientation of their (blue) world lines. The red node is not moving, therefore its (red) world line is a straight line. Upon exchange with the red node, the charge of one blue node changes sign, which is manifested by the reversed orientation of the corresponding blue world line. When brought together at a later time, the blue nodes fail to annihilate (indicated by nay symbol). The reason is that their charges do not cancel, $(+\imi)(+\imi) = -1 \neq +1$. By reinterpreting the time as third momentum component, we obtain a description of nodal lines (NL) in 3D. The orientation of a NL is reversed each time it goes under a NL of the other color (indicated by triangular arrowheads), which we further explain in Fig.~\ref{fig:orientations}.}}
 \label{fig:biaxial-points}
\end{figure*}

In this section, we discuss some theoretical properties of defects governed by the quaternion group. Our presentation to a large extent follows the discussion of Ref.~\cite{Mermin:1979} in the context of biaxial nematics, although our discussion of orientation reversal cannot be found there. We are ultimately interested in the behavior of nodal lines in 3D systems, but some arguments (such as those illustrated in Fig.~\ref{fig:conjugacy} and  Fig.~\ref{fig:biaxial-points}(A)) are more easily explained by considering \emph{point} nodes in 2D. Furthermore, 3D situations can often be conveniently interpreted as world lines for exchanging point nodes in 2D, if the third momentum component is interpreted as time.

Quite generally, one of the implications of a non-Abelian fundamental group $\pi_1(M)$ is that the charge $c$ of a topological defect is well-defined only up to conjugacy~\cite{Mermin:1979,Sethna:2006}
\begin{subequations}\label{eqn:conjuga}
\begin{equation}
c \sim g c g^{-1}\quad\textrm{with}\;\; g \in \pi_1(M). \label{eqn:up-to-conjug}
\end{equation}
The reason for this is that in the presence of additional nodes it may be possible to circumscribe the considered node with various closed paths (all based at a fixed point $\textrm{P}$) which \emph{cannot} be continuously deformed into one another. For simplicity, we illustrate this ambiguity by considering \emph{point} defects inside a two-dimensional plane as shown in Fig.~\ref{fig:conjugacy}: A point defect $\textrm{X}_1$ inside a two-dimensional plane can be enclosed by the closed path $\Gamma_{1a}$ as well as by the closed path $\Gamma_{1b}$, both based at point $\textrm{P}$. The presence of the additional point node $\textrm{X}_2$ makes it impossible to continuously deform $\Gamma_{1a}$ into $\Gamma_{1b}$. However, if we encircle node $\textrm{X}_2$ with path $\Gamma_2$ based at $\mathrm{P}$ as shown in the figure, we find that~\footnote{Here, as well as throughout the entire Supplemental Information file, we use the convention that when composing several paths as in Eq.~(\ref{eqn:path-conjugacy}), we first traverse the path on the very right, and then continue to the left. The inverse in the superscript implies that we have to traverse the path in the reverse direction.}
\begin{equation}
\Gamma_{1b} \sim \Gamma_2 \circ \Gamma_{1a} \circ \Gamma_{2}^{-1} \label{eqn:path-conjugacy}
\end{equation}
are homotopic. If we denote the representation of a pointed path $\Gamma_i$ in the fundamental group $\pi_1(M,\mathfrak{m})$ as $g_i$, we find
\begin{equation}
g_{1b}^{\phantom{1}} = g_2^{\phantom{1}} g_{1a}^{\phantom{1}} g_2^{-1},\label{eqn:group-conjugacy}
\end{equation}
\end{subequations}
which leads to the conclusion in Eq.~(\ref{eqn:up-to-conjug}). Since for Abelian groups every element forms a conjugacy class of its own, the path-dependence of the first-homotopy charge is only relevant when $\pi_1(M)$ is non-Abelian.

We now show that the noncommutativity of the quaternion charge implies \emph{non-trivial exchange} (i.e. ``braiding'') rules for point nodes in 2D systems. To see this, consider a ``node-antinode pair'' that has been pairwise created by locally adjusting the Hamiltonian, such as the ``blue'' nodes in Fig.~\ref{fig:biaxial-points}(A). We assume that the nodes have been pairwise created inside path $\Gamma_1$. Therefore, the nodes carry opposite charge $\pm \imi$ on path $\Gamma_1$, and they would pairwise annihilate if they are brought back together along a trajectory inside $\Gamma_1$ (solid orange line). However, if a ``red'' node with charge $+\imk$ is present, we may enclose the two blue nodes with path $\Gamma_2$ (also based at $\textrm{P}$) that takes a detour around the red node. On such a path, the charge of the top blue node is conjugated by the charge of the red node, i.e. it is changed to $\imk(-\imi)(-\imk) = +\imi$. The reason is the same as for the situation illustrated in Fig.~\ref{fig:conjugacy}. The total charge along  $\Gamma_2$ is therefore $n=(+\imi)(+\imi) = -1$, which is \emph{non-trivial}. Therefore, the blue nodes do \emph{not} annihilate if they are brought together along a trajectory inside $\Gamma_2$ (dashed magenta line). The fate of the two nodes upon collision depends on the trajectory used to bring them together, thus manifesting their non-Abelian topological charge. 

\edit{We remark the same argument also applies when $\Gamma_2$ is a path that winds around the BZ torus, provided that the Hamiltonian is periodic in the reciprocal lattice vectors, i.e. when $\mcH(\bs{k}) = \mcH(\bs{k} + \bs{G})$. If the Berry phase (\emph{i.e.}~also the quaternion charge) on $\Gamma_2$ is non-trivial, then the ability of the two ``blue'' nodes to annihilate depends on the choice of trajectory used to bring them together on the BZ torus \emph{even if there are no ``red'' nodes}. This, in brief, explains how the non-abelian topology discussed here explains the findings of the very recent Ref.~\cite{Ahn:2018b}.}

Let us now rephrase the described exchange of point nodes \edit{located inside a plane} by considering their \emph{world lines} inside a three-dimensional space (2 momentum components + 1 time). Such word lines for the process in Fig.~\ref{fig:biaxial-points}(A) are illustrated in Fig.~\ref{fig:biaxial-points}(B). We indicate the world line of a blue point node with charge $+\imi$ ($-\imi$) as a blue oriented line pointing in the increasing (decreasing) time direction. At an early time, two blue nodes are pairwise created (indicated by yellow star). They carry opposite $\pm\imi$ charges, which are indicated by the opposite orientation of the blue world lines. Afterwards, one blue node is exchanged with a red node. As explained in the previous paragraph, this process is accompanied by flipping the sign of the blue node. In the picture with world lines, this means that the orientation of the blue world line is reversed during the exchange with the red world line. Therefore, the two blue world lines have the \emph{same} orientation after the exchange, and they \emph{fail} to annihilate when we bring them together at a later time (indicated by the nay symbol). Note that by interpreting the time direction as third momentum component, the world lines correspond to NLs inside a 3D momentum space.

\begin{figure}[t!]
\includegraphics[width=0.4 \textwidth]{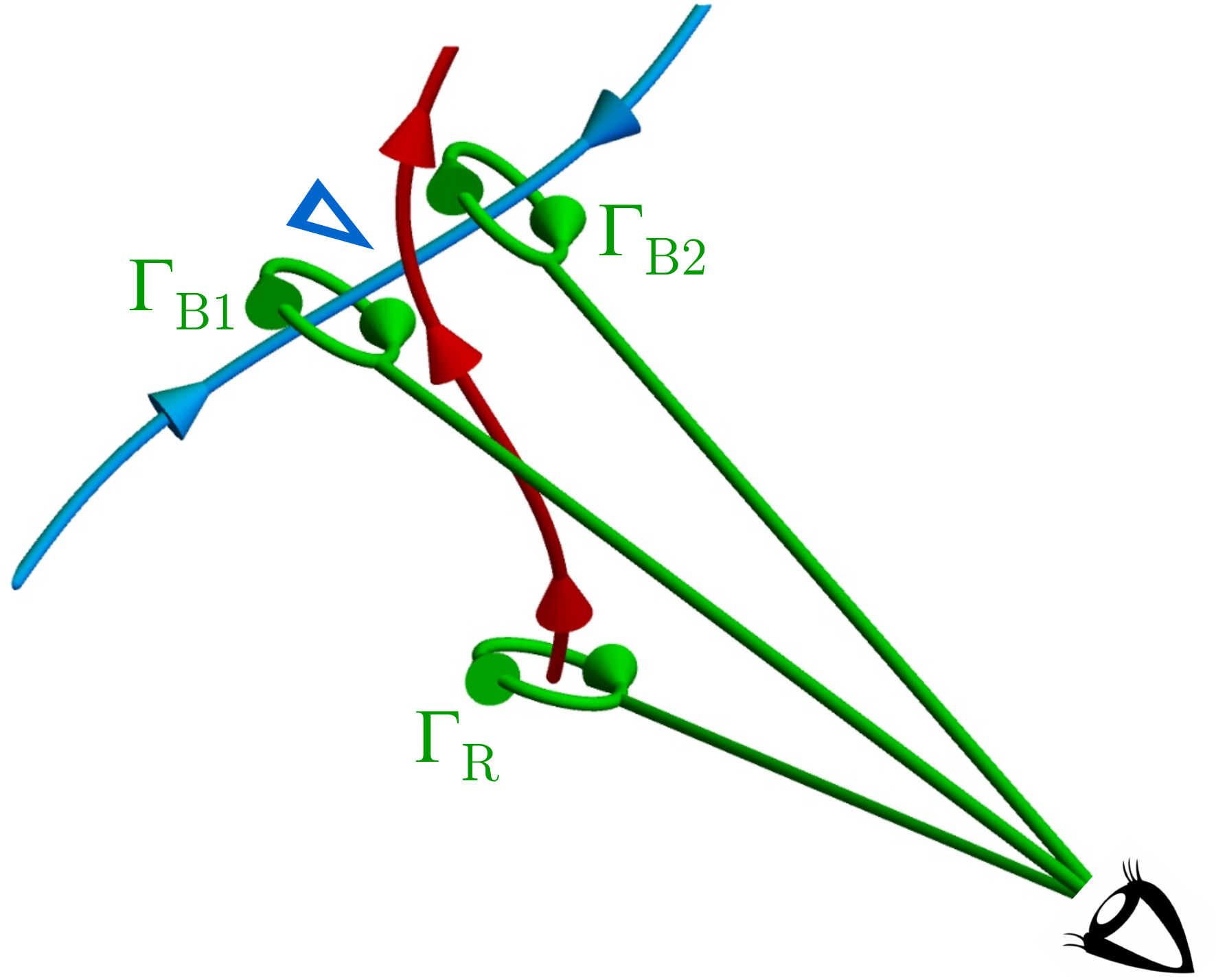}
 \caption{\textsf{To determine the orientation of nodal lines (NLs) inside a 3D momentum space, we consider the vantage point (indicated by eye) to be the base point. The closed paths $\Gamma$ which we use to calculate the charge of NLs consist of a straight line going from the eye to the vicinity of the NL, a small loop around the NL, and the same ray in the reverse direction (i.e.~returning to the eye). Two such loops enclosing the blue NL, labelled $\Gamma_{\textrm{B}1}$ and $\Gamma_{\textrm{B}2}$, are located on different sides of the red NL. Repeating the argument in Fig.~\ref{fig:conjugacy}, we find that the charge on path $\Gamma_{\textrm{B}2}$ equals the charge on path $\Gamma_{\textrm{B}1}$ (which equals $\imi$) conjugated by the charge on path $\Gamma_\textrm{R}$ (which equals $\imk$). The result of this calculation is $n = \imk \cdot \imi \cdot (-\imk) = -\imi$. According to this definition of the topological charge, the orientation of the blue NL is \emph{reversed} when it goes \emph{under} (as perceived from the vantage point) the red NL. The reversal is indicated by the triangular arrowhead.}}
\label{fig:orientations}
\end{figure}

There is a convenient geometric way to understand the reversal of nodal line (or world line) orientations, illustrated in Fig.~\ref{fig:orientations}. If we look at the NL composition from some vantage point inside the 3D space, the orientation of a NL is reversed each time it goes under a NL of the other color. To reach this conclusion, consider that your eye (your ``vantage point'') is the base point, i.e.~all the closed path begin and end in your eye. To calculate the charge of a NL at some location, we consider a closed path composed of three parts: (1) a straight line (a ``ray'') coming from your eye to the vicinity of the NL location, (2) a tight loop encircling the NL, and (3) the same ray in the reverse direction (i.e.~coming back into your eye). When perceived from the vantage point, the motion along the straight rays is projected away. Now consider a situation where a blue NL (conjugacy class $\{\pm\imi\}$) moves \emph{under} a red NL (conjugacy class $\{\pm\imk\}$), and take two locations along the blue NL located on the other side of the red NL (as perceived from the vantage point). The charge of the blue NL at these locations would correspond to the homotopy classes of paths $\Gamma_{\textrm{B}1}$ resp.~$\Gamma_{\textrm{B}2}$ in Fig.~\ref{fig:orientations}. But it follows from the argument in Fig.~\ref{fig:conjugacy} that these two charges are related by conjugacy with the homotopy class of path $\Gamma_{\textrm{R}}$ characterizing the red NL, which anticommutes with $\pm\imi$. It follows, that paths
$\Gamma_{\textrm{B}1}$ and~$\Gamma_{\textrm{B}2}$ carry \emph{opposite} charges, meaning that the orientation of the blue NL is \emph{reversed} when it goes under a red NL. 

One can construct an analogous argument with the words ``red'' and ``blue'' interchanged. Then it follows that the orientation of a red NL is reversed when it goes under a blue NL. We can summarize these two findings with the following single statement: In three-band models with NLs formed by the lower two bands (called ``red'') and NLs formed by the upper two bands (called ''blue''), the orientation of any NL is reversed when it goes under a NL of the other color.


\subsection[Constraints on admissible NL compositions]{Constraints on admissible NL compositions}\label{subsec:compositions}

In this section, we further investigate the properties of NLs in three-band models. We stick to the terminology that NLs formed by the lower/upper two bands are described as (and plotted with) red/blue. Especially, we focus on the following three properties: (1) a pair of NLs of different color cannot move across each other, (2) a closed NL \emph{ring} can only encircle an even number of NLs of the other color, and (3) the meaning of $\imi^4 = +1$, i.e.~why four NLs of the same type and the same orientation are, in certain sense, not robust.

\begin{figure*}[t!]
\includegraphics[width=0.64\textwidth]{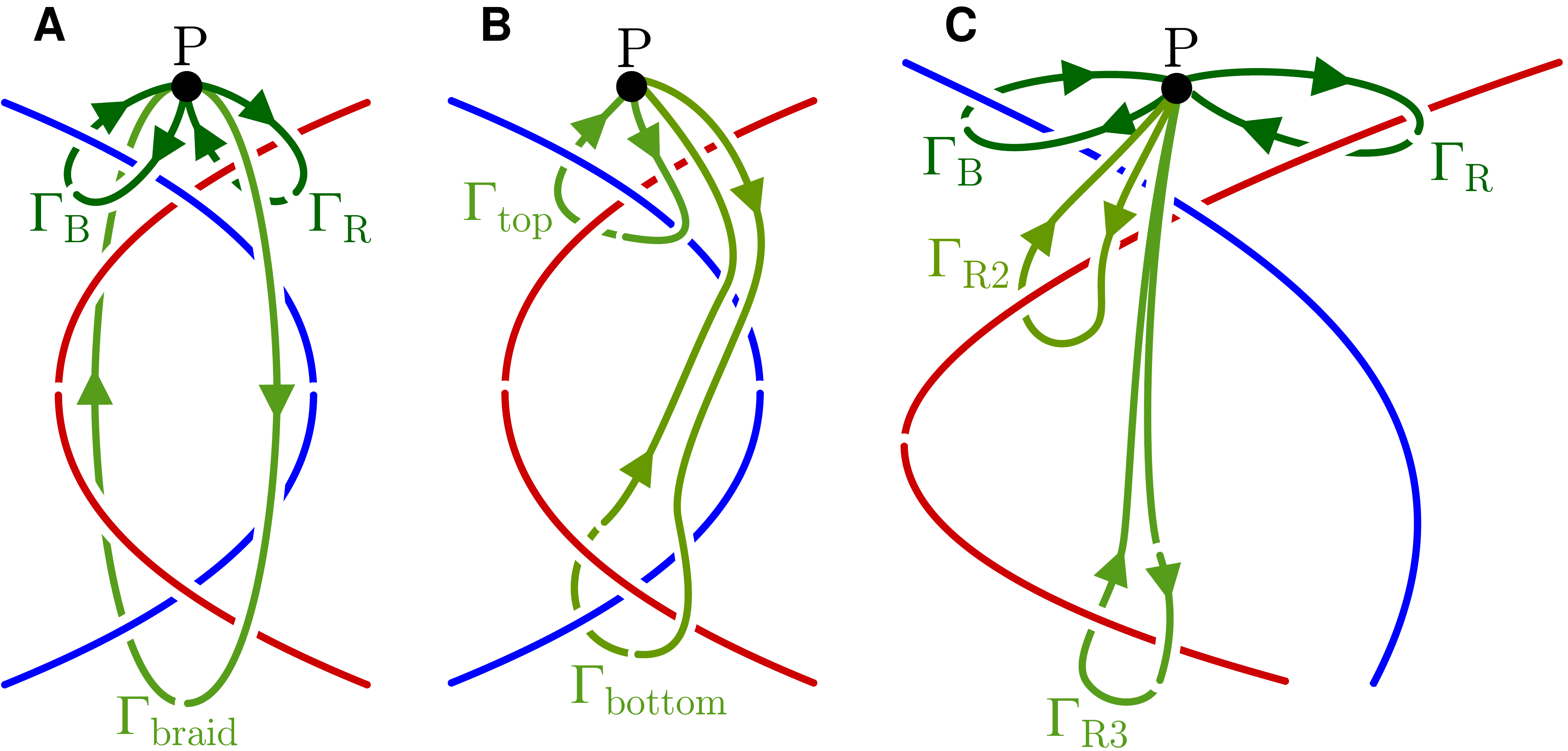}
 \caption{\textsf{The red and blue lines indicate nodal lines (NLs) formed by the lower resp.~upper two bands of a three-band Hamiltonian. The paths of different shade of green are loops based at $\textrm{P}$ which we use to encircle the NLs at various locations. We always let the paths go \emph{over} all the NLs on the way down from $\textrm{P}$. We say that a particular path $\Gamma$ encircles some set of NLs, if it goes \emph{under} those (and \emph{only} those) NLs on the way up to $\mathrm{P}$. Our goal is to express path $\Gamma_\textrm{braid}$ as an isotopy of a suitable composition of paths $\Gamma_{\textrm{R}}$ and $\Gamma_{\textrm{B}}$ encircling the red vs.~the blue NL. This goal is achieved by Eq.~(\ref{eqn:wow-it-really-braids}). For the explanation, see the text around Eqs.~(\ref{eqn:sub-braid}).}}
\label{fig:windings-biaxial-2}
\end{figure*}

We first show why a pair nodal lines with different colors cannot move across one another. We first prove this statement using an argument from Ref.~\cite{Mermin:1979}. Afterwards, we indicate how this feature can be understood as a simple consequence of the orientation reversal discussed in the previous section. Following first Ref.~\cite{Mermin:1979}, the key to prove our claim is to study the charge on path $\Gamma_\textrm{braid}$ encircling the ``braid'' of two nodal lines as shown in Fig.~\ref{fig:windings-biaxial-2}(A)~\cite{Mermin:1979,Sethna:2006}. The advantage of this path is that it becomes trivially contractible to a point if the NLs move across one another. If the charge on $\Gamma_\textrm{braid}$ is non-trivial, the path cannot be contracted to a point, which implies that the NLs \emph{cannot} move across each other.

To find the charge on path $\Gamma_\textrm{braid}$, we need to express the path as a homotopy (also called \emph{isotopy}, i.e.~a continuous deformation) of a composition of paths encircling the two nodal lines individually, namely $\Gamma_\textrm{B}$ (for ``blue'' NL) and $\Gamma_\textrm{R}$ (for ``red'' NL). In this argument, we base all the paths at point $\textrm{P}$ located \emph{above} the braid, see Fig.~\ref{fig:windings-biaxial-2}. Each path $\Gamma$ first goes down the illustration \emph{over} all the NLs. We say that the path \emph{encircles} some set of NLs, if it goes \emph{under} those (and \emph{only} those) NLs on the way back to $\textrm{P}$.  

To obtain the required expression, we first express
\begin{subequations}\label{eqn:sub-braid}
\begin{equation}
\Gamma_\textrm{braid} \sim \Gamma_\textrm{top}\circ \Gamma_\textrm{bottom},\label{eqn:path-compos}
\end{equation}
i.e.~as a composition of paths encircling the bottom and the top overlay of the two NLs as indicated by Fig.~\ref{fig:windings-biaxial-2}(B). [Recall that in expressions such as Eq.~(\ref{eqn:path-compos}) we first move along the path to the right of the composition symbol ``$\circ$'', and then along the path to the left.] Let us first zoom into the upper half of the braid, shown in Fig.~\ref{fig:windings-biaxial-2}(C). Since for the top overlay the blue NL is located \emph{above} the red one, we directly obtain
\begin{equation}
\Gamma_\textrm{top} \sim \Gamma_\textrm{R} \circ \Gamma_\textrm{B}.
\end{equation}
To express $\Gamma_\textrm{bottom}$, a few extra steps are needed. First, note that the path encircling the red NL to the left of the blue one can be expressed as
\begin{equation}
\Gamma_\textrm{R2}\sim \Gamma_\textrm{B}^{-1}\circ \Gamma_\textrm{R}\circ \Gamma_\textrm{B}.
\end{equation}
Encircling the red NL below the flection point corresponds to path $\Gamma_\textrm{R3}$. Clearly, $\Gamma_\textrm{R2}\circ \Gamma_\textrm{R3} \sim 1$ is contractible to $\textrm{P}$, therefore
\begin{equation}
\Gamma_\textrm{R3} \sim \Gamma_\textrm{R2}^{-1}.
\end{equation}
Similar inversion occurs when encircling the \emph{blue} NL below its flection point (not indicated in the figure). Since for the bottom overlay the blue NL is located \emph{below} the red one, we obtain 
\begin{equation}
\Gamma_\textrm{bottom} \sim  \Gamma_\textrm{B}^{-1} \circ \Gamma_{\textrm{R2}}^{-1}. 
\end{equation}
Composing all the partial results leads to relation 
\begin{equation}
\Gamma_\textrm{braid} \sim \Gamma_\textrm{R}^{\phantom{1}} \Gamma_\textrm{B}^{-1} \Gamma_\textrm{R}^{-1} \Gamma_\textrm{B}^{\phantom{1}}. \label{eqn:wow-it-really-braids}
\end{equation}
\end{subequations}
Assuming that the two NLs are characterized by two different imaginary units, such as $\imi$ and $\imk$, the total charge on the braid is 
\begin{equation}
n_\textrm{braid} = \imi \cdot (-\imk)\cdot (-\imi) \cdot \imk = -1 \neq 1,\label{eqn:you-shall-not-pass}
\end{equation}
i.e. \emph{non-trivial}. This result is explicitly mentioned below Eq.~(9) of the main text, and it implies that it is \emph{not possible} for the blue and red NLs in Fig.~\ref{fig:windings-biaxial-2} to cross each other. We explain below in Sec.~\ref{sec:multiband+mirror} how the presence of additional mirror symmetry leads to the appearance of the earring NLs observed in Fig.-3(F) of the main text.

\begin{figure}[b!]
\includegraphics[width=0.45 \textwidth]{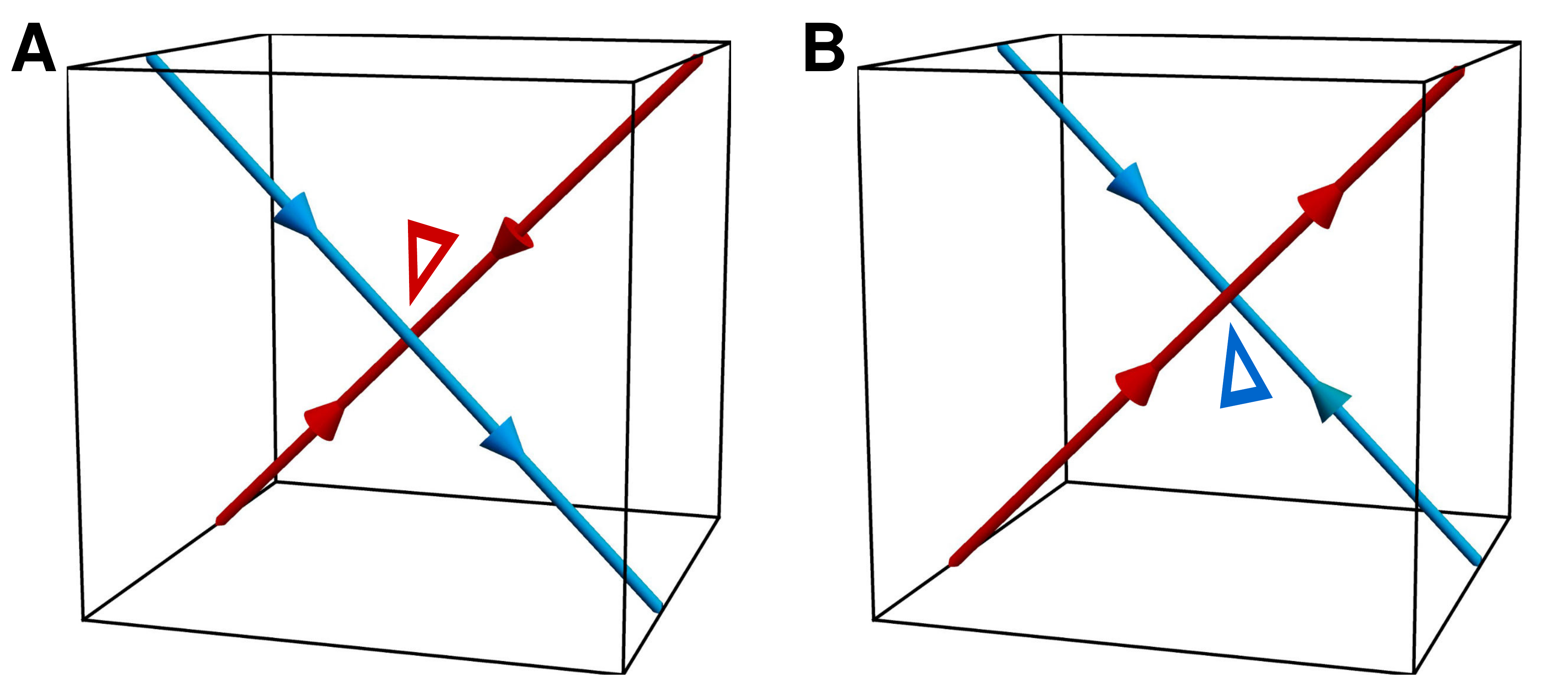}
 \caption{\textsf{(A) Red NL passing \emph{below} a blue NL in some region (the cube) of $\bs{k}$-space. The orientation of the blue NL is constant, while the orientation of the red NL is reversed at the overlay (indicated by triangular arrowhead). (B) Red NL passing \emph{over} a blue NL. In this case, the orientation of the red NL is the same everywhere, while the orientation of the blue NL is flipped at the overlay. Situation (A) cannot be evolved into situation (B) because the orientations of the NLs do not match. This implies that NLs of different color cannot move across each other.}}
\label{fig:crossing}
\end{figure}

\begin{figure*}[t!]
\includegraphics[width=1.0 \textwidth]{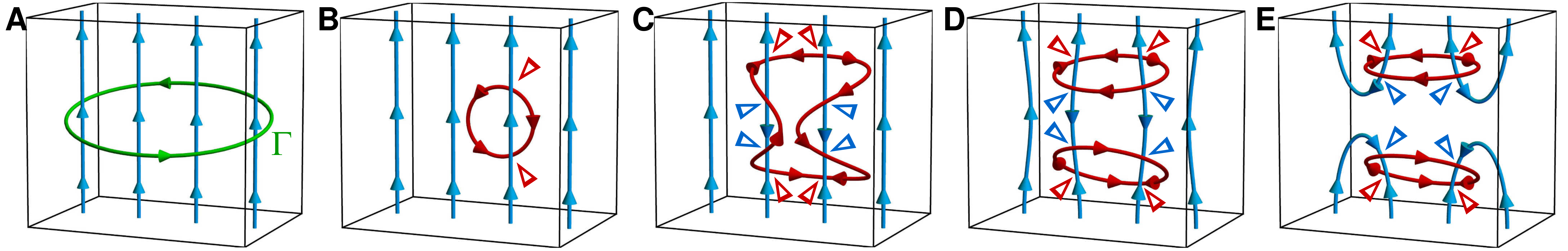}
 \caption{\textsf{Illustrating why four parallel NLs of the same type and orientation (carrying trivial charge $\imi^4 = +1$ in total) are not topologically stable. (A) We consider four blue NLs pointing vertically upward in some region (the cube) of $\bs{k}$-space. The green path $\Gamma$ carries a trivial value of the topological charge. (B) By locally distorting the Hamiltonian, we create red NL ring located \emph{behind} the blue NLs. We keep in mind that the red NL reverses orientation when passing below a blue NL (indicated by red triangular arrowheads). (C) We increase the size of the red NL ring. We fold the red NL in front of the central two blue NLs, such that it forms a narrow ``neck'' in front of the blue NLs. We indicate the orientation reversal of red/blue NLs using red/blue triangular arrowheads. (D) The two red NL segments forming the ``neck'' have opposite orientation, allowing us to reconnect then. The reconnection splits the original single red NL (which did not enclose any blue NLs) into two red NLs (each enclosing two blue NLs), which can be vertically separated. Note that the orientation of the central two blue NLs is flipped in the region between the red NL rings. (E) The reversed orientation of the blue NLs allows us to reconnect them, leading to the nodeless region at half the height of the cube. This allows us to shrink path $\Gamma$ to a point without encountering a NL, consistent with the trivial value of topological charge on $\Gamma$.}}
\label{fig:i4}
\end{figure*}

We now argue why NLs of different color cannot move across each other by referring to the orientation reversal discussed in the previous section. In this approach, we consider our vantage point (the ``eye'') to be the base point, and only work with the special paths (consisting of the straight ``rays'' composed with tight loops encircling the NLs) illustrated in Fig.~\ref{fig:orientations}. Our conclusion from the previous section is that the orientation (i.e.~the charge) of a NL reverses each time it goes under a NL of the other color. Let us first consider a situation with a red NL passing \emph{below} a blue NL, as shown in Fig.~\ref{fig:crossing}(A). The orientation of the red NL is reversed when it goes under the blue NL, while the blue NL has the same orientation everywhere. We now compare this to a situation with the red NL located \emph{in front} of the blue NL, as shown in Fig.~\ref{fig:crossing}(B). In this latter scenario, the red NL has the same orientation everywhere, while the orientation of the blue NL is reversed. One situation cannot be continuously changed into the other, because the orientation of the NLs (i.e. their charges) do not match. Therefore, NLs of different color cannot move across each other.

We now consider ring-shaped NLs. We prove that such NL rings can only enclose an \emph{even} number of NLs of the other color. The reason is that the orientation of a NL at some location is \emph{uniquely defined} -- it corresponds to the homotopy class of the Hamiltonian on paths shown in Fig.~\ref{fig:orientations}. Therefore, after traversing all along the NL ring, we must encounter an \emph{even} number of orientation reversals. In a NL composition, there are two reasons that the orientation of a NL becomes reversed: (1) There is a NL of the other color located \emph{in front of} the considered NL ring. Such situation leads to an even number of overlays, i.e.~an even number of orientation reversals. (2) There is a NL \emph{linked} with the considered NL ring. Such links lead to only a single orientation reversal of the NL ring. Therefore, for the orientation to be consistent along the whole NL ring, the ring must be linked with (i.e.~enclose) an even number of NLs of the other color. Especially, a two-color Hopf link is impossible. An example of an admissible NL composition composed of four rings is shown in Fig.~3(H) of the main text. Similar considerations in combination with the second homotopy group can be used to shed light~\cite{Tiwari:2019} on the relation of the \emph{monopole charge} of NL rings~\cite{Fang:2015} to the linking structure of the NLs~\cite{Ahn:2018}.

Finally, we discuss the meaning of $(\pm\imi)^4 = (\pm\imk)^4 = +1$. These relations suggest that four NLs of the same color and of the same orientation can mutually annihilate. This is indeed true, although not at first obvious. To explain this instability, we illustrate in Fig.~\ref{fig:i4}(A) four blue NLs pointing in the same direction, such that the green path $\Gamma$ enclosing them carries trivial value $\imi^4 = +1$ of the quaternion charge. The process of annihilating segments of the four blue NLs and of shrinking path $\Gamma$ to a point is illustrated and discussed in Fig.~\ref{fig:i4}. Here, we remark that by interpreting the vertical momentum component as time, one can see in Fig.~\ref{fig:i4}(E) world lines of non-trivially exchanged (i.e.~``braided'') point nodes of a 2D system. First, at an early time (bottom of Fig.~\ref{fig:i4}(E)), the 2D system exhibits four blue point nodes of the same charge $+\imi$. Later, a red node-antinode pair is created. The red nodes are then carried around two blue point nodes, and annihilated. While moving the red point nodes, the charge of the two blue point nodes has been flipped to $-\imi$. (The charge of the red point nodes has also been flipped from $\pm\imk$ to $\mp\imk$, but this still allows them to mutually annihilate.) This allows us to annihilate the blue point nodes with charge $+\imi$ and $-\imi$ in pairs. Therefore, at time corresponding to half the height of the cube in Fig.~\ref{fig:i4}(E), there are no point nodes in the 2D system.


\subsection[Generalized quaternion charge in multi-band models]{Generalized quaternion charge in multi-band models}\label{subsec:gen-many}

We now generalize the characterization of NLs in $\mcP\mcT$-symmetric models with $N\!=\!3$ bands to a general number $N\geq 3$ of bands. This discussion is somewhat technical. Nevertheless, our main conclusion is extraordinarily simple: Nodes formed by \emph{consecutive pairs of bands} [such as $(j-1,j)$ and $(j,j+1)$ for $1<j<N$] \emph{anticommute}, while all nodes formed by \emph{more distant pairs of bands commute}. Along the way, we base our arguments in formal mathematical terms. Following the main text, we sometimes call the homotopy group $\pi_1(M_N)$ generalizing the result in Eq.~(\ref{eqn:quat-charge}) as the \emph{generalized quaternion charge}.

By considering the frame formulation of the Hamiltonian in Eq.~(\ref{eqn:Ham-frames}), we find that the space of spectrally projected $\mcP\mcT$-symmetric Hamiltonians is equivalent to the space of all distinct orientations of a generic $N$-dimensional ellipsoid. We thus rewrite the Eq.~(\ref{eqn:M-N-space-gen}) as 
\begin{equation}
M_{N} \!=\! \mathsf{O}(N)/\mathsf{P}_{Nh} \!=\! \mathsf{SO}(N)/\mathsf{P}_N \!=\! \mathsf{Spin}(N)/\overline{\mathsf{P}}_N\label{eqn:isomorphs-N-band}
\end{equation}
where $\mathsf{P}_{Nh} \equiv \mathsf{O}(1)^N \equiv \ztwo^N$ (equivalently, the Coxeter group $N\mathsf{A}_1$) is an Abelian point group in $N$-dimensional Euclidean space generated by reflections with respect to $N$ mutually perpendicular axes.
In the second step of Eq.~(\ref{eqn:isomorphs-N-band}), we select the \emph{special} component of both groups in the coset construction. More specifically, $\mathsf{P}_N$ contains all elements of $\mathsf{P}_{Nh}$ with positive determinant, i.e. only proper rotations. One can understand $\mathsf{P}_N \cong \ztwo^{N-1}$ as being generated by $\pi$-rotations in $(N-1)$ planes $[x_1,x_2], [x_1,x_3], \ldots, [x_1,x_N]$, where $\{x_j\}_{j=1}^N$ are the coordinates of the $N$-dimensional real Hilbert space with respect to the frame $\mathsf{u}(\bs{k})$. The $\pi$-rotation in plane $[x_1,x_{j}]$ flips the sign of eigenvectors $|{u^1_{\bs{k}}}\rangle$ and $|{u^j_{\bs{k}}}\rangle$ while leaving the other $(N-2)$ eigenvectors invariant. In the last step of Eq.~(\ref{eqn:isomorphs-N-band}), we replace $\textsf{SO}(N)$ by its simply connected double cover $\textsf{Spin}(N)$, and we lift $\mathsf{P}_N$ into the corresponding double group $\overline{\mathsf{P}}_N$.

To correctly identify the double group $\overline{\mathsf{P}}_N$, note that it is generated by the same set of $(N\!-\!1)$ $\pi$-rotations as $\mathsf{P}_N$. However, within the double group, these generators \emph{anticommute} (all generators share coordinate $x_1$) and \emph{square to $-1$}. Therefore, we can identify them with the basis $\{{e}_1,{e}_2,\ldots,{e}_{N-1}\}$ of \emph{real Clifford algebra} $C\!\ell_{0,N-1}$. More precisely, we relate
\begin{equation}
\textrm{$\pi$-rotation in $[x_1,x_j]$ plane}\!\!\!\quad \leftrightarrow \!\!\!\quad \textrm{basis vector $e_{j-1}$}\label{eqn:rotation-Clifford}
\end{equation} 
where $2 \leq j \leq N$. Although elements of Clifford algebra seem to be rather abstract objects, we show in Sec.~\ref{sec:calculating-the-charge} a formulation of the basis vectors $e_{j}$ using Dirac matrices. The collection $\overline{\mathsf{P}}_N$ as a set corresponds to the $2^N$ products
\begin{equation}
\overline{\mathsf{P}}_N = \bigcup_{n_i\in\{0,1\}}\{\pm e_1^{n_1}e_2^{n_2}\ldots e_{N-1}^{n_{N-1}}\}.\label{eqn:Salingaros-def}
\end{equation}
The subset~(\ref{eqn:Salingaros-def}) of $C\!\ell_{0,N-1}$ is closed under multiplication and forms a group called the \emph{Salingaros vee group}~\cite{Salingaros:1981,Brown:2015,Ablamowicz:2017,Salingaros:1983}. Applying theorems reviewed in Sec.~\ref{eqn:homotopy-cosets} leads to
\begin{subequations}
\begin{eqnarray}
\pi_1(M_{N}) &=& \overline{\mathsf{P}}_N \label{eqn:salingaros-homotop}\\
\pi_2(M_{N}) &=& \triv
\end{eqnarray}
\end{subequations}
The groups in Eq.~(\ref{eqn:Salingaros-def}) are non-Abelian for $N \geq 3$, and can be defined recurrently using \emph{central product} of groups with coinciding centers~\cite{Salingaros:1983}. 

Let us describe properties of nodes described by group $\overline{\mathsf{P}}_N$ in more detail. For our considerations, we only need to know the \emph{conjugacy classes} of these groups, which are fairly easy to identify. Since the basis elements of the Clifford algebra $C\!\ell_{0,N-1}$ mutually anticommute, all pairs of elements of $\overline{\mathsf{P}}_N$ in Eq.~(\ref{eqn:Salingaros-def}) either commute or anticommute. Therefore, the conjugacy classes of $\overline{\mathsf{P}}_N$ consist of at most two elements that only differ in the overall sign, i.e. $\{\pm e_1^{n_1}e_2^{n_2}\ldots e_{N-1}^{n_{N-1}}\}$ with $n_i \in\{0,1\}$. The only exceptions are the following:
\begin{itemize}
\item For $N$ odd, there are two one-element conjugacy classes $\{1\}$ and $\{-1\}$, while
\item for $N$ even, there four one-element conjugacy classes $\{+1\}$, $\{-1\}$, $\{e_1 e_2 \ldots e_{N-1}\}$ and $\{- e_1 e_2 \ldots e_{N-1}\}$.
\end{itemize}
The total number of conjugacy classes is
\begin{equation}
\abs{C_{\overline{\mathsf{P}}_N}} = \left\{\begin{array}{ll}
2^{N-1} + 1     &   \textrm{for $N$ odd} \\
2^{N-1} + 2     &   \textrm{for $N$ even.} \\
\end{array}\right.\label{eqn:conj-classes-gen}
\end{equation}
We remark that the classification based on the conjugacy classes in Eq.~(\ref{eqn:conj-classes-gen}) provides more detail than the classification based on the quantization of the Berry phase of the individual bands to $0$ vs.~$\pi$~\cite{Berry:1984,Zak:1989}. Since the sum of the Berry phases of all the bands must be $0$ (mod $2\pi$), only $(N-1)$ of the phases are independent, leading to $2^{N-1}$ topological classes distinguished by the Berry phases. The fact that $\abs{C_{\overline{\mathsf{P}}_N}} > 2^{N-1}$ implies the existence of robust nodes protected by a previously unreported topological obstruction. We argue below in Sec.~\ref{sec:Wilson-failure} that this novel topological charge (corresponding to conjugacy class $\{-1\}$) survives in the stable limit of many bands. 

To get a more tangible connection between the conjugacy classes $C_{\overline{\mathsf{P}}_N}$ and the considered physics problem, let us characterize each conjugacy class in terms of the species of NLs that it characterizes. To achieve this goal, we parametrize a closed path $\Gamma \simeq S^1$ by $\alpha\in[0,2\pi)$. A node formed by consecutive bands $|{u^j_{\bs{k}}}\rangle$ and $|{u^{j+1}_{\bs{k}}}\rangle$ can be modelled by rotating the ``standard'' Hamiltonian $\mathcal{H} = \mathcal{E}$ using matrices
\begin{subequations}
\begin{eqnarray}
\alpha \mapsto R(\tfrac{\alpha}{2}) &=& \e{\tfrac{\alpha}{2} L_{j,j+1}} \nonumber \\
&=& \left(\begin{array}{cccc}
\unit_{j-1} &   0   &   0   &   0  \\
0   & \cos\tfrac{\alpha}{2}   &   -\sin\tfrac{\alpha}{2}  & 0   \\
0   & \sin\tfrac{\alpha}{2}   &   \cos\tfrac{\alpha}{2}   & 0 \\
0   &   0   &   0   & \unit_{N-j-1} 
\end{array}\right)\;\;
\end{eqnarray}
where the $N(N-1)/2$ skew-symmetric matrices 
\begin{equation}
(L_{ij})_{ab} = -\delta_{ia}\delta_{jb} + \delta_{ib}\delta_{ja} \qquad\textrm{with}\quad i<j \label{eqn:so-n-algebra-gens}
\end{equation}
span the basis of Lie algebra $\mathfrak{so}(N)$. 
Since the remaining $(N-2)$ bands are constant along $\Gamma$, we only need to keep track of the Hamiltonian block describing bands $j$ and $j+1$, which is 
\begin{eqnarray}
\mcH_\textrm{eff.}(\alpha) 
&=& R_\textrm{eff.}(\tfrac{\alpha}{2})\left(\begin{array}{cc}
\!j       & 0\!     \\
\!0       & j\!+\!1\!
\end{array}\right)R_\textrm{eff.}(\tfrac{\alpha}{2})^\top \nonumber \\
&=& \frac{1}{2}\left(\begin{array}{cc}
\cos \alpha        &   \sin \alpha \\
\sin \alpha        &   -\cos \alpha
\end{array}\right) + \left(j+\tfrac{1}{2}\right)\unit,
\end{eqnarray}
\end{subequations}
We see that the effective Hamiltonian is continuous and single-valued on path $\Gamma$, just as it should be.

Importantly, the rotation matrices $R(\pi) = \e{\pi L_{i,j}}$ obtained after completing the roundtrip along $\Gamma$ differ from the identity matrix only by two minus signs at positions $i$ and $j$ along the diagonal. Assuming $j \geq 2$ (the case $j=1$ is special), we write
\begin{subequations}
\begin{equation}
\e{\pi L_{j,j+1}} = \e{\pi L_{1,j}}\e{\pi L_{1,j+1}}. \label{eqn:so-n-compositions}
\end{equation}
meaning that we identify a NL between bands $|{u^j_{\bs{k}}}\rangle$ and $|{u^{j+1}_{\bs{k}}}\rangle$ (left-hand side) as the conjugacy class $\{\pm e_{j-1} e_{j}\}$ (right-hand side) using the assignment in Eq.~(\ref{eqn:rotation-Clifford}). The case $j=1$ does not require the additional step in Eq.~(\ref{eqn:so-n-compositions}). We thus obtain the relation between band-structure nodes and the conjugacy classes of $\overline{\mathsf{P}}_{N}$,
\begin{equation}
\textrm{node between $(j,j+1)$}\!\sim\!\left\{\!\begin{array}{ll}
\!\{\pm e_1\}             & \!\textrm{for $j=1$}\!    \\
\!\{\pm e_{j-1} e_j\}     & \!\textrm{for $j \geq 2$}\!
\end{array}\right.\label{eqn:AI-N-conj-rel}
\end{equation}
\end{subequations}
By taking the appropriate products, relation~(\ref{eqn:AI-N-conj-rel}) allows us to interpret \emph{all} conjugacy classes enumerated in Eq.~(\ref{eqn:conj-classes-gen}) as suitable collections of nodes between the individual bands.

Knowing the relation in Eq.~(\ref{eqn:AI-N-conj-rel}), we are able to analyze the non-trivial exchange rules (i.e.~``braiding'') of nodes in systems with an arbitrary number $N \geq 3$ of bands. Exchanging two band-structure nodes corresponds to commuting two conjugacy classes in Eq.~(\ref{eqn:AI-N-conj-rel}). Such an exchange reduces to commuting pairs of basis elements $e_i$. If the two classes don't share any basis element, the commutation yields an even number of minus signs, i.e. the two classes commute. On the other hand, if the two classes have one basis element in common, an odd number of minus signs is acquired during the commutation. We see from Eq.~(\ref{eqn:AI-N-conj-rel}) that the latter situation arises precisely for \emph{consecutive pairs of bands}. This completes the derivation of the  exchange rules states in the first paragraph of this section.

Without going through the details, let us summarize how some of the properties of NLs discussed in Secs.~\ref{subsec:braid} and~\ref{subsec:compositions} generalize to systems with $N\geq 3$ bands. We find that only NLs formed inside \emph{consecutive band gaps} have anticommuting charges. Therefore, NLs formed inside two consecutive band gaps \emph{cannot cross}, and they reverse orientation when passing \emph{under} each other. On the other hand, NLs formed inside more distant band gaps can trivially pass through each other, and they do \emph{not} affect each others orientation. Finally, each NL ring must be linked with an even number of NLs formed inside neighboring (i.e.~one higher and one lower) band gap. We postpone a mathematical formulation of the topological invariant defined by Eq.~(\ref{eqn:salingaros-homotop}) using parallel transport of Hamiltonian eigenstates, as well as a presentation of a numerical algorithm for computing the invariant along a given bath $\Gamma$, until Sec.~\ref{sec:calculating-the-charge} and Sec.~\ref{sec:1D-Ham} further below.


\subsection{CP transfer in the presence of mirror symmetry}~\label{sec:multiband+mirror}

In this section, we explain the transfer of crossing point (CP) of nodal lines (NLs) observed in Fig.~3(A--C) of the main text. We fix the number of bands to $N=3$ and consider the presence of $m_z$ mirror symmetry represented by $\hat{m}_z$ in Eq.~(\ref{eqn:mirror-three-bands}). The subspace $X_{m_z} \subset M_3$ consists of three connected components
\begin{subequations}
\begin{equation}
X_{m_z} = X^1\amalg X^2 \amalg X^3,\label{eqn:amalgamate-3}
\end{equation}
where the superscript indicates which one of the three bands has \emph{negative} $\hat{m}_z$ eigenvalue. 

Each of the disjoint components of $X_{m_z}$ in Eq.~(\ref{eqn:amalgamate-3}) is homeomorphic to $S^1$. To see this, recall that the most general $\mcP\mcT$-symmetric Hamiltonian commuting with $\hat{m}_z$ takes the form in Eq.~(\ref{eqn:3-band-gen-mz-Ham}). Considering as an example the component $X^2$, normalizing the eigenvalues to $\{j\}_{j=1}^3$ while taking care of the correct $\hat{m}_z$ eigenvalues of the individual eigenstates requires the coefficients $d_{0,x,z}$ and $\eps$ to fulfill
\begin{equation}
\eps = 2,\quad d_0 = 2,\quad\textrm{and}\quad d_x^2 + d_z^2 = 1
\end{equation}
which manifests the $S^1$ shape. One can similarly argue about $X^1$ and $X^3$.

We want to determine the relative homotopy group $\pi_1(M_3,X_{m_z})$, assuming as usual that both endpoints of $D^1$ are mapped to the same connected component of $X_{m_z}$. We thus consider the long exact sequence
\begin{equation}
\genfrac{}{}{0pt}{}{\pi_1(S^1)}{\intg}\genfrac{}{}{0pt}{}{\stackrel{i_1\;}{\rightarrow}}{\rightarrow}\genfrac{}{}{0pt}{}{\pi_1(M_{3})}{\mathsf{Q}}\genfrac{}{}{0pt}{}{\stackrel{j_1\;}{\rightarrow}}{\rightarrow}\genfrac{}{}{0pt}{}{\pi_1(M_{3},S^1)}{??}\genfrac{}{}{0pt}{}{\stackrel{\partial_1\;}{\rightarrow}}{\rightarrow}\genfrac{}{}{0pt}{}{\pi_0(S^1)}{\triv}\!\label{eqn:LES-AI3+mz}
\end{equation}
where we replaced $X^i \simeq S^1$, and in the second line we placed the known homotopy groups. It can be shown by repeating the arguments around Eq.~(\ref{eqn:calculate-phase}) that the image of the inclusion 
\begin{equation}
\iota^j : X^j\hookrightarrow M_{3}
\end{equation} 
\end{subequations}
winds non-trivially in $M_3$. In fact, $\iota^1$ ($\iota^3$) corresponds to enclosing a NL between the upper (lower) two bands, while $\iota^2$ corresponds to a node formed by bands $j=1$ and $j=3$ while keeping the central $j=2$ band decoupled. Therefore
\begin{subequations}\label{eqn:Z-4-too}
\begin{eqnarray}
\im \iota^{1} &=& \{1,\imi,-1,-\imi\} \cong \mathbb{Z}_4 < \mathsf{Q} \\
\im \iota^{2} &=& \{1,\imj,-1,-\imj\} \cong \mathbb{Z}_4 < \mathsf{Q} \\
\im \iota^{3} &=& \{1,\imk,-1,-\imk\} \cong \mathbb{Z}_4 < \mathsf{Q}.
\end{eqnarray}
\end{subequations}
Plugging Eqs.~(\ref{eqn:Z-4-too}) into the long exact sequence in Eq.~(\ref{eqn:LES-AI3+mz}) yields
\begin{equation}
\pi_1(M_{3},X^j) = \mathsf{Q}/\mathbb{Z}_4 \cong \ztwo\label{eqn:AI3+mz+example-quotient}
\end{equation}
for each connected component $X^j$ of $X_{m_z}$. The elements of the groups in Eq.~(\ref{eqn:AI3+mz+example-quotient}) are a pair of cosets, explicitly
\begin{subequations}\label{eqn:cosets-all}
\begin{eqnarray}
\pi_1(M_{3},X^1) &=& \left\{\{\pm 1,\pm\imi\},\{\pm\imj,\pm\imk\}\right\} \\
\pi_1(M_{3},X^2) &=& \left\{\{\pm 1,\pm\imj\},\{\pm\imi,\pm\imk\}\right\} \label{eqn:coset-j}\\
\pi_1(M_{3},X^3) &=& \left\{\{\pm 1,\pm\imk\},\{\pm\imi,\pm\imj\}\right\}.\label{eqn:coset-k}
\end{eqnarray}
\end{subequations}
The coset structure implies that the quaternion charge on an open-ended path $\gamma$ with $\partial\gamma$ lying in the $m_z$-invariant plane is well defined only up to the $\mathbb{Z}_4$ group in Eq.~(\ref{eqn:Z-4-too}).

\begin{figure}[t]
	\includegraphics[width=0.45\textwidth]{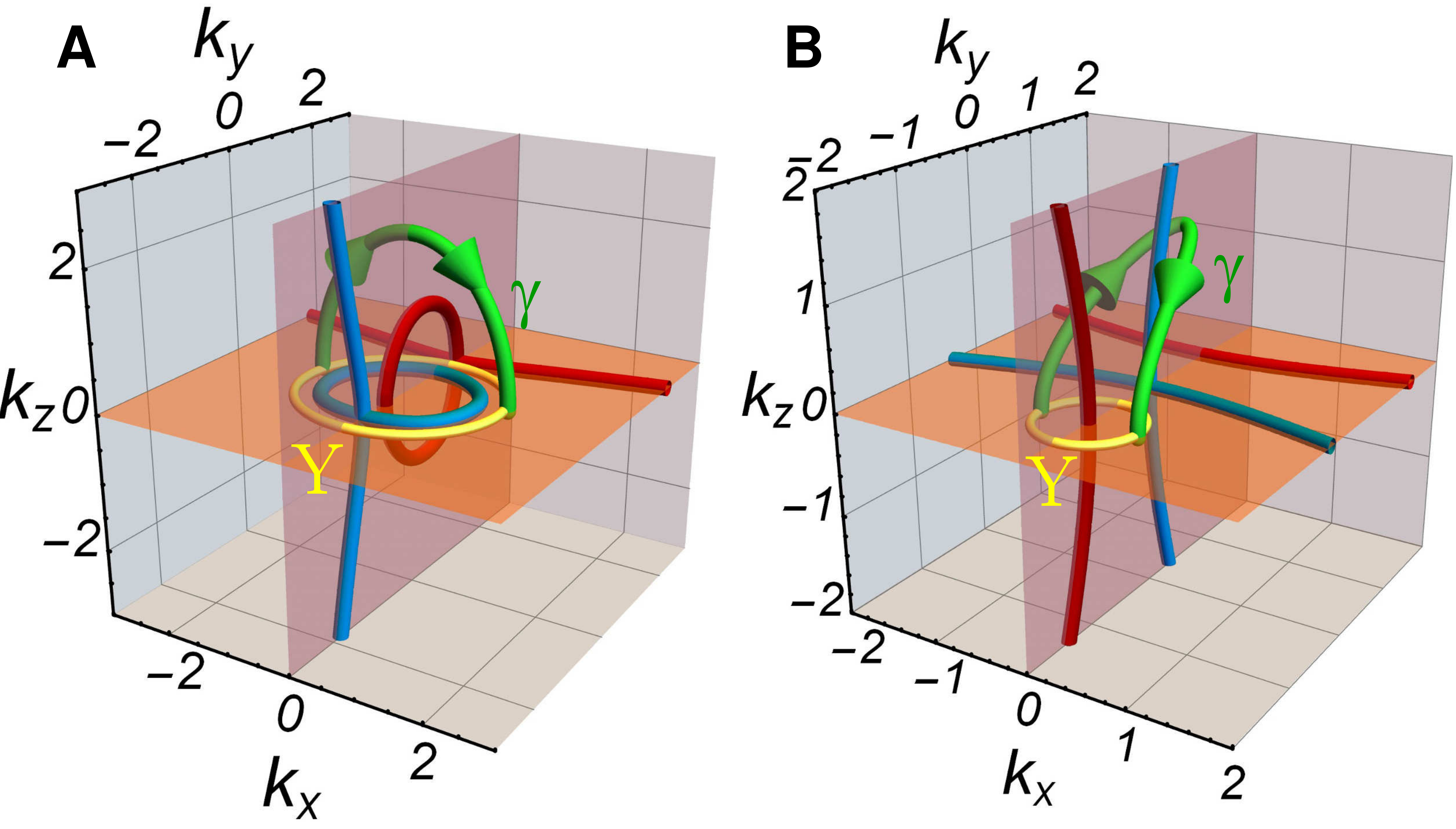}
	\caption{\textsf{The relative homotopy charge $\pi_1(M,X)<\pi_1(M)$ on an open-ended path $\gamma$ (green) with endpoints $\partial\gamma$ located inside the $m_z$-invariant plane (orange) is defined only up to a composition with elements of $\pi_1(X)$. This is represented by moving the endpoints $\partial\gamma$ along the in-plane circle (yellow). (A) Reproducing Fig.~3(F) of the main text, we find that $n_\gamma = \pm\imi$ (enclosing blue NL) or $n_\gamma= \pm\imk$ (enclosing red NL), depending on where we connect the endpoints $\partial\gamma$. (B) Reproducing Fig.~3(C) of the main text but with an altered choice of $\gamma$, we find that $n_\gamma = \pm\imi$ (enclosing blue NL) or $n_\gamma = \pm \imj$ (enclosing both NLs). The difference corresponds to the charge $n_\textrm{Y} = \pm\imk$ defined on the in-plane yellow circle.	}}
	\label{fig:cosets}
\end{figure} 

To explain the meaning of the cosets in Eq.~(\ref{eqn:cosets-all}), we consider two examples in Fig.~\ref{fig:cosets}. First, we recall the situation in Fig.~3(F) from the main text, reproduced as Fig.~\ref{fig:cosets}(A). The endpoints of the open-ended path $\gamma$ can be brought together inside the vertical red NL ``earring'', when the quaternion charge on $\gamma$ becomes $n_\gamma = \pm\imk$. On the other hand, moving $\partial\gamma$ along the indicated yellow circle allows us to connect the endpoints near the blue CP, when $n_\gamma = \pm\imi$ since now $\gamma$ encloses a \emph{blue} NL. This is consistent with the $\{\pm \imi,\pm\imk\}$ coset in Eq.~(\ref{eqn:coset-j}) and the fact that at the endpoints $\mcH\circ\iota: \partial D^1 \to X^2$, cf.~the inset to Fig.~3(F) of the main text. The difference between $\pm \imk$ and $\pm \imi$ corresponds to the charge $n_\textrm{Y} = \pm\imj$ defined on the yellow circle. Similarly, bringing together the endpoints of path $\gamma$ in Fig.~\ref{fig:cosets}(B) by sliding them along the yellow path yields either charge $n_\gamma = \pm\imi$ (blue NL) or $n_\gamma = \pm\imj$ (both NLs). This is consistent with the coset $\{\pm\imi,\pm\imj\}$ in Eq.~(\ref{eqn:coset-k}) and the fact that $\mcH\circ\iota: \partial D^1 \to X^3$, cf.~the inset to Fig.~3(C) of the main text. Note that in both examples, the semi-circular path $\gamma$ enclosing the CP corresponds to the \emph{non-trivial coset} in Eq.~(\ref{eqn:cosets-all}), i.e.~the one that does not contain the identity. We conclude that CPs are protected by a non-trivial value of the $\ztwo$ invariant corresponding to the relative homotopy group in Eq.~(\ref{eqn:AI3+mz+example-quotient}). The same topological invariant also governs the transfer of the CP between the red and blue NLs.


\section{Non-Abelian charge of 1D systems}

In this section, we discuss the geometric interpretation and the ways to numerically calculate the \emph{generalized quaternion charge}. Our presentation is structured as follows. In Sec.~\ref{sec:Wilson-failure}, we motivate the discussion by showing that the conventional approach to topological invariants based on Wilson loops (i.e. parallel transport of linear subspaces)~\cite{Yu:2011,Soluyanov:2011,Gresch:2017} is unable to detect the quaternion charge $n_\Gamma=-1$. We recognize that this ultimately stems from the two-fold universal covering of $\mathsf{SO}(N)$. 

In Sec.~\ref{sec:calculating-the-charge} we formulate the quaternion charge using parallel transport of \emph{frames} lifted into the covering space $\mathsf{Spin}(N)$. This is achieved by working explicitly inside the Lie algebra $\mathfrak{spin}(N)$. Furthermore, we find a matrix representation of the fundamental groups $\overline{\mathsf{P}}_N$. With the help of the Baker-Campbell-Hausdorff formula, we present a numerical algorithm for calculating the generalized quaternion charge accumulated along a closed path. Finally, in Sec.~\ref{sec:1D-Ham} we consider 1D $\mcP\mcT$-symmetric systems, and investigate the relation between the bulk value of the generalized quaternion invariant and the spectrum at the system edges.

\subsection{Failure of the usual Wilson-loop approach}\label{sec:Wilson-failure}

The conventional way of expressing topological invariants is the construction of Wilson-loop operators~\cite{Yu:2011,Soluyanov:2011,Gresch:2017}. Considering a collection of bands $\{u^j_{\bs{k}}\}_{j\in \mcI}$ where the index set $\mcI$ is a subset of $\{1,2,\ldots,N\}$ with $\abs{\mcI}$ elements, the Wilson operator along a closed path $\Gamma$ is defined as
\begin{subequations}\label{eqn:par-trans-vec}
\begin{equation}
\mcW^{ij}_\mcI(\Gamma) = \lim_{\Lambda\to\infty} \langle u^i_{\bs{k}_0}|\! \left[\! \ordprod_{\{\bs{k}_\alpha\}_{\alpha=1}^\Lambda\subset\Gamma} \!\!\!\!\!\!\mathbb{P}_\mcI(\bs{k}_\alpha)\right] \! | u^j_{\bs{k}_0}\rangle\label{eqn:Wilson-loop-op}
\end{equation}
where $\Lambda$ is the number of points along $\Gamma$, $\bs{k}_0$ is the basepoint of $\Gamma$, the bar over the product indicates path ordering, and 
\begin{equation}
\mathbb{P}_\mcI(\bs{k}_\alpha) = \sum_{j\in\mcI} | u^j_{\bs{k}_\alpha}\rangle \langle u^j_{\bs{k}_\alpha} |
\end{equation}
is the projector onto the linear subspace of the Hilbert space spanned by states $\{u^j_{\bs{k}}\}_{j\in \mcI}$. The expression in Eq.~(\ref{eqn:Wilson-loop-op}) can be rewritten as path-ordered exponential of the non-Abelian Berry-Wilczek-Zee (BWZ) connection~\cite{Wilczek:1984}, i.e.
\begin{eqnarray}
\mcW_\mcI(\Gamma) &=& \overline{\exp}\left[-\oint_\Gamma \mcbsA_\mcI(\bs{k})\cdot \de\bs{k}\right]\label{eqn:Wilson-using-BWZ}\\
&\phantom{=}&\textrm{with}\;\; [\mcbsA_\mcI(\bs{k})]^i_{\phantom{i}aj} = \langle u^i_{\bs{k}}| \partial_{k_a}| u^j_{\bs{k}}\rangle
\end{eqnarray}
\end{subequations}
and with $i,j\in\mcI$. It follows from $0 = \partial_{k_a}\langle u^i_{\bs{k}}| u^j_{\bs{k}} \rangle$ that components $[\mcbsA_\mcI(\bs{k})]_a$ are \emph{skew-Hermitian} matrices~\cite{Xiao:2010}, i.e.  generally (in the absence of additional symmetry) $\mcbsA_\mcI(\bs{k})$ is a $\mathfrak{u}(N)$-valued 1-form~\cite{Baez:1994}. The procedure summarized by Eqs.~(\ref{eqn:par-trans-vec}) is often called \emph{parallel transport of vectors} (or as \emph{parallel transport of linear subspaces}). 

Geometrically, the Wilson operator describes the rotation of the linear subspace spanned by $\{u^j_{\bs{k}}\}_{j\in \mcI}$ inside the Hilbert space. While the entire linear subspace returns back to itself after moving along a closed path $\Gamma$, the subspace may perform an internal rotation. More precisely, $\mcW^{ij}_\mcI(\Gamma)$ expresses the probability amplitude that state $|u^j_{\bs{k}_0}\rangle $ is evolved into state $| u^i_{\bs{k}_0} \rangle$ by the product of projectors in Eq.~(\ref{eqn:Wilson-loop-op}). Because of the reality of the states in $\mcP\mcT$-symmetric systems, $\mcW_\mcI(\Gamma)\in\mathsf{SO}(\abs{\mcI})$. 

Wilson operator depends on the gauge of $\{u^j_{\bs{k}_0}\}_{j\in \mcI}$. Performing an internal rotation among the states by  $\mathcal{X}(\bs{k})\in\mathsf{U}(\abs{\mcI})$ transforms the Wilson operator as
\begin{equation}
\mcW(\Gamma) \mapsto \mathcal{X}(\bs{k}_0) \mcW(\Gamma) \mathcal{X}(\bs{k}_0)^\dagger.\label{eqn:wilson-trsnd}
\end{equation}
The transformation in Eq.~(\ref{eqn:wilson-trsnd}) leaves the \emph{eigenvalues} of $\mcW$ invariant. Many topological invariants can be conveniently determined by studying the gauge-invariant eigenvalues of Wilson-loop operators~\cite{Gresch:2017}. Furthermore, \emph{nested} Wilson loops have been applied to describe \emph{higher-order} topological insulators and superconductors~\cite{Benalcazar:2017}.

We now show that the usual Wilson-loop approach to topological invariants, summarized by Eqs.~(\ref{eqn:par-trans-vec}), \emph{fails} to capture the (generalized) quaternion charge $n_\Gamma\!=\!-1$. To see this explicitly, recall from Sec.~\ref{subsec:gen-many} that one obtains $n_\Gamma\!=\!-1$ on a closed path $\Gamma$ by performing a $2\pi$-rotation of the Hamiltonian along the path. If we parametrize the path by $\alpha\in[-\pi,\pi)$, then
\begin{subequations}
\begin{eqnarray}
\mcH(\alpha) &=& \!\e{\alpha L_{j,j+1}} \mathcal{E} \e{-\alpha L_{j,j+1}} \nonumber \\
&=& \!\mathcal{E} \!+\! \frac{1}{2}\!\!\left(\begin{array}{cccc}
\!\triv_{j-1}\!\! &   0   &   0   &   0  \\
0   & \!1\!-\!\cos 2\alpha\!   &   -\!\sin 2\alpha  & 0   \\
0   & -\!\sin 2\alpha   &   \!\cos 2\alpha \!-\!1\!   & 0 \\
0   &   0   &   0   & \!\!\triv_{N-j-1}\!\! 
\end{array}\right)\qquad\label{eqn:2pi-rot-Ham}
\end{eqnarray}
where $\mathcal{E}$ is the diagonal matrix defined in Eq.~(\ref{eqn:standard-energies}), and the $\mathfrak{so}(N)$ basis elements $L_{i,j}$ are given by Eq.~(\ref{eqn:so-n-algebra-gens}). The eigenstates of $\mathcal{H}(k)$ in Eq.~(\ref{eqn:2pi-rot-Ham}) can be found in a globally continuous and real gauge,
\begin{equation}
| u_\alpha^i\rangle = \e{\alpha L_{j,j+1}} | u_0^i\rangle,\label{eqn:global-real-gag}
\end{equation}
where state $| u_0^i\rangle$ at position $\alpha\!=\!0$ has component $1$ at position $i$ and zeros elsewhere. The reality of the states implies that $[\mcbsA_\mcI(\bs{k})]_\alpha$ is \emph{real}, which in combination with the skew-symmetric property implies zeros on its diagonal. This manifests that for $\mcP\mcT$-symmetric systems in real gauge $\mcbsA_\mcI(\bs{k})$ is a $\mathfrak{so}(N)$-valued 1-form, In the gauge of Eq.~(\ref{eqn:global-real-gag}), the BWZ connection is globally defined on $\Gamma$, and takes the form
\begin{equation}
[\mcbsA_\forall(\alpha)]_\alpha = \left(\begin{array}{cccc}
\!\triv_{j-1}\!\! &   0   &   0   &   0  \\
0   & 0   &   -1 & 0   \\
0   & 1   &   0   & 0 \\
0   &   0   &   0   & \!\!\triv_{N-j-1}\!\! 
\end{array}\right)
\end{equation}
\end{subequations}
where the subscript ``$\forall$'' indicates that we considered all $N$ bands. 
It is easy to check that the exponentiation in Eq.~(\ref{eqn:Wilson-using-BWZ}) leads to a trivial Wilson-loop operator on path $\Gamma$ for all index sets. In other words, the quaternion charge $-1$ is completely invisible for Wilson operators.

Nevertheless, the Hamiltonian in Eq.~(\ref{eqn:2pi-rot-Ham}) \emph{cannot} be continuously deformed into the atomic limit $\mcH(k) = \mathcal{E}$ without closing the gap, because the assignment
\begin{equation}
S^1 \ni \; \alpha \mapsto \e{\alpha L_{i,j}}\; \in \mathsf{SO}(N)
\end{equation}
produces a $2\pi$-rotation inside $\mathsf{SO}(N)$, which corresponds to a \emph{non-trivial} element of $\pi_1[\mathsf{SO}(N)] = \ztwo$. Capturing the quaternion charge $n_\Gamma \!=\!-1$ thus requires distinguishing a $2\pi$-rotation from the identity. We show in the next subsection that this can be achieved by lifting the rotations into the covering space $\mathsf{Spin}(N)$. 

We note that a $2\pi$-rotation remains homotopically non-trivial upon increasing the dimension of the real Hilbert space, i.e.~upon including additional trivial bands. Therefore, the topological phase characterized by $n_\textrm{BZ}=-1$, missed by conventional Wilson-loop techniques, survives in the \emph{stable limit} of many bands.

\subsection{Parallel transport of coordinate frames}\label{sec:calculating-the-charge}

In the previous subsection we concluded that identifying the charge $\overline{\mathsf{P}}_N$ acquired along a closed path $\Gamma$ requires describing the rotation of frames $\mathsf{u}(\bs{k})$ inside the covering space $\textsf{Spin}(N)$. Furthermore, since the group $\overline{\mathsf{P}}_N$ is non-commutative, it is not possible to express the value of the charge using $\reals$ or $\cmplx$ numbers. Instead, we express the generalized quaternion charge as a \emph{matrix group}. In this subsection, we develop these ideas within a solid mathematical framework.

Assuming we have fixed the basis $\{\ket{i}\}_{i=1}^N$ of the real Hilbert space, we encode the Hamiltonian $\mcH: \textrm{BZ}\backslash{\mathcal{N}}_\mcH \to M_N$ using a (right-handed and orthonormal) frame field $\mathsf{u}(\bs{k})$. However, a globally defined continuous gauge for $\mathsf{u}(\bs{k})$ may not exist due to the presence of NLs. Therefore, it is appropriate to use the language of fiber bundles. Then $\mathsf{u}(\bs{k})$ can be seen as a local section of an orthonormal frame bundle on $\textrm{BZ}\backslash{\mathcal{N}}_\mcH$. However, it turns out that an easier description is possible. The reason is that the non-uniqueness of $\mathsf{u}(\bs{k})$ for a fixed standard frame is fully captured by a \emph{discrete} point group $\mathsf{P}_{N}$, cf.~Sec.~\ref{subsec:gen-many}. This should be contrasted with the usual $\mathsf{SO}(N)$ gauge group of a typical orthonormal frame bundle. Especially, the {discreteness} of the \emph{gauge group} $\mathsf{P}_{N}$ implies that the only continuous gauge transformations of $\mathsf{u}(\bs{k})$ are locally \emph{constants} in $\mathsf{P}_{N}$. Therefore, the easier description of a given band-structure is via a \emph{principal bundle}
\begin{equation}
E \to \textrm{BZ}\backslash{\mathcal{N}}_\mcH
\end{equation}
with a discrete gauge group ${\mathsf{P}}_{N}$.

To describe the rotation of a frame $\mathsf{u}(\bs{k})$ along a closed path $\Gamma$, we introduce the \emph{affine connection}~\footnote{The components of the connection are called \emph{Christoffel symbols} in the related problem of a metric connection on a tangent bundle~\cite{Carroll:2003}. They are usually indicated by $\Gamma$. We write them with $\mathrm{F}$ while keeping $\Gamma$ exclusively for closed paths.}
\begin{subequations}
\begin{equation}
\nabla_a \bs{\mathsf{u}}_j(\bs{k}) = \Gamma^i_{\phantom{i}aj}(\bs{k}) \bs{\mathsf{u}}_i(\bs{k}) \label{eqn:connection-def}
\end{equation}
where indices $a,b,\ldots$ refer to components of momenta, while indices $i,j,\ldots$ indicate eigenvectors of $\mcH(\bs{k})$ ordered into the frame $\mathsf{u}(\bs{k})$. Although we are now using a different language than in Sec.~\ref{sec:Wilson-failure}, we have just reformulated the same mathematical considerations using slightly different objects. It can be easily shown that for $\mcP\mcT$-symmetric systems
\begin{equation}
\Gamma^i_{\phantom{i}aj}(\bs{k}) = [\mcA_\forall(\bs{k})]^i_{\phantom{i}aj},
\end{equation}
\end{subequations}
i.e. the affine connection is just the BWZ connection over \emph{all} the bands. The reality condition on $\mcH(\bs{k})$ and $\mathsf{u}(\bs{k})$ implies certain analogies to parallel transport of frames in general relativity. Throughout the remainder of this section, we use the $\mcA_\forall$ notation, and we call it the \emph{affine BWZ connection}. We remark that the discreteness of the gauge group makes the components $[\mcA_\forall(\bs{k})]^i_{\phantom{i}aj}$ of the connection \emph{gauge invariant}.

The rotation of the frame along a closed path $\Gamma$ can be expressed as
\begin{equation}
R(\Gamma) = \overline{\textrm{exp}}\left(\oint_\Gamma \mcbsA_\forall(\bs{k})\cdot \de\bs{k} \right)\in\mathsf{P}_N.\label{eqn:par-transpo-so-n}
\end{equation}
We remark that although a \emph{global} real and continuous gauge for $\bs{\mathsf{u}}_i(\bs{k})$ may be absent in the presence of NLs, we have just argued that $\mcbsA_\forall(\bs{k})$ is gauge invariant. The connection is therefore continuous and single-valued even when we describe eigestates on $\Gamma$ using multiple sections. The quaternion charge $n_\Gamma \!=\! -1$ corresponds to a $2\pi$-rotation of the frame, implying $R(\Gamma) = \unit$. The affine BWZ connection allows us to perform the lift to $\overline{\mathsf{P}}_N$ and thus distinguish the $2\pi$-rotation from the identity. This is achieved by replacing in the decomposition of the connection the basis of Lie algebra $\mathfrak{so}(N)$ by the basis of $\mathfrak{spin}(N)$, while keeping the same coefficients, i.e.~\cite{Steane:2013} 
\begin{subequations}
\begin{equation}
L_{i,j} \mapsto t_{i,j} = -\tfrac{1}{4}[\mathbb{\Gamma}_i,\mathbb{\Gamma}_j] \in\mathfrak{spin}(N),\label{eqn:lie-algebra-lift}
\end{equation}
where $\{\mathbb{\Gamma}_i\}_{i=1}^N$ are Dirac matrices of dimensions $2^{\lfloor{N}/{2}\rfloor}\times 2^{\lfloor {N}/{2}\rfloor}$ which mutually anticommute, i.e. $\{\mathbb{\Gamma}_i,\mathbb{\Gamma}_j\} = 2 \delta_{ij}$. Decomposing the affine BWZ connection into the $\mathfrak{so}(N)$ basis as
\begin{equation}
[\mcA_\forall(\bs{k})]_a = \sum_{i<j} \beta^{ij}_a(\bs{k}) L_{i,j},
\end{equation}
we obtain the $\mathfrak{spin}(N)$-valued 1-form, which we call \emph{spin BWZ connection}, as
\begin{equation}
[\overline{\mcA_\forall}(\bs{k})]_a = \sum_{i<j} \beta^{ij}_a(\bs{k}) t_{i,j}.
\end{equation}
Especially, the lift of $\pi$-rotations that generate the group $\overline{\mathsf{P}}_N$ in Eq.~(\ref{eqn:Salingaros-def}) is obtained by
\begin{equation}
e_{j-1} : = \e{-\tfrac{\pi}{4}[\mathbb{\Gamma}_1,\mathbb{\Gamma}_j]} = \tfrac{1}{2}[\mathbb{\Gamma}_j,\mathbb{\Gamma}_1]\label{eqn:Dirac-spin-Clifford}
\end{equation}
for $2 \leq j \leq N$. It is easy to check that the assignment in Eq.~(\ref{eqn:Dirac-spin-Clifford}) indeed produces the basis of Clifford algebra $C\!\ell_{0,N-1}$. We have thus described the group $\overline{\mathsf{P}}_N$ as a matrix algebra. The generalized quaternion charge acquired along a closed path $\Gamma$ can be expressed as
\begin{equation}
n_\Gamma \equiv \overline{R}(\Gamma) = \overline{\textrm{exp}}\left(\oint_\Gamma \overline{\mcbsA_\forall}(\bs{k})\cdot \de\bs{k}\right)\in \overline{\mathsf{P}}_N. \label{eqn:gene-quat-charge}
\end{equation}
\end{subequations}
We remark that in Eq.~(\ref{eqn:gene-quat-charge}) the horizontal bar in ``$\overline{\textrm{exp}}$''  indicates path-ordering, while all the other horizontal bars indicate objects that have been lifted from $\mathsf{SO}(N)$ to $\mathsf{Spin}(N)$.

\begin{figure}[t!]
	\includegraphics[width=0.25\textwidth]{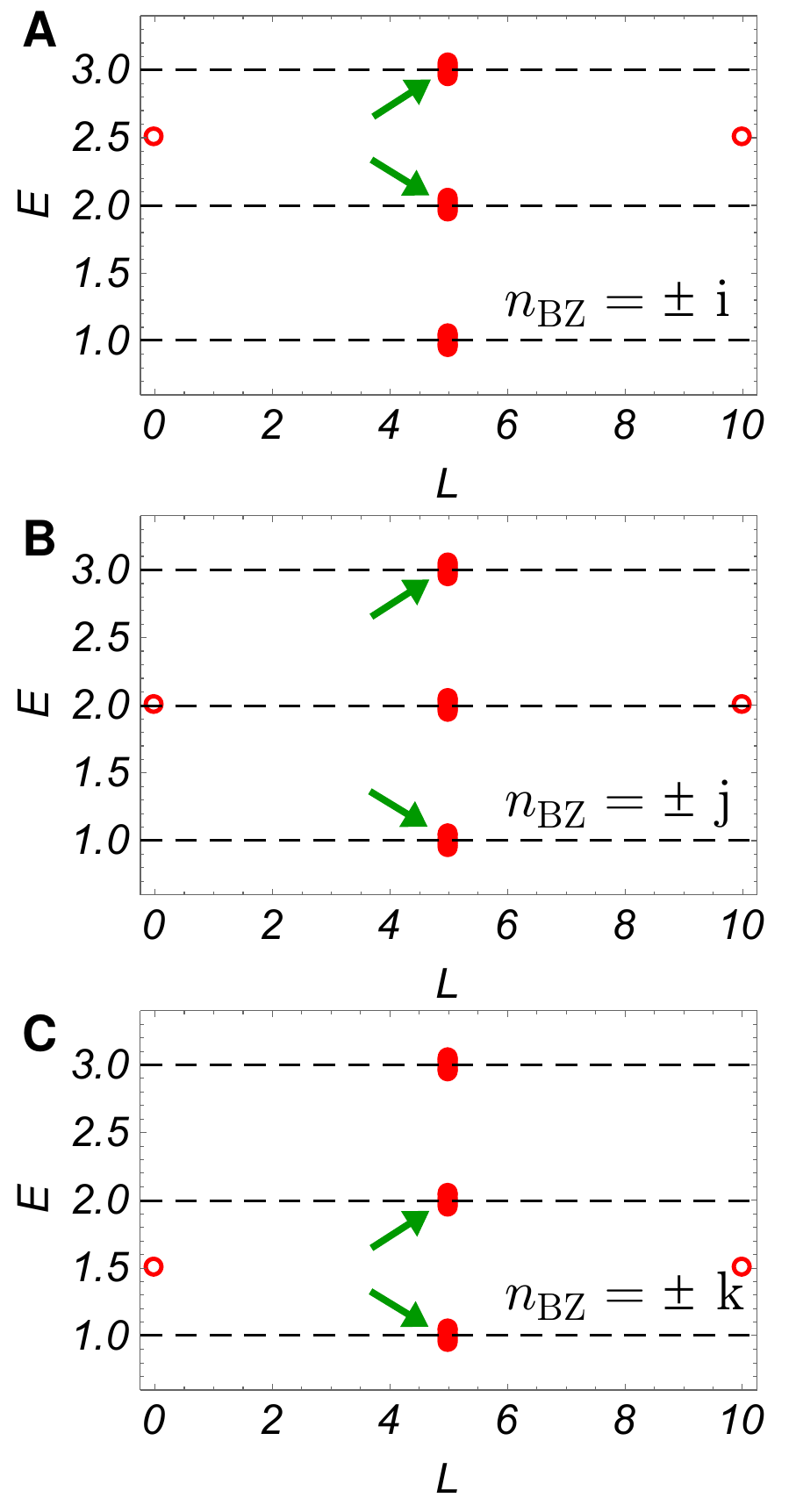}
	\caption{\textsf{Spectrum of a tight-binding Hamiltonian obtained from the $k$-space Hamiltonian in Eq.~(\ref{eqn:i-j-k-hams}). We considered a system with 11 sites and open boundary conditions. The value of the bulk quaternion invariant is indicated by $n_\textrm{BZ}$. Each red dot indicates one eigenstate. The vertical (horizontal) position of each dot represents the energy (center of mass) of the state. The pair of arrows (green) in each panel indicate the two bands that perform the $\pi$-rotation along the 1D Brillouin zone. We observe the formation of one mid-gap boundary state at each end of the 1D system. The horizontal dashed lines indicate the eigenvalues of the three bulk bands. }}
	\label{fig:i-j-k}
\end{figure} 

\begin{figure*}[t!]
	\includegraphics[width=0.99\textwidth]{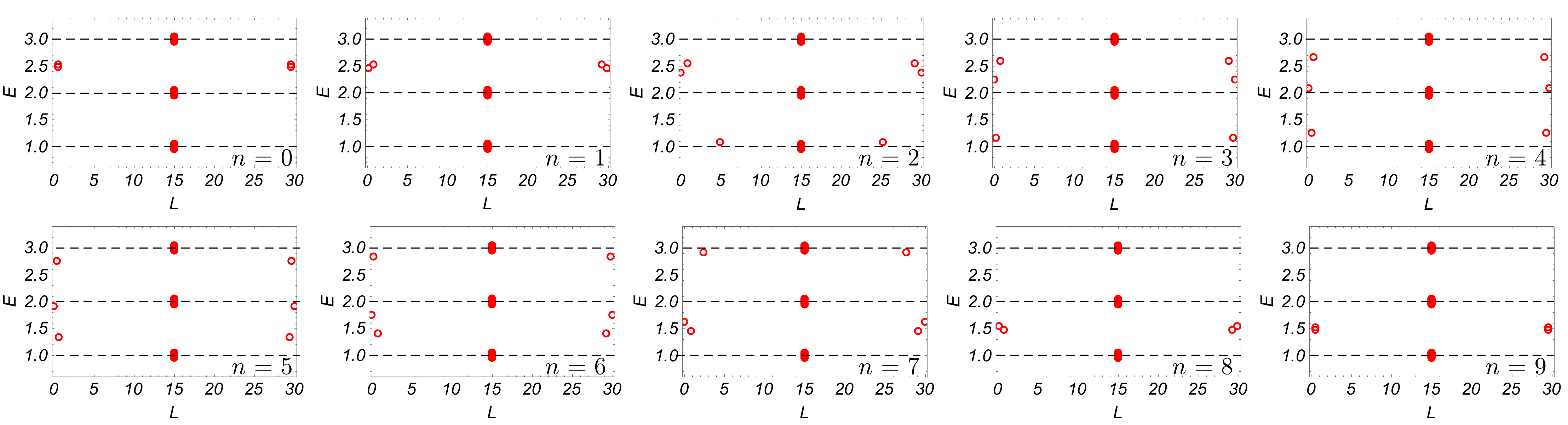}
	\caption{\textsf{Spectrum of a real space Hamiltonian $\hat{\mcH}$ with open boundary conditions, which was obtained from Eq.~(\ref{eqn:interpolate-i2-k2}) for $t=n\pi/18$ with integer $0 \leq n \leq 9$. Each read circle represents one eigenstate. The vertical (horizontal) coordinate of each dot indicates the energy (center-of-mass position) of the corresponding wave function. The dashed lines indicate the bulk band energies. For the discussion of the boundary states, see the text below Eq.~(\ref{eqn:interpolate-i2-k2}).}}
	\label{fig:i2-to-k2}
\end{figure*} 

For the case $N=3$, we numerically implemented the formula in Eq.~(\ref{eqn:gene-quat-charge}) by first rewriting the path-ordered exponential as a path-ordered product
\begin{subequations}
\begin{equation}
n_\Gamma = \! \ordprod_{\{\bs{k}_\alpha\}_{\alpha=1}^\Lambda} \!\!\!\!\e{\overline{\mcbsA_\forall}(\bs{k}_\alpha)\cdot \Delta\bs{k}} \! \label{eqn:BCH-numerically}
\end{equation}
where $\{\bs{k}_\alpha\}_{\alpha=1}^\Lambda$ corresponds to a sequence of $\bs{k}$-points ordered along path $\Gamma$, $\Delta\bs{k}$ is the (oriented) distance between two consecutive $\bs{k}$-points, and the number $\Lambda$ of $\bs{k}$-points should be large enough. The path-ordered product in Eq.~(\ref{eqn:BCH-numerically}) can be obtained iteratively. If we define
\begin{equation}
 \log\left( \! \ordprod_{\{\bs{k}_\alpha\}_{\alpha=1}^N} \!\!\!\!\e{\overline{\mcbsA_\forall}(\bs{k}_\alpha)\cdot \Delta\bs{k}} \! \right) \equiv X_N
\end{equation}
for $1 \leq N \leq \Lambda$, then one obtains a recurrence relation
\begin{eqnarray}
X_{N+1} \!&=&\! \log\left( \e{\overline{\mcbsA_\forall}(\bs{k}_{N+1})\cdot\Delta\bs{k}}\e{X_{N}} \right) \equiv \log\left( \e{\Delta X}\e{X_{N}} \right) \\
\!&\approx&\! \Delta X + X_{N} + \frac{1}{2}[\Delta X, X_{N}]\\
\!&\phantom{=}&\!\!+\frac{1}{12}\!\left(\left[\Delta X,\![\Delta X,X_N]\right]\!+\!\left[X_N,\![X_N,\Delta X]\right]\right)\!+\ldots \nonumber
\end{eqnarray}
\end{subequations}
where in the first step we simplified the notation by writing $\overline{\mcbsA_\forall}(\bs{k}_{N+1})\cdot\Delta\bs{k} \equiv \Delta X$, and in the second step we used the Baker-Campbell-Hausdorff formula to a desired order to approximate the product of exponentials of non-commuting matrices.

\subsection{Topological edge modes in 1D systems}~\label{sec:1D-Ham}

In this section, we numerically study the relation between the bulk quaternion invariant $n_\textrm{BZ}$ of 1D $\mcP\mcT$-symmetric three-band models, and the boundary modes of such systems. As discussed in Sec.~\ref{sec:calculating-the-charge} a Hamiltonian $\mcH$ (with standard energies $\mathcal{E}$) is uniquely fixed by a choice of an orthonormal frame $\mathsf{u}$, cf.~Eq.~(\ref{eqn:Ham-frames}). The frame can be further understood as an $\mathsf{SO}(3)$ rotation of the \emph{standard} frame $\hat{\mathsf{u}}=\unit$. In this section, we will conveniently express the Hamiltonian $\mcH$ using the $\mathfrak{so}(3)$ algebra,
\begin{equation}
\mcH(k) = R(k) \mathcal{E} R(k)^\top \quad \textrm{with} \quad R(k) = \e{\bs{\beta}(k)\cdot \bs{L}}
\end{equation}
where $k\in(-\pi,\pi]\equiv\textrm{BZ}$ (the first Brillouin zone), and $\bs{\beta}= (\beta_x,\beta_y,\beta_z)$ are components of the $\mathfrak{so}(3)$ elements in the basis $\bs{L}=(L_x,L_y,L_z)$ given in Eq.~(\ref{eqn:gens-so-3}).

Let us first consider 1D systems with the quaternion invariant with value $\imi$, $\imj$, $\imk$. This is achieved by taking
\begin{equation}
R(k) = \e{k L_{x,y,z}/2}.\label{eqn:i-j-k-hams}
\end{equation}
Assuming a system with a finite number of sites, we transform the Hamiltonian $\mcH(k)$ to real space using Fourier transformation while assuming open boundary conditions. We label the resulting real-space Hamiltonian as $\hat\mcH$. To avoid a huge degeneracy of the bulk flat bands (which causes certain problems for numerical calculations), we \emph{always} add a small bulk dispersion
\begin{equation}
\mcH'(k) = \tfrac{1}{20}\sin k \cdot \unit\label{eqn:flatness-perturb}
\end{equation}
to $\mcH(k)$ before performing the Fourier transformation.

We plot the numerically obtained spectrum of the real-space Hamiltonian $\hat\mcH$ in Fig.~\ref{fig:i-j-k}. Each eigenstate is represented by a red circle inside a two-dimensional plane. The vertical coordinate indicates the energy of the eigenstate, while the horizontal position represents the center-of-mass position of the corresponding wave function along the 1D system. To obtain the data presented in Fig.~\ref{fig:i-j-k}, we considered a system with $11$ sites, such that the center-of-mass positions lie inside the interval $L\in[0,10]$. We observe that invariant value $\imi$/$\imj$/$\imk$ leads to one boundary mode at each edge of the 1D system, located at energy $\tfrac{5}{2}$/$2$/$\tfrac{3}{2}$. This is easily understood from the fact that for the Hamiltonian given by Eq.~(\ref{eqn:i-j-k-hams}) one of the orbitals is decoupled from the rest, while the other pair of orbitals perform a $\pi$-rotation in the sense of Sec.~\ref{subsec:A11}. The $\pi$-rotation leads to the appearance of a mid-gap state at each edge, where by ``mid-gap'' we mean energy $\tfrac{1}{2}(j_1 + j_2)$ if the pair of bulk bands performing the $\pi$-rotation have energy $j_1$ and $j_2$.

\begin{figure*}[t!]
    \centering
    \includegraphics[width=17.5cm]{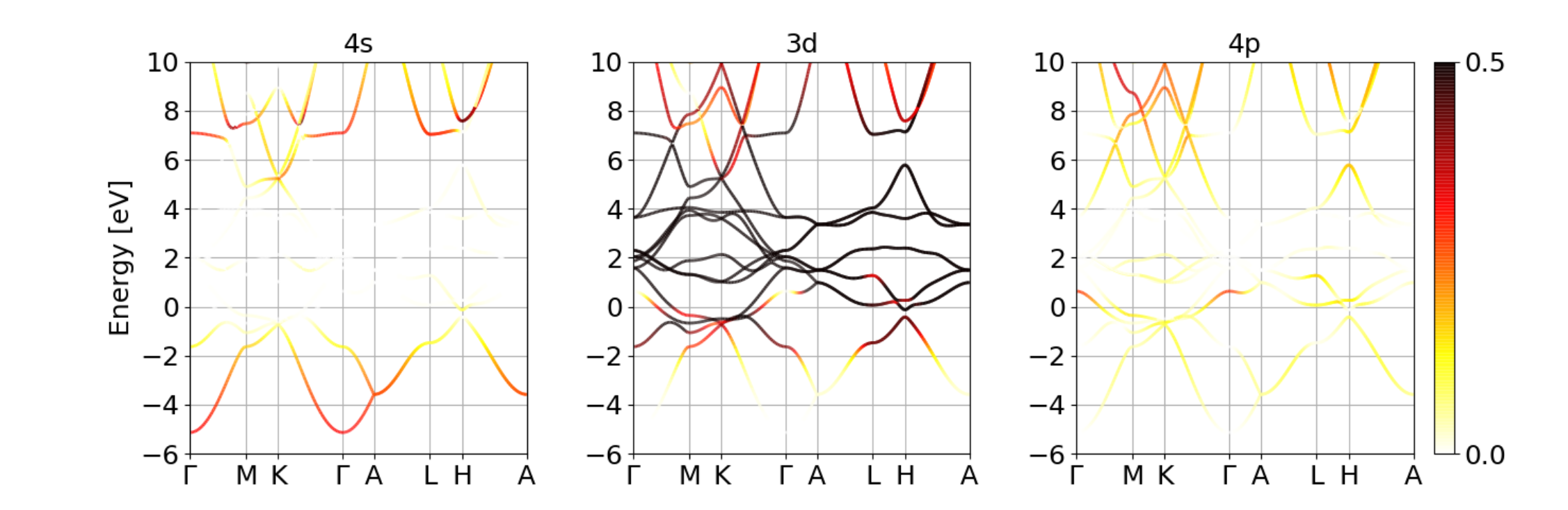}
    \caption{\textsf{
    The $4s$, $3d$, $4p$ atomic-orbital projected energy bands.}}
    \label{fig:fatband}
\end{figure*} 

One can similarly understand the edge modes appearing for Hamiltonians given by 
\begin{equation}
R(k) = \e{n k L_{i}/2}.\label{eqn:i-j-k-hams-2}
\end{equation}
where $i\in\{x,y,z\}$ and $n\in\intg$. Since for such Hamiltonians one of the orbitals is decoupled from the other two, Eq.~(\ref{eqn:i-j-k-hams-2}) describes an effectively two-band model. The effective two-band model has an additional \emph{chiral} symmetry (represented by $\sigma_y$ for $\mcP\mcT=\mcK$) anticommuting with $\mcH(k)$, implying that such systems belong to Altland-Zirnbauer class $\textrm{AIII}$ supporting a $\intg$ bulk invariant in 1D~\cite{Ryu:2010}. Therefore, the Hamiltonian given by Eq.~(\ref{eqn:i-j-k-hams-2}) supports $n$ mid-gap states. However, these states are not robust against including terms that couple \emph{all three} orbitals, since such terms break the accidental chiral symmetry. Nevertheless, we know from the theory of polarization of crystalline solids~\cite{Zak:1989,King-Smith:1993} that a \emph{single} edge state can still be stabilized by a non-trivial quantized value of the Zak-Berry phase $\phi_\textrm{ZB} = \pi$.

By virtue of the quaternion invariant $n_\textrm{BZ} = -1$ identified in our work, we find that a \emph{pair} of edge modes can also be stabilized at both edges even when the Zak-Berry phase of each band is trivial. To demonstrate this property, we consider a family of Hamiltonians given by
\begin{equation}
R(k;t) = k \left( L_x \cos t  + L_z \sin t \right)\label{eqn:interpolate-i2-k2}
\end{equation}
where $t\in[0,\tfrac{\pi}{2}]$ is a tunable parameter. The special case $t = 0$ ($t= \tfrac{\pi}{2}$) corresponds to a $2\pi$-rotation around the $x$-axis ($z$-axis), and is naturally interpreted as $n_\textrm{BZ} = \imi^2$ (as $n_\textrm{BZ} = \imk^2$). In accordance with the discussion in the previous paragraph, we expect such a Hamiltonian to exhibit a \emph{pair} of mid-gap states on each boundary between the upper (lower) two bands.

Importantly, the parametrization in Eq.~(\ref{eqn:interpolate-i2-k2}) continuously interpolates between the two special $\imi^2$ and $\imk^2$ interpretations of the $n_\textrm{BZ}=-1$ value of the quaternion charge. This allows us to demonstrate the stability of the two pairs of edge states even in the \emph{absence} of the accidental chiral symmetry. For a system with $31$ sites, the Hamiltonian $\hat\mcH$ obtained from Eq.~(\ref{eqn:interpolate-i2-k2}) with $t = n \pi/18$ and integer $0 \leq n \leq 9$ exhibits the eigenstates plotted in Fig~\ref{fig:i2-to-k2} [we again include the small perturbation given by Eq.~(\ref{eqn:flatness-perturb})]. Let us summarize our observations. For $t=0$, we find a pair of mig-gap boundary states at each edge of the 1D system between the \emph{upper} two bands. Increasing the value of $t$ splits the energy of the two states: The boundary state with \emph{increasing} energy eventually merges with the highest-energy band, while the state with \emph{decreasing} energy crosses the central band at $t \approx \tfrac{\pi}{4}$ and moves into the lower bulk gap. Simultaneously, the lowest-energy band ejects a boundary state at each edge of the system. The energy of this state moves upward with further increasing $t$. Ultimately, the two boundary states existing on each edge converge to mid-gap energy between the \emph{lower} two bands at $t=\tfrac{\pi}{2}$.

\begin{figure*}[p!]
    \centering
    \includegraphics[width=16cm]{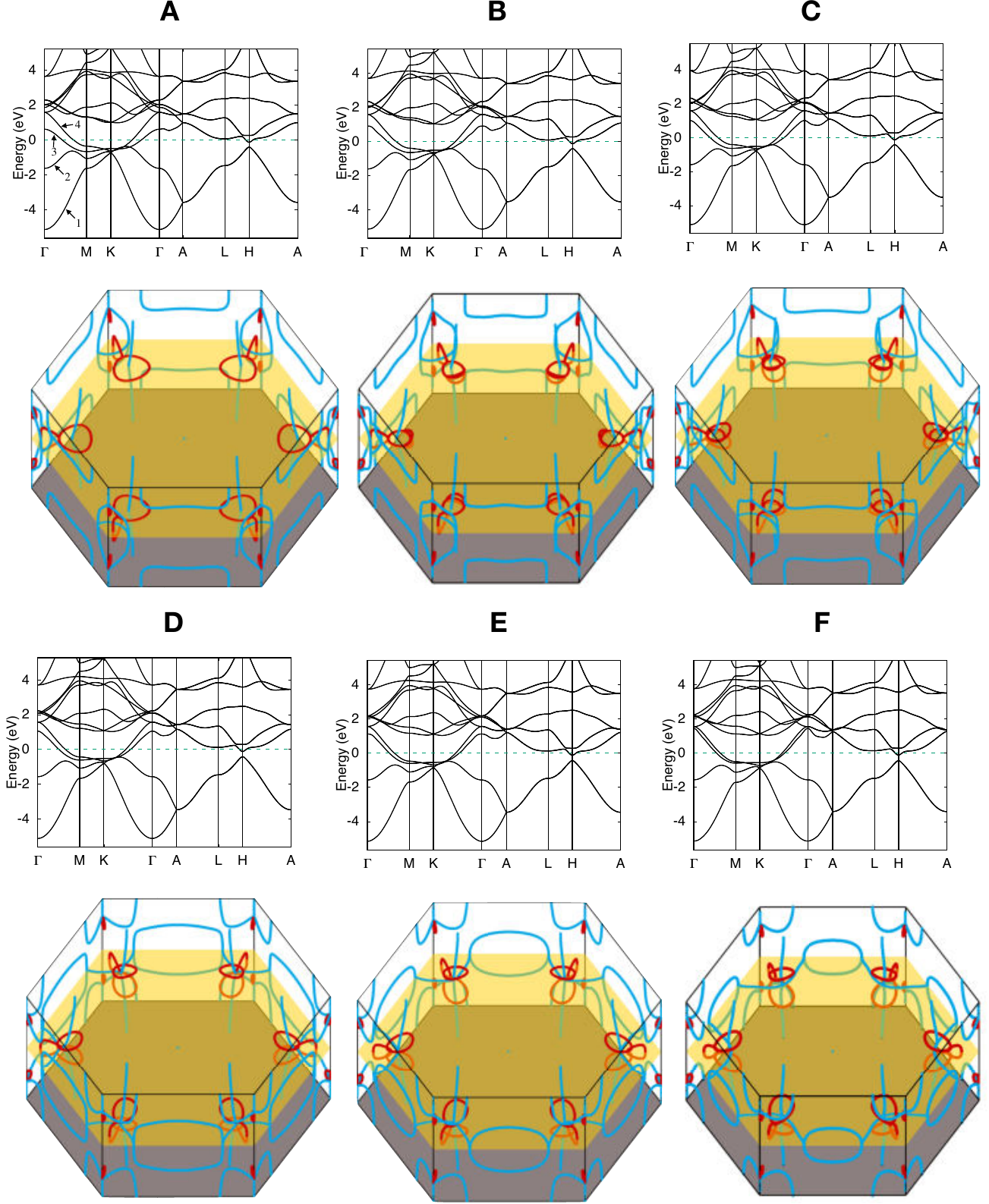}
    \caption{\textsf{Nodal lines evolution due to [0001] uniaxial strain. The band structure and the nodal line plot are combined as a pair. The $\varepsilon_{zz}$ strain values corresponding to panels (A--F) are 0\%, -1.5\%, -2\%, 
    -3\%, -4\%, -5\%, respectively. The band indices according to the energy values are indicated in panel (A). The red lines in the colored plots indicate band degeneracies (i.e.~nodal lines) between the 2nd and the 3rd energy band, while the blue lines correspond to nodes between the 3rd and the 4th energy band. The $k_z=0$ plane is indicated with a yellow sheet. The grey color plane in the bottom (top) of the BZ is a nodal surface due to non-symmorphic symmetry.}}
    \label{fig:evolve}
\end{figure*}   

\begin{figure*}[p!]
    \centering
    \includegraphics[width=0.99\textwidth]{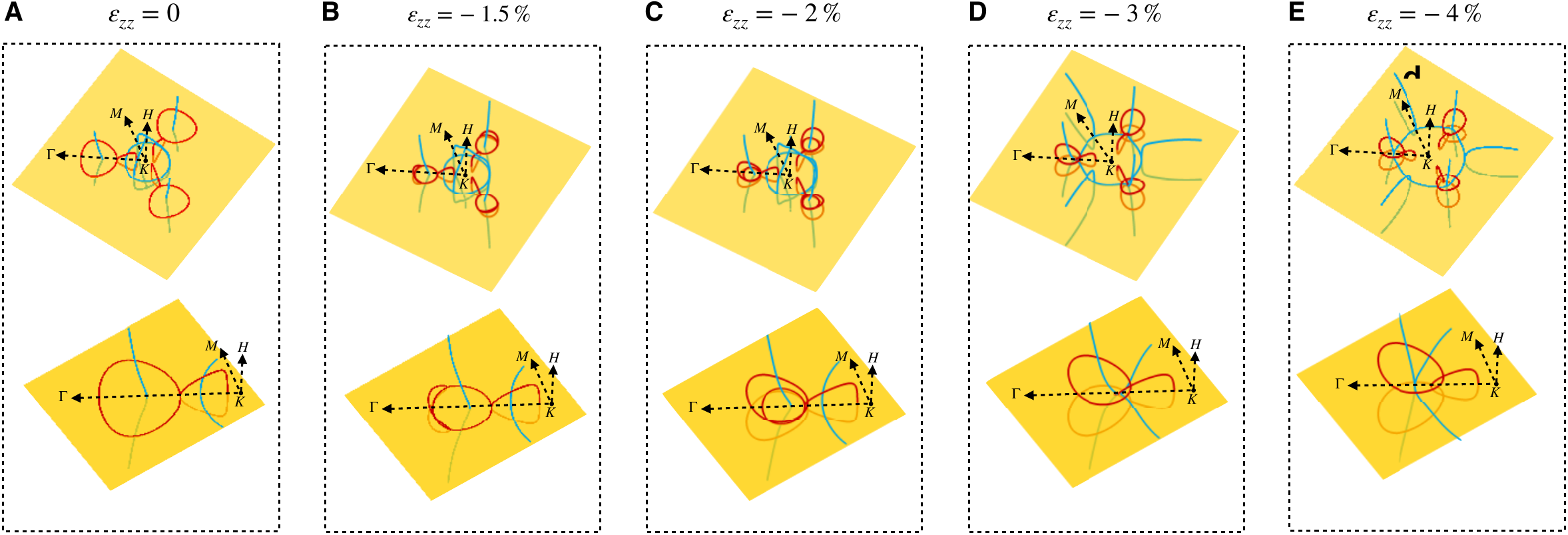}
    \caption{\textsf{Zoom-in of the nodal lines around the K point for $\varepsilon_{zz}$ strain values 0\%, -1.5\%, -2\%, -3\% and -4\%. For each dashed box, the top plot is the zoom-in plot around the K point, while the bottom one is a further zoom-in of the top plot.}}
    \label{fig:evolve-zoom}
\end{figure*} 

\begin{figure*}[p!]
    \centering
    \includegraphics[width=16cm]{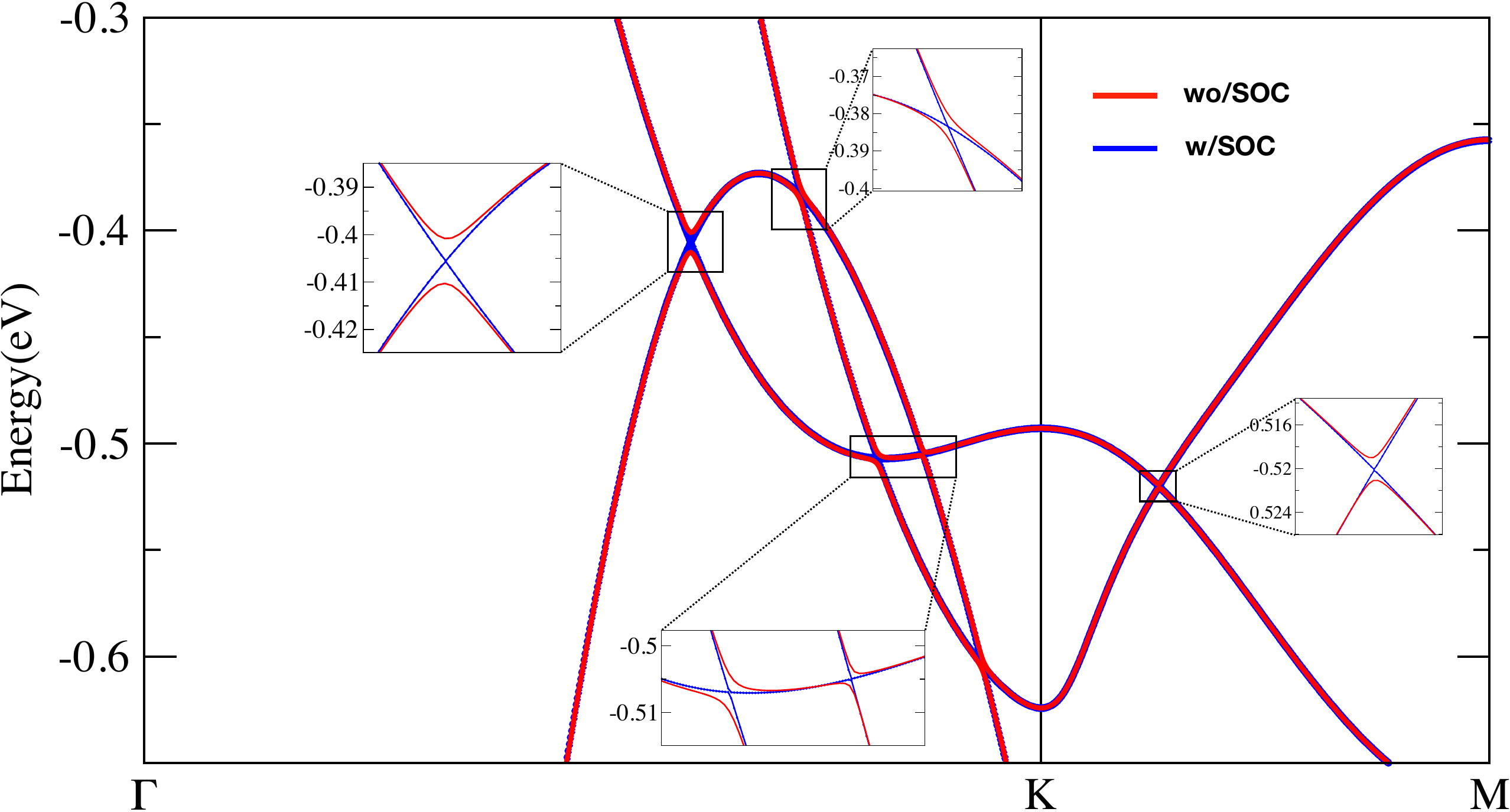}
    \caption{\textsf{Band structures of hcp-Sc with and without SOC. The inset figures show the detailed gap opening due to the SOC.}}
    \label{fig:bs-soc-nsoc}
\end{figure*}  

\begin{figure*}[p!]
    \centering
    \includegraphics[width=17cm]{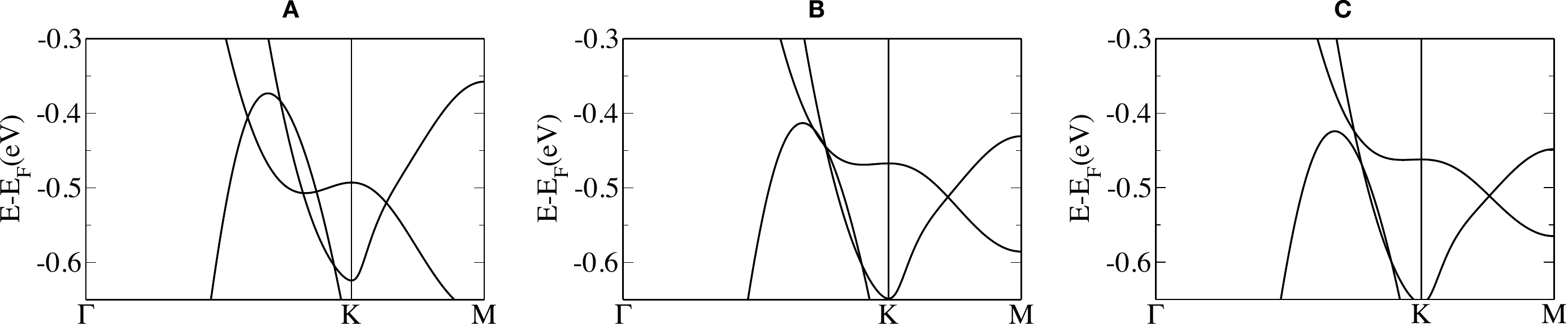}
    \caption{\textsf{\edit{Band structures of hcp-\textsf{Sc} under biaxial tensile strain without SOC. The lateral strains $\epsilon_{xx}=\epsilon_{yy}$ in panels (A), (B), (C) are 0\%, 2\%, and 2.5\%,  respectively.}}}
    \label{fig:bs-biaxial-nsoc}
\end{figure*}  

\begin{figure*}[t!]
    \centering
    \includegraphics[width=17.5cm]{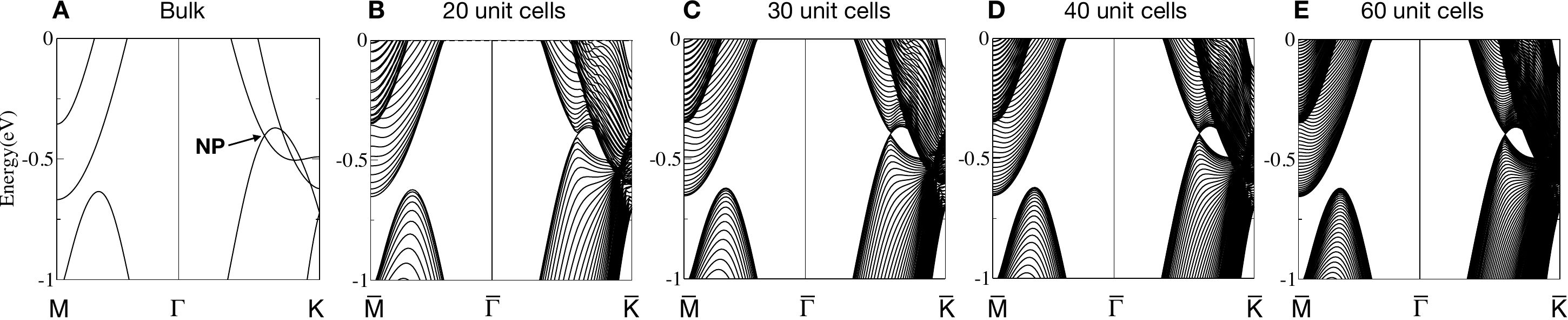}
    \caption{\textsf{\edit{Band structures of the bulk and of thin films of hcp-\textsf{Sc} without SOC calculated from VASP. The thickness of the thin films is indicated by the number of conventional unit cells in the headers of the subplot. All thin films are stacked along the [0001] direction of the bulk hcp-$\textsf{Sc}$. The thickness of each conventional unit cell (which corresponds to two atomic layers of $\mathsf{Sc}$) is 5.157~\AA. The typical energy gaps of the thin films near the nodal point (NP), which lies along a nodal line of the bulk bands, are 22~meV (for 20 cells), 13~meV (for 30 cells), 9~meV (for 40 cells), and 5~meV (for 60 cells).}
    }}
    \label{fig:bs-thinfilm}
\end{figure*}  

We remark that although we observe two robust boundary states at each edge of a 1D system characterized by quaternion charge $n_\textrm{BZ} = -1$, we are currently not able to explain their appearance using a physical interpretation of the topological invariant. A direct connection between the topological invariant and the electronic properties, similar to the relation between the Zak-Berry phase and the bulk polarization~\cite{Zak:1989,King-Smith:1993}, is missing. Further research is required to elucidate the evolution of boundary states observed in Fig.~\ref{fig:i2-to-k2}.


\section{Details of \emph{ab initio} calculations}

Hexagonal close packed (hcp) scandium (also called $\alpha$-Sc \cite{Kammler2008}) has two Sc atoms per primitive unit cell. Our first-principles calculations are performed within the density functional theory framework using VASP (Vienna \emph{Ab initio} Simulation Package)~\cite{PhysRevB.54.11169, PhysRevB.59.1758}. The approach relies on all-electron projector augmented wave (PAW) basis sets~\cite{PhysRevB.50.17953} combined with the generalized gradient approximation (GGA) with exchange-correlation functional of Perdew, Burke and Ernzerhof (PBE)~\cite{PhysRevLett.77.3865}.
The cutoff energy for the plane wave expansion was set to 600~eV and a $\bs{k}$-point mesh of $24\times24\times12$ was used in the bulk calculations. The lattice constants are full relaxed to $a=3.3212 \textrm{\AA}$ and $c=5.1567 \textrm{\AA}$. The WannierTools code~\cite{wanniertools} was used to search the nodal line in the Brillouin zone based on the maximally localized Wannier functions tight-binding model~\cite{RevModPhys.84.1419} that was constructed by using the Wannier90 package~\cite{wannier90} with Sc $s,p,d$ atomic orbitals as projectors. Since our theory applies for systems with weak spin-orbit coupling (SOC), we didn't take SOC effects into consideration in our calculation. Such an approximation is justified by the small atomic number of scandium. Our result for the unstrained Sc was verified to be consistent with previously reported calculations~\cite{Das1976, Sichkar2012}. The band structure projected onto the $4s$, $3d$ and $4p$ atomic orbitals is shown in Fig.~\ref{fig:fatband}. We find that electrons close to the Fermi level mainly come from the $3d$ orbitals. However, there is a non-negligible $4p$ orbital weight hybridized with the $3d$ orbitals.

\begin{figure}[b!]
    \centering
    \includegraphics[width=0.485\textwidth]{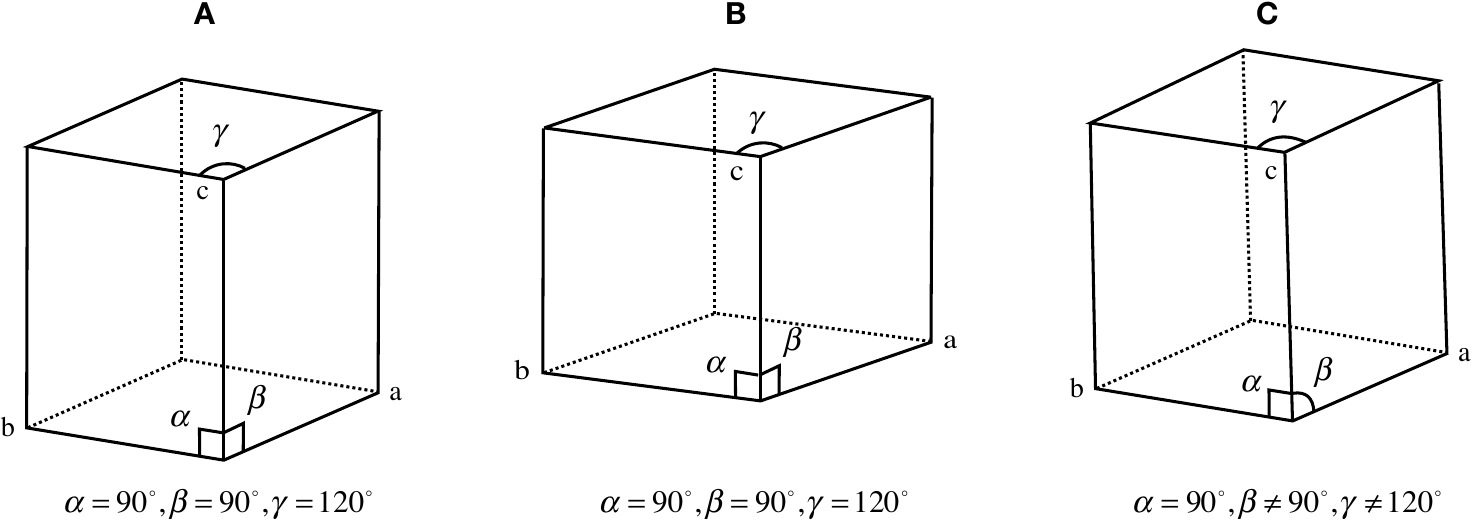}
    \caption{\textsf{(A) Lattice without strain. (B) [0001]-uniaxial strain which preserves symmetries of the space group $P6_3/mmc$. (C) A typical strain which only keeps the inversion symmetry. \edit{Since in the latter case we only want to demonstrate the crossing-point breaking effect, in the first-principles calculations we simply adjust the lattice angles to $\alpha=\ang{90}$, $\beta=\ang{90.3}$ and $\gamma=\ang{120.3}$}.}}
    \label{fig:strain}
\end{figure}  

\edit{To study the nodal lines of hcp-$\mathsf{Sc}$ under the influence of a symmetry-preserving strain, we consider two geometries.} First, a symmetry-preserving deformation is realized by applying [0001] \emph{uniaxial compressive strain} \edit{($\epsilon_{zz} <0$). In the first-principles simulations, we adjust the values of $\epsilon_{xx},\epsilon_{yy}<0$ such that the $\sigma_{xx},\sigma_{yy}$ components of the stress tensor vanish.} The evolution of the nodal lines due such a deformation strain is shown in Fig.~\ref{fig:evolve}. A zoom-in view of the evolution around the $\textrm{K}$ point is shown in Fig.~\ref{fig:evolve-zoom}. In order to reach the critical point where the ``blue'' and the ``red'' nodal chains reconnect, we need to apply approximatelly $\varepsilon_{zz} \approx -3\%$ uniaxial strain along the [0001] direction. 

\FloatBarrier

\edit{Alternatively, the reconnection of the nodal lines would in experiment be more easily achieved using a \emph{biaxial tensile strain} in the $x,y$-plane. In the first-principles simulations, we achieve this by fixing the magnitude of the lateral strains $\epsilon_{xx}=\epsilon_{yy}>0$ and by adjusting the value of $\epsilon_{zz}<0$ such that the $\sigma_{zz}$ component of the stress tensor vanishes. As we explain in the main text, such a geometry corresponds to thin films grown on a substrate with a lattice mismatch. In this setup, we find that the critical value of the biaxial tensile strain is approximately  $\epsilon_{xx}=\epsilon_{yy} \approx 2\%$. Such values of elastic strain are commonly achieved in experiments~\cite{King:2014,Burganov:2016,Catalano:2014,Ivashko:2019,Flototto:2018}. In Fig.~\ref{fig:bs-biaxial-nsoc}, it is showed that the effect of a $\epsilon_{zz} = -4\%$ uniaxial compressive strain in the [0001] direction is similar to a $\epsilon_{xx}=\epsilon_{yy} = 2.25\%$ biaxial tensile strain in the $x,y$-plane. We also study the band structure of thin films with different values of the finite thickness, see Fig.~\ref{fig:bs-thinfilm}). We find that for 60~conventional unit cells (i.e. 120 atomic layers, approximately 31~nm thick) the sub-band energy gap induced by finite size in the $z$-direction is approximately 5~meV, which is below the experimental resolution of the best existing ARPES instruments.}

We also study a deformation which \emph{breaks all the mirror and rotation symmetries} of hcp-$\textsf{Sc}$ while preserving the spatial inversion is obtained by applying a [$10\bar{1}1]$ strain. Since we only want to demonstrate the crossing-point breaking effect, for the sake of simplicity we just modify the lattice angles to $\alpha=\ang{90}$, $\beta=\ang{90.3}$ and $\gamma=\ang{120.3}$. The sketch of the resulting primitive unit cell is shown in Fig.~\ref{fig:strain}(C), and the resulting nodal lines are shown in Fig.~4(F--G) of the main text.

The nodal lines discussed in this work are obtained without the consideration of SOC. The comparison of the band structures with and without SOC from first-principle calculation is shown in Fig.\ref{fig:bs-soc-nsoc}. It is shown that the nodal lines will be gaped out under the consideration of SOC. The size of the gap is different for different band touching points. The gap we obtained is less than 10~meV.

It is worthy to mention that the nodal line structures discussed in Sc could also be found in the hexagonal-close-packed elemental metal crystal of Beryllium (Be), Magnesium (Mg), Calcium (Ca), Titanium (Ti), Zirconium (Zr) \emph{etc.}. However, the nodal lines in Be and Mg are above the Fermi level~\cite{Jain2013}; the hcp-Ca is stable only at high temperatures above 700K~\cite{Schottmiller1958}, and while the nodal lines in Ti and Zr are below the Fermi level~\cite{Jain2013}, the strain needed to realize the three-bands touching is too large. Nonetheless, there are many other materials not discussed here, that might provide the band structure optimally suited for simpler experimental verification of the predictions presented in this work.

\putbib

\end{bibunit}
\end{document}